\documentclass[superscriptaddress,aps,twocolumn,amsmath,amssymb,amsthm, dsfont,prb]{revtex4-1}
\usepackage{graphicx}
\usepackage{mathrsfs}
\usepackage{dcolumn}
\usepackage{bm}
\usepackage[colorlinks=true,citecolor=blue,linkcolor=magenta]{hyperref}
\usepackage{tikz}
\usepackage{hhline}
\usepackage{float}
\usepackage{scalerel}
\usepackage{amsfonts}
\usepackage{bbm}
\usepackage{color}
\usepackage{tabularx}
\usepackage{multirow}
\usepackage[caption=false]{subfig}
\newcolumntype{L}{>{\centering\arraybackslash}m{3cm}}
\newcolumntype{K}{>{\centering\arraybackslash}m{2.2cm}}
\usetikzlibrary{decorations.markings}

\makeatletter
\def\l@subsection#1#2{}
\def\l@subsubsection#1#2{}
\makeatother
\DeclareMathOperator{\im}{im}
\begin{document}

%\preprint{APS/123-QED}

\def\bra#1{\mathinner{\langle{#1}|}}
\def\ket#1{\mathinner{|{#1}\rangle}}
\newcommand{\Ket}[1]{\vcenter{\hbox{$\displaystyle\stretchleftright{|}{#1}{\bigg\rangle}$}}}
\newcommand{\comment}[1]{{\color{red} (#1)}}

\title{Searching for Fracton Orders via Symmetry Defect Condensation}

\author{Nathanan Tantivasadakarn}
\affiliation{Department of Physics, Harvard University, Cambridge, MA 02138, USA}
\author{Sagar Vijay}
\affiliation{Department of Physics, Harvard University, Cambridge, MA 02138, USA}

\begin{abstract}
We propose a set of constraints on the ground-state wavefunctions of fracton phases, which provide a possible generalization of the string-net equations \cite{LevinWen2005} used to characterize topological orders in two spatial dimensions.  Our constraint equations arise by exploiting a duality between certain fracton orders and quantum phases with ``subsystem" symmetries, which are defined as global symmetries on lower-dimensional manifolds, and then studying the distinct ways in which the defects of a subsystem symmetry group can be consistently condensed to produce a gapped, symmetric state.
We numerically solve these constraint equations in certain tractable cases to obtain the following results: in $d=3$ spatial dimensions, the solutions to these equations yield gapped fracton phases that are distinct as conventional quantum phases, along with their dual subsystem symmetry-protected topological (SSPT) states.  For an appropriate choice of subsystem symmetry group, we recover known fracton phases such as Haah's code \cite{Haah2011}, along with new, symmetry-enriched versions of these phases, such as non-stabilizer fracton models which are distinct from both the X-cube model and the checkerboard model in the presence of global time-reversal symmetry, as well as a variety of fracton phases enriched by spatial symmetries.  In $d=2$ dimensions, we find solutions that describe new weak and strong SSPT states, such as ones with both line-like subsystem symmetries and global time-reversal symmetry.  In $d=1$ dimension, we show that any group cohomology solution for a symmetry-protected topological state protected by a global symmetry, along with lattice translational symmetry necessarily satisfies our consistency conditions.
\end{abstract}

\maketitle

\tableofcontents
\section{Introduction}\label{intro}
Fracton phases are exotic, zero-temperature states of matter with gapped, point-like excitations that have fundamentally restricted mobility, despite the presence of translational symmetry.  As a consequence, completely gapped fracton phases also have a ground-state degeneracy that scales subextensively with the system size, and that remains robust against the addition of local perturbations \cite{VijayHaahFu2016}.  Fracton phases have attracted much recent interest as an exotic kind of fractionalization in higher dimensions, and for their potential use as a quantum memory \cite{BravyiHaah2013}.  

A diverse array of techniques have been successfully used to discover and characterize new fracton orders, including algebraic methods to search for stabilizer codes with exotic properties \cite{Haah2011,Haah2013,VijayHaahFu2015}, generalized gauge theories \cite{VijayHaahFu2016}, and tensor gauge theories to understand fracton phases with gapless excitations \cite{Pretko2017}.  Importantly, many fracton phases have been shown to be dual descriptions of quantum systems with ``subsystem" symmetries, which are defined as global symmetries along lower-dimensional manifolds \cite{VijayHaahFu2016}.  This connection has led to constructions of fracton phases using layers of two-dimensional, topologically-ordered states of matter \cite{MaLakeChenHermele2017,Vijay2017}.  Certain fracton phases are now known to exist on any three-manifold with a specified set of intersecting, sub-dimensional manifolds, which specify a so-called ``foliation" structure \cite{ShirleySlagleWangChen2018}.

\begin{figure}
    \centering
\includegraphics[]{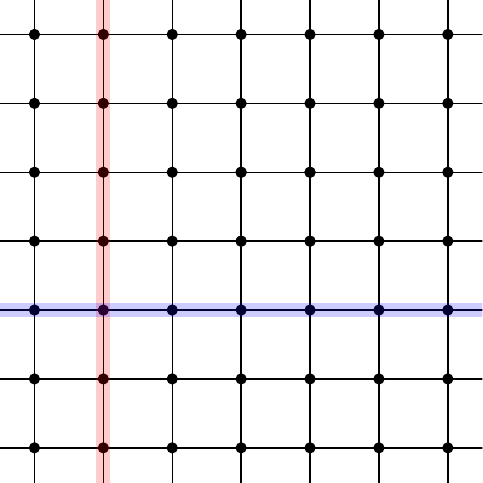}
    \caption{Subsystem symmetries act on lower-dimensional regions of the full lattice. Here, a two-dimensional square lattice is shown, with subsystem symmetries acting as spin flips over individual horizontal (blue) or vertical (red) lines. In three spatial dimensions, many  quantum phases with subsystem symmetries are known to be dual to models with fracton topological order \cite{VijayHaahFu2016}.}
    \label{fig:sspt}
\end{figure}

In this work, we exploit this duality between ($i$) fracton phases and ($ii$) quantum phases with subsystem symmetries, in order to derive a set of equations that constrain the wavefunctions of these exotic phases.  We then use these conditions -- termed the \emph{defect homology equations} -- as a starting point for finding new fracton orders in three spatial dimensions, as well as new kinds of quantum phases that are protected by subsystem symmetries, in both two and three spatial dimensions.  The latter -- subsystem symmetry protected topological (SSPT) phases -- are of interest in their own right as a generalization of conventional symmetry-protected topological (SPT) orders, and as a platform for measurement-based quantum computation, with certain quantum gates that can be implemented in a robust, protected manner throughout the entire phase\cite{Raussendorfetal2019,DevakulWilliamson2018,Stephenetal2019,DanielAlexanderMiyake2019}. Subsystem symmetries are also known to naturally arise in materials with both spin and orbital degrees of freedom \cite{KugelKhomski1982}.  An example of a subsystem symmetry in two spatial dimensions is illustrated in Fig. \ref{fig:sspt}.

\begin{figure}[t]
\centering
\includegraphics[scale=0.6]{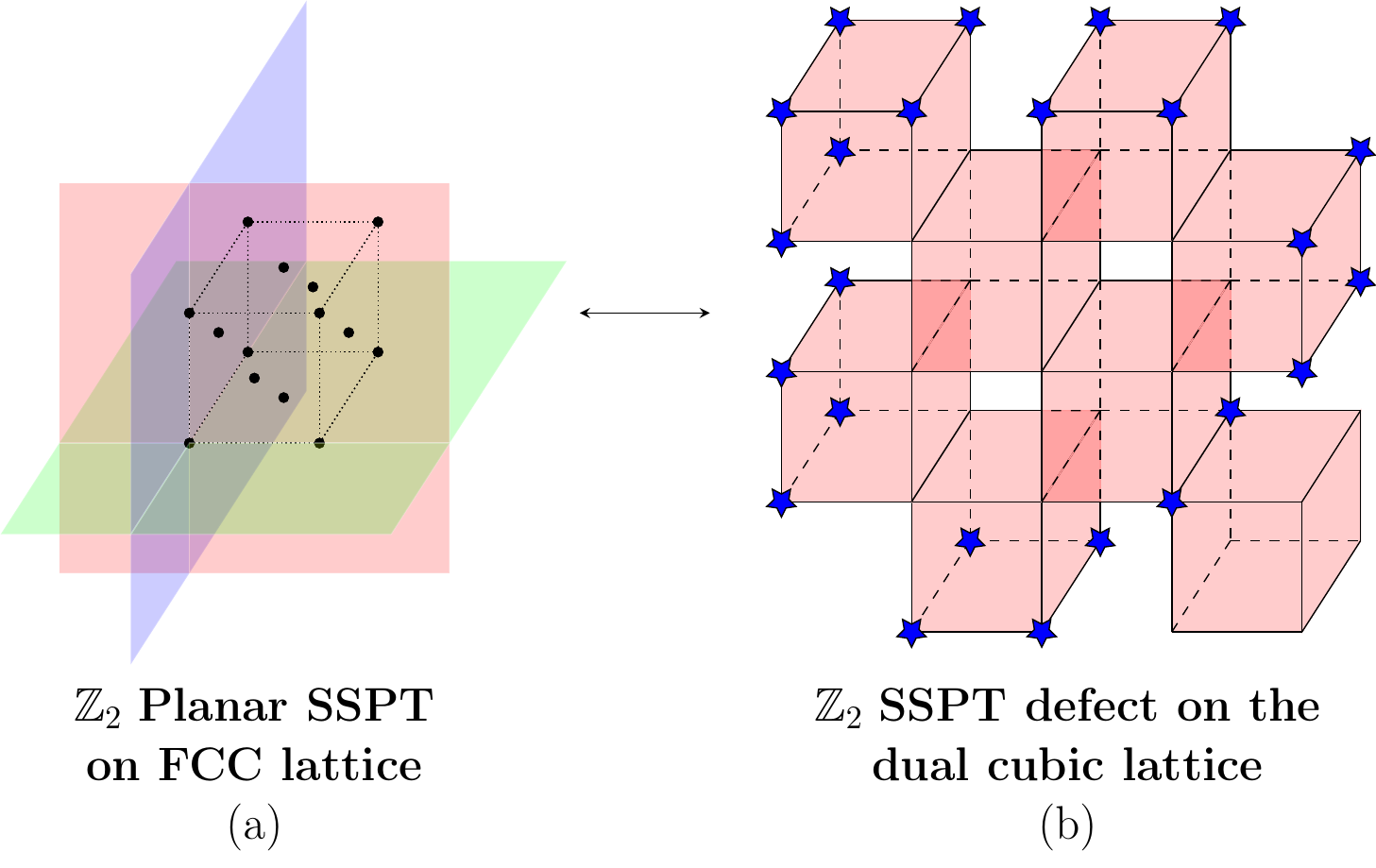}
\caption{{\bf Symmetry Defects in an SSPT with Planar Symmetries:}  We consider an SSPT state protected by planar $\mathbb{Z}_{2}$ symmetries along the (100), (010), and (001) planes along with the translation group of the face-centered cubic (FCC) lattice in (a).  The defects of this symmetry group live on the sites of a dual cubic lattice, and a possible configuration of defects is shown in (b).  These defects are constrained so that they can only be created in groups of eight at the corners of each of the colored cubes in (b), and quantum fluctuations hop such defects around the lattice. The number of defects in any configuration is always a multiple of four. For the time-reversal enriched checkerboard model, which is given by the non-stabilizer Hamiltonian Eq. \eqref{equ:Tenrichedcheckerboard}, each defect appears with a phase factor $e^{\pi i/4}$.} 
\label{fig:defectconfig}
\end{figure}

\begin{figure}[t]
\centering
\includegraphics[scale=0.6]{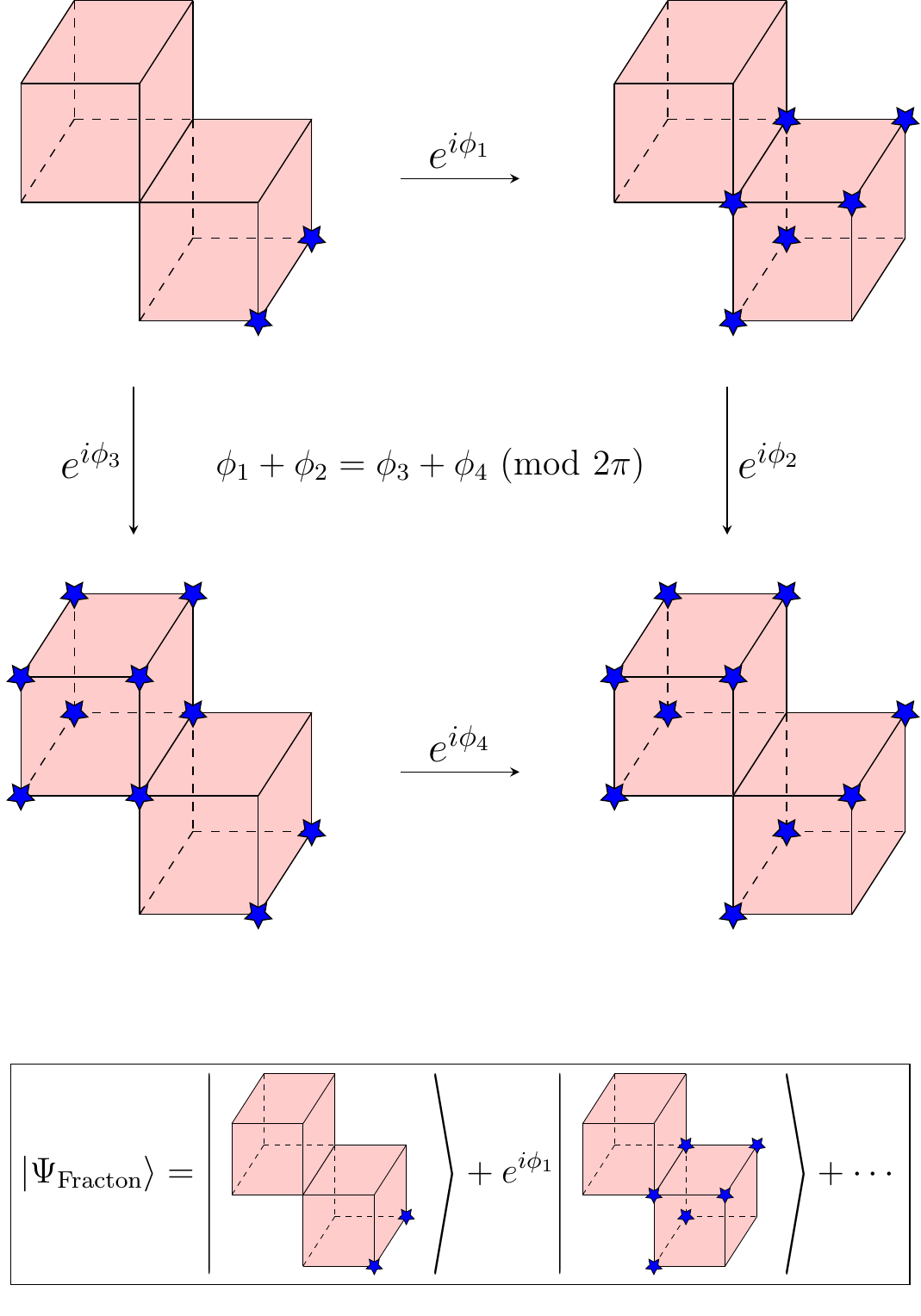}
\caption{{\bf Consistency Conditions:} An example of one of the consistent conditions for fluctuating defects in $\mathbb{Z}_{2}$ planar SSPT with symmetries along the (100), (010), and (001) planes in the FCC lattice, considered in Fig. \ref{fig:defectconfig}. The defect homology equations requires that $e^{i(\phi_1+\phi_2)} = e^{i(\phi_3+\phi_4)}$. The wavefunction for the dual fracton phase, given by a superposition of allowed defect configurations with these relative amplitudes, is shown schematically.}
\label{fig:defectconsistency}
\end{figure}

Our approach for finding fracton and SSPT orders is best understood by considering similar techniques that have been used to characterize SPT phases and conventional topological orders in two spatial dimensions. For (2+1)-dimensional SPT's that are protected by unitary global symmetries, the ground-state wavefunction can be viewed as a superposition of fluctuating domain walls of the symmetry group. After gauging this symmetry, the domain walls become closed loops of the fractionalized excitations in a topologically-ordered phase  \cite{LevinGu2012, LevinWen2005}. The relative amplitudes in the wavefunction for the topologically-ordered phase are naturally constrained by the requirement that these fractionalized excitations are deconfined.  These ``string-net" equations \cite{LevinWen2005} provide a starting point for searching for topological orders, and for understanding the essential data that characterizes distinct phases. Our defect homology equations arise from similar considerations, now applied to quantum phases that are protected by arbitrary subsystem symmetry groups. An example of a symmetry defect condensate and consistency conditions that gives rise to the checkerboard model is shown in Figures \ref{fig:defectconfig} and \ref{fig:defectconsistency}, respectively. The abstract relation between our approach and the conventional approach to understand SPT orders and their dual topological orders, is summarized in Table \ref{tab:generalizestringnet}.

After deriving the defect homology equations, we numerically solve these equations to find fracton phases and their dual SSPT states for various subsystem symmetry groups. This approach recovers known fracton orders, while also discovering new fracton phases, and provides a first step towards ``bootstrapping" fracton orders from a small set of input data.  The bottleneck in our numerical search arises from the fact that the defect homology equations are formulated on the lattice, with a translation symmetry group.  As a result, the number of equations to be solved scales exponentially with number of
neighbors.
A further reduction in the number of equations, by exploiting space group symmetries, may permit wider searches that may uncover new fracton models; this remains an ongoing topic of study.

We emphasize that the defect homology equations provide a constructive approach for searching for SSPT states, and their dual fracton orders. Nevertheless, like the solutions to the string-net equations, our solutions require additional interpretation in order to reveal the detailed topological properties of the phase. In particular, we find that many of the fracton solutions admit a defect decoration construction\cite{ElseThorngren2019,SongFangQi2018,Songetal2018,ShiozakiXiongGomi2018} akin to usual topological phases. This provides a complementary view to the solutions of the defect homology equations, which we thoroughly detail in subsequent sections.

\begin{table} 
    \caption{Summary of the generalization of the mathematical framework for (2+1)-dimensional SPT orders to SSPT orders.}
    \begin{tabular}{|K|L|L|}
    \hline
            & \textbf{SPT Order} & \textbf{SSPT Order}\\
\hline
    \textbf{Input Data} & Symmetry group & (Sub)system symmetry group \& lattice translation group  \\
    \hline
    \textbf{Result after gauging} & Twisted quantum double\cite{HuWanWu2013} & Topological order/Fracton order\\
    \hline
    \textbf{Mathematical Framework} & Group cohomology\cite{DijkgraafWitten,Chenetal2013} & Symmetry defect homology (Eq. \eqref{equ:defecthomology})\\
\hline
    \end{tabular}

    \label{tab:generalizestringnet}
\end{table}

\begin{table*}
\caption{ {\bf Summary of SSPT phases found using symmetry defect homology:} The onsite representation of the symmetry group is indicated in the ``symmetry" column. The number next to each ``type" of symmetry indicates the number of lines/planes intersecting a site, in cases where the subsystem symmetries act along lines or planes, respectively.  In all cases, translational symmetry is assumed, with an appropriate choice of unit cell, and time-reversal ($\mathcal T$) acts as complex conjugation. Color denotes new models found using the search, and red indicates non-stabilizer solutions. The dual fracton order of the \emph{trivial} SSPT is listed, and otherwise, the trivial SSPT is dual to a symmetry-breaking state. A detailed discussion of all of the fracton orders discovered is provided in Sec. \ref{sec:solving_defect_homology}.}
\begin{tabular}{|c|c|c|c|c|c|c|c|c|c|}
\hline
\multirow{ 2}{*}{$d$} & \multirow{ 2}{*}{Lattice} & \multirow{ 2}{*}{Symmetry} & \multirow{ 2}{*}{Type} & \multicolumn{2}{c|}{Without $\mathcal T$} & \multicolumn{2}{c|}{With $\mathcal T$}  &Dual fracton order of &\multirow{2}{*}{Section}\\
\cline{5-8}
&&&& Weak &Strong & Weak & Strong & trivial SSPT&\\
\hline
2 & Square &$\mathbb Z_2$ & Line $(\times 2)$ & $\mathbb Z_2 $ & $\mathbb Z_2 $ &  $\mathbb Z_2 $ &\color{blue}{$\mathbb Z_2^2 $}  & -- & \ref{2Dsquare}\\
\multirow{3}{*}{$\vdots$}  & Square & $\mathbb Z_2\times \mathbb Z_2$ & Line $(\times 2)$ & $\mathbb Z_2^2$ & \color{blue}{$\mathbb Z_2^5$}  &  $\mathbb Z_2^2$ & \color{blue}{$\mathbb Z_2^7$} & --&\ref{2DsquareZ2Z2}\\
 & Triangular & $\mathbb Z_2$ & Line $(\times 3)$&$\mathbb Z_2$ & \color{blue}{$\mathbb Z_2^3$}  & $\mathbb Z_2$ & \color{blue}{$\mathbb Z_2^4$} &--&\ref{2Dtriangle}\\
 & Triangular & $\mathbb Z_2$ & Fractal & 0 & 0 & 0 & 0 & --&\ref{NewmanMoore}\\
  & Square & $\mathbb Z_2$ & Fractal & $\mathbb Z_2$ & 0 & $\mathbb Z_2$ &\color{blue}{$\mathbb Z_2$}  & --&\ref{Fibonacci}\\
 \hline
3 & Cubic & $\mathbb Z_2$ & Line $(\times 3)$&$\mathbb Z_2$ &\color{blue}{$\mathbb Z_2^3$} & $\mathbb Z_2$ & \color{blue}{$\mathbb Z_2^4$} & --& \ref{3Dline}\\
\multirow{4}{*}{$\vdots$} & Cubic& $\mathbb Z_2$ & Plane $(\times 3)$& \color{blue}{$\mathbb Z_2^4$} &0 &  \color{blue}{$\mathbb Z_2^7$} & \color{red}{$\mathbb Z_2^2$} & X-cube\cite{VijayHaahFu2016}&\ref{Xcube}\\
 & FCC & $\mathbb Z_2$ & Plane $(\times 3)$& \color{blue}{$\mathbb Z_2^4$} & 0 & \color{blue}{$\mathbb Z_2^4$} & $\color{blue}\mathbb Z_2$ $\times$ \color{red}{$\mathbb Z_2$} & Checkerboard\cite{VijayHaahFu2016}&\ref{checkerboard}\\
 & Cubic & $\mathbb Z_2$ & Plane $(\times 4)$& \color{blue}{$\mathbb Z_2^3$} & 0 & \color{blue}{$\mathbb Z_2^5$}& 0  &Chamon\cite{Chamon2005,BravyiLeemhuisTerhal2011}(double model)&\ref{Chamon}\\
 & Cubic &$\mathbb Z_2$  & Fractal & $\mathbb Z_2$ & 0 & $\mathbb Z_2$ & \color{blue}{$\mathbb Z_2^2$} & Haah's code\cite{Haah2011}&\ref{Haah}\\
\hline
\end{tabular}
\label{tab:summary}
\end{table*}

\subsection{Summary of main results}
We now provide an outline of this work, and a detailed summary of our main results.  We begin, in Sec. \ref{sec:SPT_states} by reviewing how the classification of a simple SPT state in one spatial dimension, with a global Ising symmetry and translation, can be viewed in two complementary ways by ($i$) applying well-established techniques in group cohomology \cite{Chenetal2013} or ($ii$) by studying the consistent, symmetric wavefunctions that can be written as  condensates of the defects of this symmetry group.  We explicitly show how these approaches yield identical results.  The latter method, however, generalizes to other symmetry groups of interest, and to higher spatial dimensions.  

In Section \ref{sec:homology}, we derive the defect homology equations in generality.  We start from a quantum system with an onsite, unitary subsystem symmetry group $G$ along with translation and possible time-reversal symmetries.  Considering the self-consistent and symmetric wavefunctions that can be written as condensates of the {defects} of these symmetries yields the defect homology equations. These equations exhibit a gauge redundacy, due to the fact that certain relative amplitudes in a wavefunction are not observable; therefore, the solutions to these equations are to be studied modulo a gauge equivalence relation, which lies in the homology of a chain complex, which we term the \emph{symmetry defect homology}.   Furthermore, since $G$ is a subsystem symmetry group, our defect homology equations are naturally written with an underlying spatial lattice. Solving these equations is a numerically tractable exercise, if we introduce a unit cell in the lattice, and a translation symmetry group.  We also analytically demonstrate that in 1+1 dimensions, all SPT phases with a finite, unitary symmetry group that are classified by group cohomology \cite{Chenetal2013} also appear as solutions to our defect homology equations.

The solutions that we have obtained to the defect homology equations are discussed in detail in Sec. \ref{sec:solving_defect_homology}.  {These solutions yield } new kinds of fracton phases, along with their dual SSPT's that are protected by a combination of subsystem and global symmetries, and the input translational symmetry group, as summarized in Table \ref{tab:summary}.  In many cases of interest, we are able to show that the translational symmetry is not necessary to protect either the fracton phase or the SSPT.   In (3+1)-dimensions, the solutions to the defect homology equations yield new SSPT's which are dual to fracton models enriched by translational symmetry or time-reversal symmetry, and are discussed in detail in Sec. \ref{sec:solving_defect_homology}. We emphasize that the defect homology equations are exhaustive, in that they allow us to list all possible translationally-invariant fracton models with a certain spatial range of interactions. 

\subsection{Weak and Strong SSPT phases}
Finally, and before proceeding with the details of our procedure, we clarify some terminology pertaining to ``weak" and ``strong" subsystem symmetry protected topological phases that will be used in the remainder of this work. 
Given an onsite global symmetry $G$ and translational symmetry $\mathbb Z^d$, a crystalline SPT phase is defined as the equivalence class of $G \times \mathbb Z^d$-symmetric states under finite depth local unitaries that respect $G \times \mathbb Z^d$.

A natural extension of the above definition for a subsymmetry group $G$ is the following: a crystalline SSPT phase is the equivalence class of $G \rtimes \mathbb Z^d$-symmetric states under finite depth local unitaries that respect $G \rtimes \mathbb Z^d$. Here, the semidirect product denotes the fact that translation can shift the action of the subsystem symmetries.

Accordingly, we call a crystalline SSPT \textit{weak} if the wavefunction is related to a trivial product state via a finite depth local unitary, which necessarily breaks translation symmetry. Otherwise, we call the phase \textit{strong}. These definitions are a natural generalization of similar definitions of weak and strong phases, for crystalline phases with global symmetries.

We note that there exists a different definition of weak and strong SSPT's\cite{Youetal2018,DevakulWilliamsonYou2018,DevakulShirleyWang2019}, which uses an equivalence of states under linear-depth quantum circuits. Furthermore, the states and the quantum circuits do not need to be translationally-invariant. In (3+1)-dimensions, this definition is closely related to the definition of foliated fracton order \cite{ShirleySlagleWangChen2018} of the dual phase. In contrast, we believe that our definition of strong and weak SSPT's more likely matches the definition of translationally-invariant fracton phases\cite{PaiHermele2019}, whose classification differs from the former. In particular, we find that all of our weak SSPT's are considered trivial in that definition, while our strong SSPT's can either be considered strong or weak according to Refs. \onlinecite{Youetal2018,DevakulWilliamsonYou2018,DevakulShirleyWang2019}.

Table \ref{tab:summary} summarizes the group structure of the SSPT phases we find from calculating the symmetry defect homology. We would like to emphasize that although the number of phases we find is finite, under our definition of crystalline SSPT phases, we are unable to rule out the possibility that the complete classification under this definition might scale with the system size.

\section{Self-Consistent Wavefunctions for SPT phases}\label{sec:SPT_states}
To motivate our study, we first review the self-consistent wavefunction approach for SPT's with global symmetry. We focus on a particular example of a $\mathbb Z_2$ global symmetry, and show how our formalism coincides with the group cohomology classification. A general discussion for a finite unitary group $G$ is given in Appendix \ref{app:1DgeneralG}.

\subsection{Example: classification of translational invariant $\mathbb Z_2$ SPT's in 1+1D}

An important question that we would like address in this paper is the following: given an on-site unitary symmetry $G$, what are the possible wavefunctions with consistent rules for the creation, annihilation, fusing, and hopping of domain walls of the symmetry group? For a global Ising symmetry ($G \cong \mathbb{Z}_{2}$), one obvious solution is the product state
\begin{equation}
    \ket{\psi_0} = \bigotimes_i \ket{\rightarrow}_i,
\end{equation}
which in the $Z$ basis, is an equal-weight superposition of all $\ket{\uparrow}$ and $\ket{\downarrow}$ at every site. Thus, all the domain wall configurations do not have any relative sign.

The gapped Hamiltonian whose ground-state is the above product state is the paramagnet Hamiltonian
\begin{equation}
H = -\sum_i X_i.
\label{equ:1DHam}
\end{equation}
with $\mathbb Z_2$ symmetry given by $\prod_i X_i=1$.

In order to directly analyze the domain walls themselves rather than the spins, we perform the following duality on the Hamiltonian
\begin{align}
Z_iZ_{i+1} &\longleftrightarrow Z_{i+1}\\
X_i &\longleftrightarrow X_{i}X_{i+1}
\label{equ:KW}
\end{align}
Formally, this is the usual Kramers-Wannier duality followed by a transversal Hadamard rotation, but we will colloquially refer to it as the ``gauging map" and refer to the original model and its dual as the ``ungauged" and ``gauged" side of the duality, respectively\cite{Wegner1971,Kogut1979,VijayHaahFu2016,KubicaYoshida2018,Williamson2016}.  The duality is not defined on the entire Hilbert space. It only maps states where $\prod_i X_i=1$ in the ungauged side to those where $\prod_i Z_i=1$ in the gauged side.

The result of the gauging map is that we have swapped the roles of the spins and domain walls. The symmetry defects on the ungauged side are domain walls (violations of $Z_iZ_{i+1}$), which map to pairs of charges (or spin flips) of the dual $\mathbb Z_2$ symmetry. The fact that domain walls must be created in pairs in the ungauged side corresponds to the $\mathbb Z_2$ charge conservation in the gauged side. Hence, in the following discussion, ``domain walls" will refer to the spin flips in the gauged side, and are not to be confused with domain walls of the dual symmetry $\prod_i Z_i$. %Therefore like the 2+1D case, the fluctuation of domain walls can be viewed as the fluctuation of charge-conserving configurations in the gauged side.

As an example, gauging the paramagnet Hamiltonian \eqref{equ:1DHam} gives
\begin{align}
H_\text{gauged} = -\sum_i  X_{i}X_{i+1}.
\end{align}
The operator $X_{i}X_{i+1}$ allows pairs of domain walls to be created or annihilated, and allows them to individually hop around the chain. The ground-state wavefunction of this Hamiltonian is thus an equal weight superposition of an even number of domain walls. This is compactly written in the $X$ basis as
\begin{equation}
\ket{\psi_\text{gauged}} =\frac{1}{\sqrt{2}} \left( \ket{\rightarrow \rightarrow \rightarrow \cdots \rightarrow } + \ket{\leftarrow \leftarrow \leftarrow \cdots \leftarrow}  \right).
\end{equation}

Other symmetric states maybe built using the domain wall variables in the gauged side. Let 0 and 1 denote the absence and presence of domain wall configurations, respectively. We can define a more general ``hopping matrix"
\begin{align}
M_{i,i+1}=& e^{i \phi_{00}} \ket{11}\bra{00} + e^{i \phi_{01}} \ket{10}\bra{01} + \nonumber \\
&e^{i \phi_{10}} \ket{01}\bra{10}+ e^{i \phi_{11}} \ket{00}\bra{11}.
\label{equ:M1D}
\end{align}
This matrix acts on a two-site nearest-neighbor Hilbert space $i$ and $i+1$. It creates, annihilates, and hop domain walls, but up to an independent phase factor for each individual process. Note that when all angles are zero, $M_{i,i+1}=X_iX_{i+1}$ . We will assume that these phase factors are translationally invariant and that they only depend on the occupancy of the domain walls at these two sites. The corresponding gauged Hamiltonian is
\begin{align}
H'_\text{gauged} = -\sum_i  M_{i,i+1},
\label{equ:1DHamprime}
\end{align}
We further constrain the hopping matrix with the following conditions:

1. The following processes must accrue no phase: a creation followed by immediate annihilation of pairs, and hopping back and forth of a domain wall. This is enforced via $M_{i,i+1}^2=1$ to get rid of this sign and results in the following consistency equations\footnote{In general, $M_{i,i+1}^2=e^{i\theta}$, but we can redefine $M_{i,i+1} \rightarrow M_{i,i+1}e^{i\theta/2}$}
\begin{align}
\label{equ:con1}
\phi_{00} +\phi_{11}&= 0,\\
\phi_{10} + \phi_{01}&= 0.
\end{align}

2. $M_{i-1,i}M_{i,i+1}= M_{i,i+1}M_{i-1,i}$.  Using the first constraint, this is equivalent to $(M_{i-1,i}M_{i,i+1})^2=1$, which corresponds to the process of creating a pair of domain walls, hopping them both to the left or right, and annihilating the pair. This gives
\begin{align}
\phi_{00} + 2\phi_{10} + \phi_{11} &= 0,\\
\phi_{00} + 2\phi_{01} + \phi_{11} &= 0.
\label{equ:con2}
\end{align}

\begin{figure}
    \centering
    \includegraphics[scale=0.3]{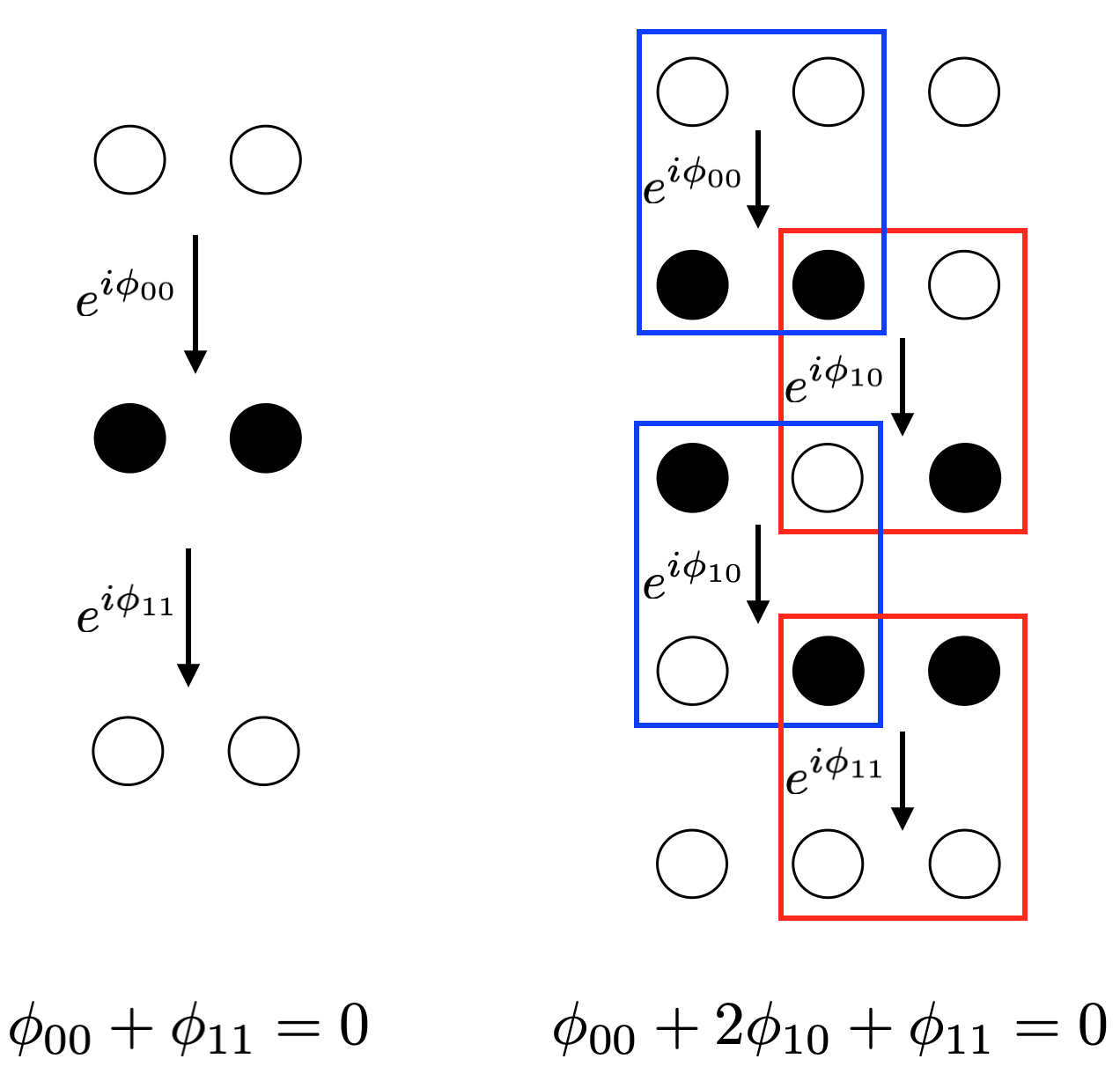}
    \caption{Examples of consistency conditions coming from $M_{i,i+1}^2=1$ (left) and $(M_{i-1,i}M_{i,i+1})^2=1$ (right). White and black circles denotes the states 0 and 1, respectively.}
    \label{fig:hopping}
\end{figure}
An example of such process is illustrated in Figure \ref{fig:hopping}. The consistency conditions above are naturally the conditions for a stabilizer Hamiltonian when $M_i$ are Pauli. In general, they describe a commuting projector Hamiltonian with terms generated by $\frac{1+M_i}{2}$. Therefore, the ground-state wavefunction of the Hamiltonian \eqref{equ:1DHamprime} is

\begin{equation}
\ket{\psi'_\text{gauged}} \sim \prod_i \left ( \frac{1+M_{i,i+1}}{2}\right ) \ket{000 \cdots 0},
\label{equ:gaugedwavefunction1D}
\end{equation}
where $\ket{000 \cdots 0}$ denotes the no domain wall configuration, and we have ignored the normalization factor. Because of the second constraint, the product is order independent.

We now solve the above consistency equations \eqref{equ:con1}- \eqref{equ:con2}, we find the following solutions:
\begin{align}
\phi_{01} =& \phi_{10} = 0;\pi & \phi_{00}&= -\phi_{11}.
\end{align}
In other words,
\begin{align}
\boldsymbol \phi=
\begin{pmatrix}
\phi_{00}\\
\phi_{01}\\
\phi_{10}\\
\phi_{11}
\end{pmatrix}
= n\pi 
\begin{pmatrix}
0\\
1\\
1\\
0
\end{pmatrix}
+ \theta \begin{pmatrix}
-1\\
0\\
0\\
1
\end{pmatrix}.
\end{align}
Since $n \in \mathbb Z_2$ and $\theta \in U(1)$, it ostensibly seems that we have a $\mathbb Z_2 \times U(1)$ worth of distinct solutions. However, not all solutions are inequivalent. In particular, on the gauged side, we are allowed to change the parameters $\phi$ that make up $M$ by conjugating with any finite depth circuit whose gates respect the $\mathbb Z_2$ symmetry $\prod_i Z_i$. Since we are assuming translational symmetry, it is necessary (but also sufficient) to restrict ourselves to consider unitaries that are product of onsite phase gates.

Returning to the example, the phase gate $R_\varphi = \exp \left(i\varphi \frac{1-Z}{2} \right)$ has the following onsite action:
\begin{align}\
\ket{0} &\rightarrow \ket{0}, & \ket{1} &\rightarrow e^{i\varphi}\ket{1}.
\end{align}
So by conjugating $M$ in Eq. \eqref{equ:M1D} with $R_\varphi$, we find the following equivalence relation
\begin{align}\label{equ:gauge1d}
\boldsymbol \phi
\sim
\boldsymbol \phi + \varphi
\begin{pmatrix}
-2\\
0\\
0\\
2
\end{pmatrix}.
\end{align}
And so we can always pick $\varphi=-\theta/2$ to set $\phi_{00}=\phi_{11} =0$. That is, we can always quotient out the continuous $U(1)$ part of the solution. However, the discrete $\mathbb Z_2$ part with non-trivial solution $\phi_{01}=\phi_{10}=\pi$ remains distinct.

Let us further analyze the interpretation of this non-trivial solution. Without loss of generality, we chose $\phi_{00}=\phi_{11} =\pi$. Writing $M_{i,i+1}$ in terms of Pauli operators, we find
\begin{equation}
M_{i,i+1}=-X_iX_{i+1},
\label{equ:crystallinesol1D}
\end{equation}
We have found a consistent hopping matrix for the domain walls, which is inequivalent to the trivial one. To return to the spin variables, we use the duality \eqref{equ:KW} to ``ungauge" the hopping matrix to $\tilde X_i = - X_i$. This is precisely the weak SPT protected by $\mathbb Z_2$ and translation where there is a $\mathbb Z_2$ charge sitting in every unit cell\cite{HuangSongHuangHermele2017}. Indeed, if either the $\mathbb Z_2$ symmetry or translation is broken, then it can be trivialized by conjugating with a transversal $Z$ gate without breaking the remaining symmetry.

Now consider imposing an extra time-reversal symmetry $\mathcal T=K$. The matrix $M$ now must be real, which means that the continuous variable $\theta$ is now restricted to $0,\pi$. Therefore,
\begin{align}
\boldsymbol \phi^\mathcal{T}
= n\pi 
\begin{pmatrix}
0\\
1\\
1\\
0
\end{pmatrix}
+ m\pi \begin{pmatrix}
1\\
0\\
0\\
1
\end{pmatrix}.
\end{align}

On the other hand, the phase gates $R_\varphi$ must also be real, so we are only allowed to conjugate $M$ with $R_\pi$, which acts as
\begin{align}\
\ket{0} &\rightarrow \ket{0}, & \ket{1} &\rightarrow -\ket{1}.
\end{align}
However, this gate leaves the hopping matrix invariant, so there is no non-trivial equivalence relation. As a result, we have two new solutions:
\begin{align}
\boldsymbol \phi^{\mathcal T}
=\pi \begin{pmatrix}
1\\
0\\
0\\
1
\end{pmatrix}, 
\pi \begin{pmatrix}
1\\
1\\
1\\
1
\end{pmatrix},
\end{align}
which cannot be trivialized as long as time-reversal is also preserved. We focus on the first solution, since both are in the same phase if we break translation. In terms of Pauli operators,
\begin{align}
M_{i,i+1} =  Y_{i}Y_{i+1}.
\end{align}
Indeed, without time-reversal, the unitary that turns $Y_{i}Y_{i+1}$ into $X_iX_{i+1}$ is just the transversal $R_{\pi/2}$ gate (the $S$ gate). We now prove that they cannot be connected without breaking time reversal by showing that they belong to different (symmetry broken) phases. This is most clearly seen from the action of time-reversal on the degenerate ground-state \eqref{equ:gaugedwavefunction1D}. For $M_{i,i+1}=Y_{i}Y_{i+1}$, complex conjugation maps a ground-state to an orthogonal state, while it does not for $M_{i,i+1}=X_iX_{i+1}$.

Ungauging $M$ now gives us the following stabilizer
\begin{align}
\tilde X_i = -Z_{i-1}X_iZ_{i+1}.
\end{align}
This is the negative of the 1D cluster state stabilizer, which realizes the ``Haldane phase" protected by the diagonal symmetry of $\mathbb Z_2 \times \mathbb Z_2^T$.

A few remarks are in order. First, the $\mathbb Z_2 \times \mathbb Z_2^T$ SPT we have constructed can in some sense be thought of as a decorated domain wall construction \cite{ChenLuVishwanath2014}, where domain walls of the $\mathbb Z_2$ symmetry are decorated with a phase $i$ (although there is no corresponding K\"unneth formula). This is equivalent to the fact that $\phi_{00} = \phi_{11}=\pi$ i.e. creating a pair of domain walls comes with a minus sign.

Second, we do not obtain the SPT phase protected purely by $\mathcal T=K$, since it requires considering fluctuation time-reversal domain walls, which is outside the scope of this search.

\subsection{Comparison to group cohomology}

Having obtained these results, it is insightful to compare our results to the classification using group cohomology\cite{Chenetal2013}. Here, we will show in this example that for the case without time-reversal, our data exactly matches the group cohomology data, which we will first review.

\subsubsection{1+1D SPT's using group cohomology}
Bosonic SPT's with symmetry group $G$ in 1+1D are classified by the second cohomology group $H^2(G,U(1))$\cite{ChenGuWen2011,ChenGuWen2011-2,Schuchelat2011,Chenetal2013}. More precisely, on an open chain, a non-trivial SPT phase will exhibit an anomalous symmetry action on the boundary, in the sense that the symmetry can only be realized via a projective representation of $G$ instead of a linear representation \cite{ElseNayak2014}. We will review this classification in a slightly different viewpoint, which is when the system has no boundary.

Given a closed 1D chain which is symmetric under a group $G$ (which for simplicity, we will assume to be unitary and onsite), we can simulate a system with boundary by acting the symmetry $g\in G$ only in a certain region in the 1D chain. This will create two domain walls (i.e. symmetry defects) at the two edges of the region. Without loss of generality, we label the left and right domain walls $R(g^{-1})$ and $R(g)$, respectively, and focus on the right edge of the region.

Consider acting in the region with the symmetry $h$ followed by $g$. This creates domain walls $R(g)R(h)$ on the right boundary. We can compare it to the case where we instead act $gh$ in the region, which creates $R(gh)$ on the right boundary. In general, fusing the domain walls $R(g)$ and $R(h)$ can differ from $R(gh)$ up to some $U(1)$ phase factor $\omega(g,h)$:
\begin{equation}
R(g)R(h) = \omega(g,h)R(gh).
\end{equation}
This is precisely the statement that the domain walls need only carry a projective representation of $G$.

The phase factor $\omega(g,h)$ must satisfy a certain consistency condition: fusing domain walls $R(g)$, $R(h)$, $R(k)$ in different orders must be associative. This gives
\begin{equation}
\omega(g,h)\omega(gh,k)=\omega(h,k)\omega(g,hk),
\label{equ:cocyclecondition}
 \end{equation}
called the cocycle condition. Furthermore, an equivalent projective representation can be obtained by independently modifying each domain wall $R(g)$ by a phase factor $\mu(g)$. Thus, this defines an equivalence class called a coboundary transformation
 \begin{equation}
\omega(g,h) \sim \omega(g,h)\frac{\mu(g)\mu(h)}{\mu(gh)}
\label{equ:coboundary}
 \end{equation}
for the function $\omega$. This equivalence class classifies inequivalent types of domain walls that a 1D chain with symmetry group $G$ can have, and is denoted by the group $H^2(G,U(1))$.

In addition to global symmetries, a 1D system on a lattice can be protected by a discrete translational symmetry. Consider a domain wall $R(g)_i$ in the unit cell labeled by $i$. Under the action of a unit-cell translation denoted by $T$, we are allowed to have
\begin{equation}
    T \cdot R(g)_i = \alpha(g) R(g)_{i+1}
\end{equation}
For some phase $\alpha(g)$. The consistency condition for $\alpha(g)$ comes from the commutativity of fusing two domain walls $R(g)$, $R(h)$ and translating them
\begin{equation}
     \alpha(g) \alpha(h) = \alpha(gh).
     \label{equ:cocyclealpha}
\end{equation}
Therefore, in addition to a projective representation $[\omega] \in H^2(G,U(1))$, there is an additional data given by a one-dimensional representation of $G$, which is classified by $[\alpha] \in H^1(G,U(1))$\cite{ChenGuWen2011}.

\subsubsection{Equivalence to group cohomology for $\mathbb Z_2$ SPT}
We will now show that the phases $\phi$ from our self-consistent wavefunction matches the data from group cohomology. First, we fix a certain gauge for $\omega$. Here, we will represent $G =\mathbb Z_2=\{1,g\}$. The cocycle conditions \eqref{equ:cocyclecondition} give us
\begin{align}
\omega(1,1)=\omega(1,g)=\omega(g,1),
\end{align}
and the coboundary transformations \eqref{equ:coboundary} are
\begin{align}
  \omega(1,1) &\sim \omega(1,1)\mu(1),\\
  \omega(1,g) &\sim \omega(1,g)\mu(1),\\
    \omega(g,1) &\sim \omega(g,1)\mu(1),\\
      \omega(g,g) &\sim \omega(g,g)\frac{\mu(g)^2}{\mu(1)}.
\end{align}
Thus, we see that we can always set $\omega(1,1)=\omega(1,g)=\omega(g,1)=1$. Physically, this means that we can always redefine the domain walls so that there is no phase associated to fusing the ``identity" domain wall $R(1)$ with other domain walls. Similarly, the cocycle condition \eqref{equ:cocyclealpha} for $\alpha$ gives
\begin{align}
\alpha(1)&=1, & \alpha(g)= \pm 1,
\end{align}
which is just the choice of the trivial or sign representation of $\mathbb Z_2$. Now, we can associate the phases $\phi$ with the cocycles from group cohomology. For example, $\phi_{11}$ is the phase associated with annihilating two domain walls at nearest-neighboring sites. This is equivalent to translating the left domain wall to the right site, which gives a phase $\alpha(g)$ and fusing them together, which gives a phase $\omega(g,g)$. Similarly, we can work out the relations for the other three phase factors. The full correspondence is
\begin{align}
  e^{i\phi_{00}}&= \alpha(g)^{-1}\omega(g,g)^{-1},\\
    e^{i\phi_{01}}&= \alpha(g),\\
      e^{i\phi_{10}}&= \alpha(g)^{-1},\\
    e^{i\phi_{11}}&= \alpha(g)\omega(g,g).
\end{align}
We see that the consistency equations  \eqref{equ:con1} - \eqref{equ:con2} are automatically satisfied. Furthermore the gauge transformation corresponds to the coboundary transformation $\mu(g)$. It is worth noting that to express the crystalline SPT protected by $\mathbb Z_2$ in Eq. \eqref{equ:crystallinesol1D}, we chose $\phi_{00} = \phi_{11} =\pi$. This can be justified in this correspondence by choosing $\alpha(g)=-1$ and $\omega(g,g)=1$. To conclude, our formalism contains the same data as group cohomology for $\mathbb Z_2$.

In Appendix \ref{app:1DgeneralG}, we show that for a general finite unitary symmetry group $G$, all choices of $[\omega] \in H^2(G,U(1))$ and $[\alpha]\in H^1(G,U(1))$ also satisfy our consistency equations.

\section{Symmetry Defect Homology Formalism}\label{sec:homology}
In this section, we describe the general formalism that can be used to solve for phases protected by more general types of symmetries, including subsystem symmetries. We assume that  after applying a duality transformation to an (S)SPT state, the symmetry defects are described by excitations that are created/annihilated from the vacuum and are allowed to hop via the hopping matrix $M$. The consistency conditions can be written in the form
\begin{equation}
\boldsymbol{C} \boldsymbol{\phi} =0,
\label{equ:defecthomology}
\end{equation}
where $\boldsymbol{C}$ is an integer matrix, and $\boldsymbol{\phi}$ is a column vector of the phase factors. To simplify the computation, we take the argument of these phase factors and restrict them to $\mathbb R/\mathbb Z$ (the factors of $2\pi$ can later be restored). The solutions $\boldsymbol{\phi}$ have an equivalence relation, coming from conjugating $M$ with a symmetric phase gate. We can express such equivalence relation via a matrix $\boldsymbol{G}$ as
\begin{equation}
\boldsymbol{\phi} \sim \boldsymbol{\phi} + \boldsymbol{G \varphi}
\end{equation}
for any choice of vector $\boldsymbol{\varphi}$. Now, such conjugation always preserves the consistency conditions of the matrix. Namely, $M_i^2=1$ implies $(UM_iU^\dagger)^2=1$ and $[M_i,M_j]=0$ implies $[UM_iU^\dagger,UM_jU^\dagger]= 0$. Therefore, the new phase factors $\boldsymbol{\phi}'=\boldsymbol{\phi} + \boldsymbol{G \varphi}$ will also satisfy $\boldsymbol{C} \boldsymbol{\phi}' =0$ for any $\boldsymbol \varphi$, which implies
\begin{equation}
\boldsymbol{C} \circ \boldsymbol{G} =0.
\end{equation}
Thus, we have the following chain complex
\begin{equation}
 (\mathbb R/\mathbb Z)^{n_G}  \xrightarrow[]{\boldsymbol G} (\mathbb R/\mathbb Z)^{ n_\phi}   \xrightarrow[]{\boldsymbol C} (\mathbb R/\mathbb Z)^{ n_C}
 \label{equ:symdefecthomology}
\end{equation}
where $n_G, n_\phi$, and $n_C$ are the number of independent ``gauge transformations", variables, and consistency conditions, respectively. The classification (or ``gauge inequivalent" solutions) is then simply $\ker \boldsymbol {C}/\im \boldsymbol{G},$ the homology group of this chain complex. We will refer to this homology group as our \textit{symmetry defect homology}.

Let us now outline how to numerically compute such solutions. First, the kernel of $\boldsymbol C$ can be computed from its Smith decomposition, which can be thought of a singular value decomposition for integer matrices. More formally, there exist integer matrices $\boldsymbol{U},\boldsymbol{D},\boldsymbol{V}$ such that $\boldsymbol{U},\boldsymbol{V}$ are invertible unimodular, $\boldsymbol{D}$ is non zero only along the diagonal, and
\begin{equation}
 \boldsymbol{UC V} = \boldsymbol{D}.
  \end{equation}
Using such decomposition, we see that 
\begin{equation}
    \ker \boldsymbol C = \ker \boldsymbol{D}\boldsymbol{V}^{T}.
\end{equation}
Hence, if we denote the diagonals of $\boldsymbol{D}$ as $d_i$, and the column vectors of $\boldsymbol{V}$ as $\boldsymbol {V}^i$, the kernel can be computed as
\begin{equation}
\ker \boldsymbol{C} =  \text{span} \left.
\begin{cases}
\frac{1}{d_i} \boldsymbol {V}^i &; d_i \ne 0\\
\boldsymbol {V}^i &; d_i = 0
\end{cases} \right \}, 
\label{equ:kerC}
\end{equation}

The desired homology group can then be computed from $\ker \boldsymbol {C}/\im \boldsymbol{G}$.

\subsection{The 1D example revisited}
To illustrate how this machinary works, the 1D example is recomputed in this formalism. First, the consistency conditions \eqref{equ:con1} - \eqref{equ:con2}, and the gauge transformations \eqref{equ:gauge1d} are written as
\begin{align}
\boldsymbol C &=\begin{pmatrix}
1& 0& 0 & 1\\
0&1& 1&0\\
1 &2 & 0 & 1\\
1& 0 & 2 & 1
\end{pmatrix},
&
\boldsymbol G &=\begin{pmatrix}
-2\\
0 \\
0 \\
2 \end{pmatrix},
\end{align}
which satisfies $\boldsymbol{C} \circ \boldsymbol{G} =0.$ Computing the Smith decomposition of $\boldsymbol C$, we obtain
\begin{align}
\boldsymbol{D}&=\left(
\begin{array}{cccc}
 1 & 0 & 0 & 0 \\
 0 & 1 & 0 & 0 \\
 0 & 0 & 2 & 0 \\
 0 & 0 & 0 & 0 \\
\end{array}
\right)
&\boldsymbol{V}&=\left(
\begin{array}{cccc}
 1 & 0 & 0 & -1 \\
 0 & 1 & -1 & 0 \\
 0 & 0 & 1 & 0 \\
 0 & 0 & 0 & 1 \\
\end{array}
\right)
\end{align}
We have omitted $\boldsymbol{U}$, as it is not needed in further computations. From Eq. \eqref{equ:kerC}, we find that $\boldsymbol \phi \in \ker \boldsymbol{C}$ has the form
\begin{align}
\boldsymbol \phi
= n
\begin{pmatrix}
0\\
1/2\\
1/2\\
0
\end{pmatrix}
+ \theta \begin{pmatrix}
-1\\
0\\
0\\
1
\end{pmatrix},
\end{align}
where $n=0,1$ and $\theta \in \mathbb R/\mathbb Z$. Next, since the image of $\boldsymbol G$ is simply

\begin{align}
 \varphi \begin{pmatrix}
-2\\
0\\
0\\
2
\end{pmatrix},
\end{align}
we can quotient out the continuous part by setting $\theta=\varphi/2$, and we are left with a $\mathbb Z_2$ classification.

To add time-reversal, we also demand that phase factors need to satisfy $e^{2\pi i  \phi} =\pm 1$ (where again, $\phi \in \mathbb R/\mathbb Z$). In other words, they must satisfy $2\phi=0$. To capture this fact, we append a diagonal of $2$'s to $\boldsymbol C$ and repeat the calculation.

\begin{equation}
\boldsymbol C_{\mathcal T} =\begin{pmatrix}
1& 0& 0 & 1\\
0&1& 1&0\\
1 &2 & 0 & 1\\
1& 0 & 2 & 1\\
2& 0& 0 & 0\\
0&2& 0&0\\
0 &0 & 2 & 0\\
0& 0 & 0 & 2
\end{pmatrix}.
\end{equation}

With this, we find only discrete solutions

\begin{align}
\boldsymbol{\phi}_{\mathcal T}&= n\begin{pmatrix} 0\\1/2\\1/2\\0\end{pmatrix} + m \begin{pmatrix} 1/2\\0\\0\\1/2\end{pmatrix}.
\end{align}

The gauge transformations are also restricted. Since they need to be real, the vector $\varphi$ can only take values $0$ or $\frac{1}{2}$, meaning that we can consider the matrix $\boldsymbol G$ mod 2, and delete any duplicate columns. In this case, there are no such transformations left, which leaves us with $\mathbb Z_2^2$ solutions. The new generator protected by an extra time-reversal is the ``Haldane phase" described earlier.

As a sanity check, we also reproduce known results for other 1D and 2D SPT's with global symmetries. These are discussed in Appendix \ref{globalresults}.
\section{Solutions to the Defect Homology Equations}\label{sec:solving_defect_homology}

We now discuss in detail the non-trivial SSPT's we found using symmetry defect homology, which is summarized in Table \ref{tab:summary}. 

\begin{table*}[t]
\caption{Generators of SSPT phases with line subsystem symmetry on a square lattice. Without time-reversal, the two generators are a crystalline phase, and a cluster state on a triangular lattice. This is mapped to the Wen-plaquette model under the Xu-Moore duality. With time-reversal, there is an extra generator which is two cluster states on each checkerboard tile of the square lattice.}
\begin{tabular}{|c|c|c|c|}
\hline
 SSPT($\tilde X$) & Cluster State & SSPT Dual($M$) & Comment \\
\hline
 $-X$ &
N/A (weak) &
 $-\begin{array}{cc}
X&X\\
X&X
\end{array}$  & weak SSB
\\
\hline
$\begin{array}{ccc}
& Z & Z\\
Z &X&Z\\
Z&Z&
\end{array}$ & 
\raisebox{-.5\height}{\includegraphics[]{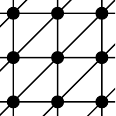}}
&
 $\begin{array}{cc}
X&Y\\
Y&X
\end{array}$ & Wen-Plaquette (SSB). Has topological order. \\
\hhline{====}
 $\begin{array}{ccc}
Z&  & Z\\
 &X&\\
Z&&Z 
\end{array}$ &
\raisebox{-.5\height}{\includegraphics[]{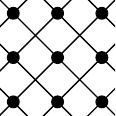}}
& $\begin{array}{cc}
Y&Y\\
Y&Y
\end{array}$ & $\mathbb Z_2^\text{line} \times \mathbb Z_2^T$ SSB
 \\
 \hline
\end{tabular}
\label{tab:2D}
\end{table*}

\subsection{2D square lattice with line symmetry}\label{2Dsquare}
The duality we consider is the self-duality considered in the Xu-Moore model\cite{XuMoore2004} followed by a Hadamard transformation. Consider a square lattice with qubits living on vertices. The trivial Hamiltonian is
\begin{equation}
H = -\sum_{i,j} X_{ij},
\end{equation}
which has line symmetries $\prod_{j} X_{ij}$ for each $i$ and $\prod_{i} X_{ij}$ for each $j$. The symmetries flips spins on each individual rows and columns as shown in Figure \ref{fig:sspt}. From now on, we will depict the lattice visually. Under the duality,

\begin{align}
\begin{array}{cc}
Z&Z\\
Z&Z
\end{array} &\longleftrightarrow 
\begin{array}{cc}
I&I\\
I&Z
\end{array},
\\
\begin{array}{cc}
X&I\\
I&I
\end{array} &\longleftrightarrow\begin{array}{cc}
X&X\\
X&X
\end{array}.
\label{equ:XuMooreDuality}
\end{align}
So the gauged Hamiltonian is
\begin{equation}
H_\text{gauged} = -\sum \begin{array}{cc}
X&X\\
X&X
\end{array} %+ g Z
\end{equation}
with symmetries $\prod Z$ on each row and column. The gauged side describes the creation of symmetry defects in groups of four, and the dual symmetry denotes the conservation of symmetry defects mod $2$ in every row and column.

Calculating the symmetry defect homology group, we find $\mathbb Z_2^2$ SSPT phases. The generators are shown in Table \ref{tab:2D}. The first generator is a weak phase, which corresponds to placing a charge under both line symmetries in each unit cell. The second generator is a 2D cluster state on a triangular lattice, obtained by adding diagonal lines in one direction to the original square lattice. This 2D cluster state is dual to the Wen-Plaquette model\cite{Wen2003}.

To argue that this cluster state is a strong SSPT, we can show that the restriction of adjacent line symmetries to the boundary anticommute. Alternatively, we can show that their duals realize different subsymmetry broken phases. Indeed, on a torus of size $L_x \times L_y$, the dual of the product state breaks the subsystem symmetry extensively, with ground-state degeneracy $2^{L_x+L_y-1}$. On the other hand, the dual of the cluster state, the Wen-plaquette model, has ground-state degeneracy $4$. Since there is no symmetric finite depth local circuit to connect the two symmetry-broken models even in the absence of translation, we conclude that there is also no symmetric finite depth local circuit that connects the 2D cluster state to the product state.

We remark that the Wen-plaquette model has an emergent 1-form symmetry in the degenerate ground-state subspace, which is spontaneously broken\cite{Gaiottoetal2015}. In this model, the ground-state degeneracy still exists even when we explicitly break the 1-form symmetry to the line symmetry subgroup.

With time reversal symmetry, we find an extra generator. It is a cluster state which connects vertices of only next-nearest (diagonal) neighbors of the square lattice. The wavefunction also has a nice interpretation in terms of symmetry defects. The SSPT phase corresponds to attaching each symmetry defect with a phase $i$. Since the number of symmetry defects in any configuration is always even, the overall wavefunction is always real.

The dual of this phase is distinct from the usual symmetry broken phase when time-reversal is present. Although they have the same ground-state degeneracy, the action of time-reversal maps a ground-state to an orthogonal state, while the dual of the trivial SSPT does not.

\begin{table*}[t]
\caption{Generators of SSPT phases protected by $\mathbb Z_2\ \times \mathbb Z_2$ line symmetry on a square lattice.}
\begin{tabular}{|c|c|c|c|c|}
\hline
\multicolumn{2}{|c|}{SSPT} &\multirow{2}{*}{Comment} & \multicolumn{2}{c|}{SSPT Dual}\\
\cline{1-2} \cline{4-5}
$\tilde X^{(a)}$ & $\tilde X^{(b)}$ & &  $M^{(a)}$ & $M^{(b)}$ \\
\hline
 $-XI$ & $IX$ & $\mathbb Z_2^{(a)}$ weak &
 $-\begin{array}{cc}
XI&XI\\
XI&XI
\end{array}$ & 
$\begin{array}{cc}
IX&IX\\
IX&IX
\end{array}$\\
\hline
$XI$ & $-IX$ & $\mathbb Z_2^{(b)}$ weak &
$\begin{array}{cc}
XI&XI\\
XI&XI
\end{array}$ & 
$-\begin{array}{cc}
IX&IX\\
IX&IX
\end{array}$ \\
\hline
$\begin{array}{ccc}
& ZI & ZI\\
ZI &XI&ZI\\
ZI&ZI& 
\end{array}$ & $IX$ & $\mathbb Z_2^{(a)}$ cluster state &$\begin{array}{cc}
XI&YI\\
YI&XI
\end{array}$ & 
$\begin{array}{cc}
IX&IX\\
IX&IX
\end{array}$\\
\hline
$XI$ & $\begin{array}{ccc}
& IZ & IZ\\
IZ &IX&IZ\\
IZ&IZ&
\end{array}$ & $\mathbb Z_2^{(b)}$ cluster state & 
$\begin{array}{cc}
XI&XI\\
XI&XI
\end{array}$ & 
$\begin{array}{cc}
IX&IY\\
IY&IX
\end{array}$\\
\hline
$\begin{array}{cc}
IZ&IZ\\
IZ&XZ
\end{array}$ & 
$\begin{array}{cc}
ZX&ZI\\
ZI&ZI
\end{array}$ & $\mathbb Z_2^{(a)} \times \mathbb Z_2^{(b)}$ cluster state & $\begin{array}{cc}
XZ&XI\\
XI&XI
\end{array}$ & 
$\begin{array}{cc}
IX&IX\\
IX&ZX
\end{array}$\\
\hline
$\begin{array}{cc}
IZ &IZ \\
XI &\\
IZ& IZ
\end{array}$
&
$\begin{array}{cc}
ZI &ZI \\
 &IX\\
ZI& ZI
\end{array}$&  & $\begin{array}{cc}
XZ&XI\\
XZ&XI
\end{array}$& $\begin{array}{cc}
IX&ZX\\
IX&ZX
\end{array}$\\
\hline
$\begin{array}{ccc}
IZ&XI & IZ\\
IZ& & IZ
\end{array}$
&
$\begin{array}{ccc}
ZI& & ZI\\
ZI& IX & ZI
\end{array}$
&  & 
$\begin{array}{cc}
XI&XI\\
XZ&XZ
\end{array}$&
$\begin{array}{cc}
ZX&ZX\\
IX&IX
\end{array}$\\
\hhline{=====}
$\begin{array}{ccc}
ZI&  & ZI\\
 &XI&\\
ZI&&ZI 
\end{array}$ &
 $IX$ & $\mathbb Z_2^{(a)}$ time-reversal & 
 $\begin{array}{cc}
YI&YI\\
YI&YI
\end{array}$ & 
$\begin{array}{cc}
IX&IX\\
IX&IX
\end{array}$\\
\hline
 $XI$ &
$\begin{array}{ccc}
IZ&  & IZ\\
 &IX&\\
IZ&&IZ 
\end{array}$ & $\mathbb Z_2^{(b)}$ time-reversal &
$\begin{array}{cc}
XI&XI\\
XI&XI
\end{array}$ & 
$\begin{array}{cc}
IY&IY\\
IY&IY
\end{array}$\\
\hline
\end{tabular}
\label{tab:Z2xZ2line2D}
\end{table*}

\subsection{Square Lattice with $\mathbb Z_2\ \times \mathbb Z_2$ line symmetry}\label{2DsquareZ2Z2}
We repeat the calculation on the square lattice, but now each site forms a representation of $\mathbb Z_2 \times \mathbb Z_2$. Here, we will use the notation of adjoining the two operators in the same unit cell. The product state Hamiltonian consists of two terms $ X^{(a)}=XI$ and $ X^{(b)}=IX$.

We find that the symmetry defect homology group is $\mathbb Z_2^7$ without time reversal. The generators are listed in Table \ref{tab:Z2xZ2line2D}.  The first two are weak phases, corresponding to putting a charge of each $\mathbb Z_2$ in each unit cell. The next two generators are the 2D cluster states from the previous subsection for each $\mathbb Z_2$. The fifth generator, is a cluster state which mixes the two $\mathbb Z_2$'s. Its non-trivialness comes from the self-duality under the Xu-Moore map. The final two generators are also strong SSPT's under our definition. As an example, let us look at the stabilizer in row 6 of Table \ref{tab:Z2xZ2line2D}:
\begin{align}
\tilde X^{(a)}&=\begin{array}{cc}
IZ &IZ \\
XI &\\
IZ& IZ
\end{array}
&
\tilde X^{(b)}&=\begin{array}{cc}
ZI &ZI \\
 &IX\\
ZI& ZI
\end{array}
\end{align}

First, we consider the line symmetries in the vertical direction. By assuming the ground-state of the SSPT is satisfied by $\tilde X^{(a)}=1$ and $\tilde X^{(b)}=1$, we can restrict each line symmetry to its top and bottom edges
On the top edge, we find
\begin{align}
S^{(a)}_T&=\begin{array}{cc}
XZ & IZ  \\
IZ & IZ 
\end{array}
&
S^{(b)}_T&=\begin{array}{cc}
ZI & ZX \\
ZI & ZI
\end{array},
\end{align}
which mutually commute. However, when we restrict the horizontal line symmetries to the left and right edges, we find that on the left edge,
\begin{align}
S^{(a)}_L&=\begin{array}{cc}
IZ  \\
 \\
IZ
\end{array}
&
S^{(b)}_L&=\begin{array}{cc}
ZI \\
IX\\
ZI
\end{array}.
\end{align}
Hence, we see that each horizontal line symmetry, when restricted to the edge, anticommutes with the other species in the two adjacent rows.  Since the anticommutation of the horizontal symmetries at the edge cannot be removed even when translation is broken, it is a strong SSPT. The final generator can be similarly argued since it is just a rotated version of the former.

We remark that under the equivalence of linear-depth circuits as in Ref. \onlinecite{DevakulWilliamsonYou2018}, the last two generators would be considered weak, while generators 3, 4, and 5 will still be considered strong, and are the generators of the strong SSPT's given in their classification.

With time-reversal, we find two more strong generators, which are just the time-reversal SSPT's for each $\mathbb Z_2$ given in the previous subsection.

\begin{table*}[t]
\caption{Triangular Lattice with line symmetries}
\begin{tabular}{|c|c|c|}
\hline
 SSPT($\tilde X$) & Comment & SSPT Dual ($M$)  \\
 \hline
 $-$\raisebox{-.5\height}{\includegraphics[scale=0.9]{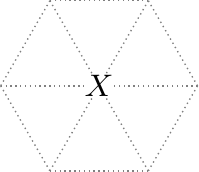}}& Weak &$-$\raisebox{-.5\height}{\includegraphics[scale=0.9]{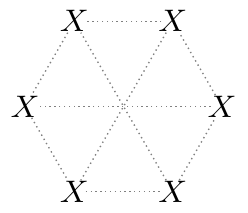}}\\
\hline
\raisebox{-.5\height}{\includegraphics[scale=0.9]{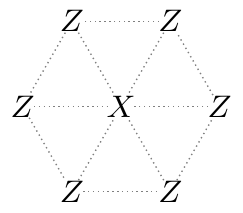}}
&Adjacent lines anticommute in 3 directions&
\raisebox{-.5\height}{\includegraphics[scale=0.9]{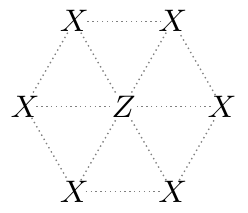}}
\\
\hline
\raisebox{-.5\height}{\includegraphics[scale=0.9]{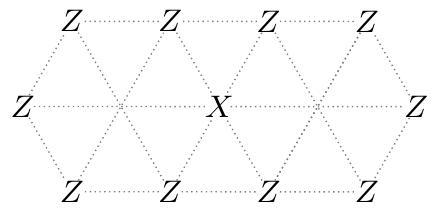}}
&Next nearest lines anticommute in 2 directions&
\raisebox{-.5\height}{\includegraphics[scale=0.9]{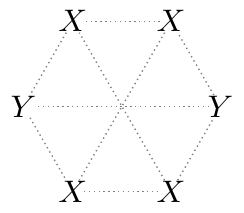}}
\\
\hline
\raisebox{-.5\height}{\includegraphics[scale=0.9]{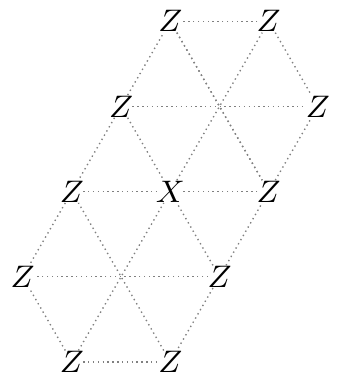}}
&Next nearest lines anticommute in 2 directions&
\raisebox{-.5\height}{\includegraphics[scale=0.9]{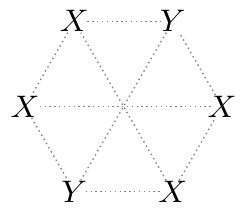}}
\\
\hhline{===}
\raisebox{-.5\height}{\includegraphics[scale=0.9]{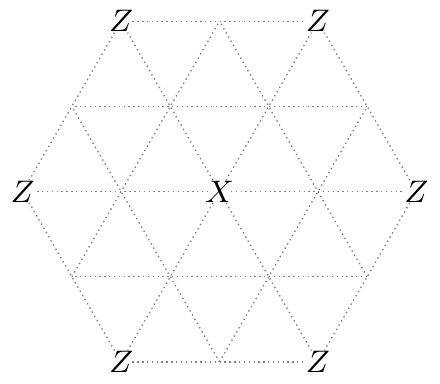}}
&Time-reversal&
\raisebox{-.5\height}{\includegraphics[scale=0.9]{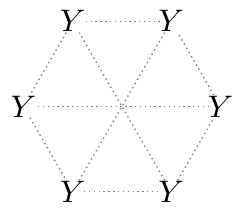}}
\\
\hline
\end{tabular}
\label{tab:2Dtriangular}
\end{table*}

\subsection{Triangular Lattice with line symmetry}\label{2Dtriangle}
We consider the triangular lattice with qubits on vertices, and three line symmetries. The associated duality is
\begin{align}
\raisebox{-.5\height}{\includegraphics[]{2Dtriangle1.pdf}}
&\longleftrightarrow \raisebox{-.5\height}{\includegraphics[]{2Dtriangle2.pdf}}, 
\\
\raisebox{-.5\height}{\includegraphics[]{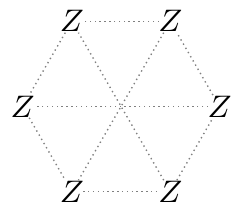}}
&\longleftrightarrow \raisebox{-.5\height}{\includegraphics[]{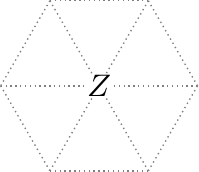}},
\end{align}
where a spin flip creates six symmetry defects in the dual picture. 
Calculating the symmetry defect homology gives $\mathbb Z_2^4$ phases without time reversal, summarized in Table \ref{tab:2Dtriangular}. The first generator is a weak phase obtained by putting charges of the three line symmetries in each unit cell. The second generator is a cluster state on this triangular lattice. We notice that by ignoring one of the line symmetries, it is the same as the strong SSPT on the square lattice, which means that it is a strong SSPT. Alternatively, one can also argue this from the self-duality of the SSPT, or from the fact that by restricting the action of the line symmetries to the boundary, adjacent lines anticommute for all three directions.

Next consider the third generator in Table \ref{tab:2Dtriangular}. Restricting the horizontal line symmetries to the boundary, we find that adjacent lines commute. However, the restriction of the lines in the other two directions anticommute for next nearest-neighbor lines. Thus, it is a strong phase. The last generator is just a $60^\circ$ rotation of the previous one.

With time reversal, we find an additional generator, which is four copies of the 2D cluster state in a doubled unit cell.

We remark that although rows 3 and 4 are $60^\circ$ rotations of each other, we have not included the $120^\circ$ rotation. To see why, we note that combining the unitaries that generate these three models would have no anticommuting line operators at the boundary. Furthermore, it is exactly the SSPT given in the last row, which is trivial without time-reversal.

\subsection{Sierpinski Fractal Symmetry}\label{NewmanMoore}
Consider the square lattice with Ising interaction
\begin{equation}
 \begin{array}{cc}
Z & \\
Z&Z
\end{array}
\end{equation}
The model has a fractal symmetry where spin flips form the pattern of the Sierpinski cellular automaton\cite{Devakuletal2018}. This model can also be considered on a triangular lattice, where the Ising terms only up on right-side up ($\Delta$) triangles\cite{NewmanMoore1999}.

We find only the trivial solution, which is consistent with Ref. \onlinecite{Devakul2019}. In particular, we also don't find any crystalline solutions. This might seem surprising, since naively decorating each unit cell with a charge can be trivialized. However this is due to the fact that the Ising interaction has an odd number of $Z$'s. Thus, conjugating $-X$ with a product of all Ising terms disentangles it to $X$. 

With time-reversal, there are no new solutions. This is also because the Ising interaction has an odd number of $Z$'s.

\subsection{Fibonacci}\label{Fibonacci}
Consider the square lattice with Ising interaction
\begin{equation}
 \begin{array}{ccc}
& Z & \\
Z &Z&Z
\end{array}
\end{equation}
The model has a fractal symmetry where spin flips form the pattern of the classical Fibonacci cellular automaton\cite{Devakuletal2018} (also known as Rule 150).

The duality is given by
\begin{align}
        \begin{array}{ccc}
& X & \\
I &I&I
\end{array} &\longleftrightarrow       \begin{array}{ccc}
X&X  &X \\
 &X&
\end{array}\\ 
     \begin{array}{ccc}
& Z & \\
Z &Z&Z
\end{array} &\longleftrightarrow       \begin{array}{ccc}
I& I &I \\
 &Z&
\end{array}
\end{align}

The results are shown in Table \ref{tab:fibonacci}. Apart from the crystalline phase, we find an SSPT protected by time-reversal. Similarly to the square lattice case, the number of symmetry defects is always even, so each defect can carry a phase factor $i$.
\begin{table}[t]
\caption{Generators of SSPT phases protected by the fractal symmetry generated by the Fibonacci cellular automaton}
\begin{tabular}{|c|c|}
\hline
SSPT ($\tilde X$) & SSPT Dual (M)\\
\hline
$
- \begin{array}{ccc}
& X & \\
I &I&I
\end{array}
$&$ -\begin{array}{ccc}
X&X  &X \\
 &X&
\end{array}$\\
\hhline{==}
$
\begin{array}{ccccc}
&Z&Z&Z&\\
Z& & X & &Z \\
&Z&Z&Z&
\end{array}$& $\begin{array}{ccc}
Y&Y  &Y \\
 &Y&
\end{array}$\\
\hline
\end{tabular}
\label{tab:fibonacci}
\end{table}

\subsection{3D Cubic lattice with line symmetry}\label{3Dline}
We study possible SSPT's for the Cubic Ising model\cite{Youetal2018}. The Ising term consists of eight $Z$ operators at the corners of each cube. In the dual theory, eight $X$ at the corners of the cube create symmetry defects, which are conserved modulo 2 for each line. The duality is

\begin{align}
\raisebox{-.5\height}{\includegraphics[]{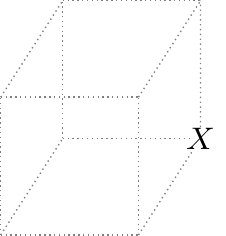}}
&\longleftrightarrow 
\raisebox{-.5\height}{\includegraphics[]{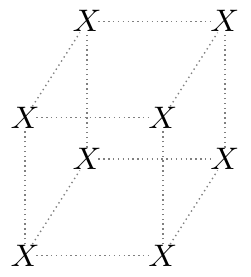}},
\end{align}
\begin{align}
\raisebox{-.5\height}{\includegraphics[]{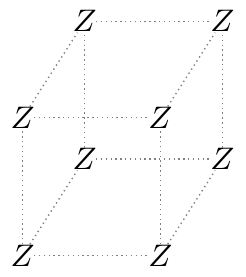}}
&\longleftrightarrow
\raisebox{-.5\height}{\includegraphics[]{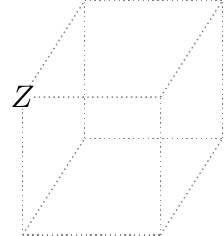}}.
\end{align}

Apart from the weak solution, we find three generators of 3D cluster states as shown in Table \ref{tab:3Dline}. To show that the cluster states realize different SSPT phases, we look at the ground-state degeneracy of the duals on an $L\times L \times L$ torus. The usual symmetry breaking phase has ground-state degeneracy  $\log_2(GSD) \propto L^2$ at leading order. On the other hand, the ground-state degeneracy of duals of the three SSPT generators is calculated in Appendix \ref{app:3Dlinecalc} to be $\log_2(GSD) =3L-2$. 

We remark that these cluster states, when protected by planar symmetries (which is a subgroup of the line symmetries of this model), is dual to the Semionic X-Cube model\cite{MaLakeChenHermele2017} (in particular, the Hamiltonian discussed in Ref. \onlinecite{WangShirleyChen2019}).

With time-reversal, we find an extra solution. It consists of two copies of a 3D cluster state, each on a checkerboard tiling of the cubic lattice.

\begin{table}[!t]
\caption{Cubic lattice with three line symmetries}
\begin{tabular}{|c|c|}
\hline
SSPT ($\tilde X$) & SSPT Dual($M$)  \\
\hline
$-$\raisebox{-.5\height}{\includegraphics[scale=0.99]{3Dline1.pdf}} &
$-$\raisebox{-.5\height}{\includegraphics[scale=0.99]{3Dline2.pdf}}
\\
\hline
\raisebox{-.5\height}{\includegraphics[scale=0.99]{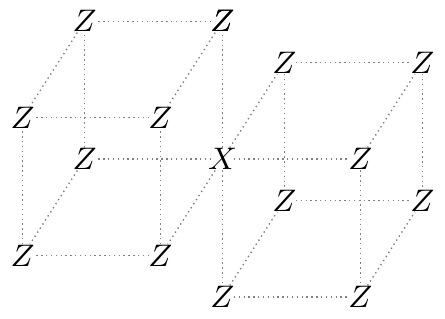}}
& 
\raisebox{-.5\height}{\includegraphics[scale=0.99]{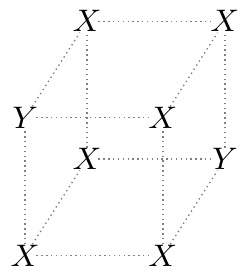}}
\\
\hline
\raisebox{-.5\height}{\includegraphics[scale=0.99]{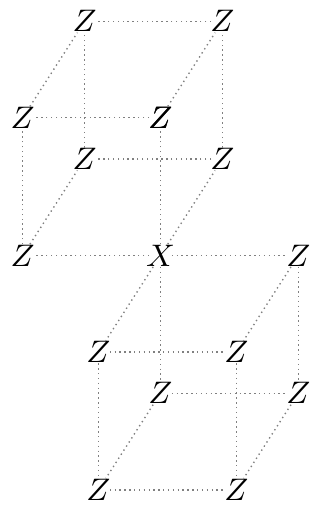}}
& 
\raisebox{-.5\height}{\includegraphics[scale=0.99]{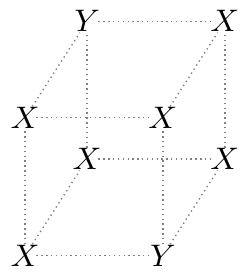}}
\\
\hline
\raisebox{-.5\height}{\includegraphics[scale=0.99]{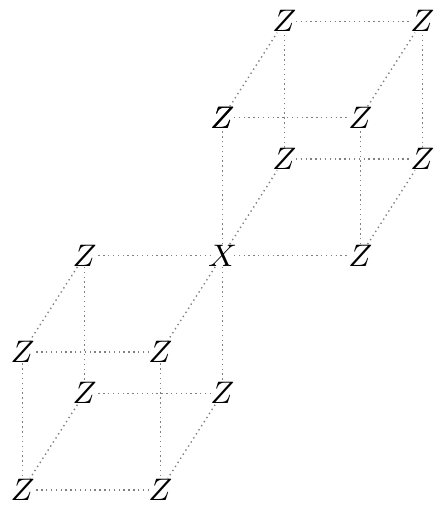}}
&
\raisebox{-.5\height}{\includegraphics[scale=0.99]{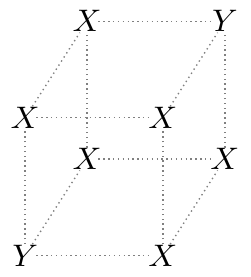}}
\\
\hhline{==}
\raisebox{-.5\height}{\includegraphics[scale=0.99]{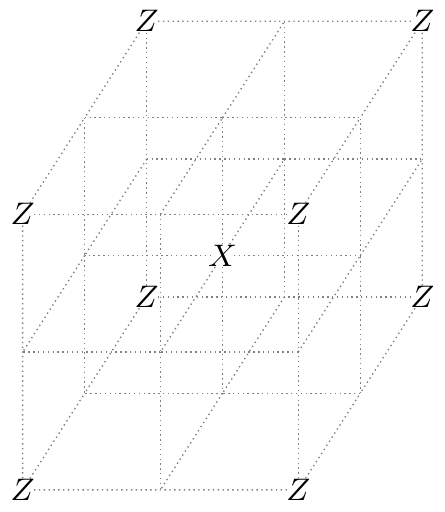}}
&
\raisebox{-.5\height}{\includegraphics[scale=0.99]{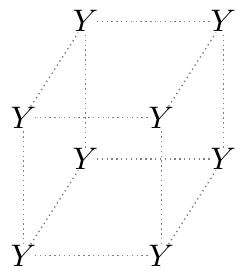}}
 \\
\hline
\end{tabular}
\label{tab:3Dline}
\end{table}

\subsection{3D Cubic lattice with planar symmetry}\label{Xcube}
We study the dual of the X-Cube model\cite{VijayHaahFu2016}. This is a 3D cubic lattice with planar symmetries in the $(100)$, $(010)$, and $(001)$ directions. The Ising terms are a product of four $Z$ operators on each face of the cubic lattice. The duality is given by

\begin{align}
X &\longleftrightarrow
\raisebox{-.5\height}{\includegraphics[]{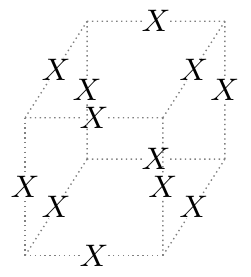}}
\\
\raisebox{-.5\height}{\includegraphics[]{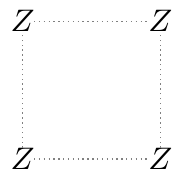}}
& \longleftrightarrow
\raisebox{-.5\height}{\includegraphics[]{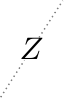}},
\\
\raisebox{-.5\height}{\includegraphics[]{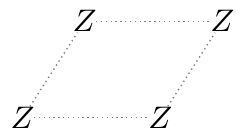}}
& \longleftrightarrow
\raisebox{-.5\height}{\includegraphics[]{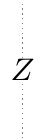}},
\\
\raisebox{-.5\height}{\includegraphics[]{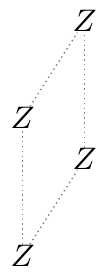}}
& \longleftrightarrow
\raisebox{-.5\height}{\includegraphics[]{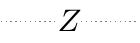}},
\end{align}
where the gauged side is drawn on the dual lattice.
Physically, flipping spin on a single site excites twelve symmetry defects, which corresponds to the twelve $X$ operators in the X-Cube model. Thus the superposition of all possible symmetry defect configuration can be thought of as a cage-net condensate, a picture introduced in Ref. \onlinecite{PremHuangSongHermele2019}. Nevertheless, we would like to emphasize that the models we present here should not be thought of as a result of some coupled layer construction\cite{MaLakeChenHermele2017,Vijay2017}, but only as a superposition of cage configurations with a possible relative sign between them.

\begin{table}[t]
\caption{SSPT's protected by planar symmetry on a cubic lattice. All generators are weak.}
\begin{tabular}{|c|c|}
\hline
 SSPT($\tilde X$) & Fracton ($M$)  \\
 \hline
$-X$ & 
\raisebox{-.5\height}{\includegraphics[scale=0.9]{Xcube1.pdf}}
\\
\hline
\raisebox{-.5\height}{\includegraphics[scale=0.9]{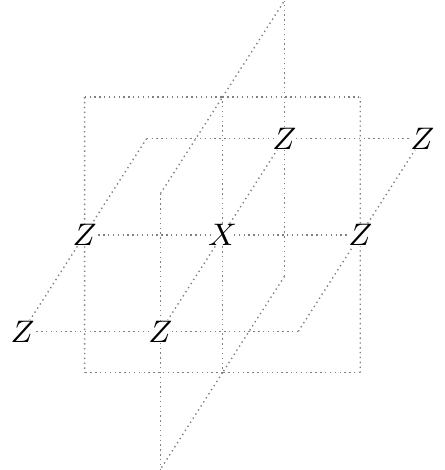}}
&
\raisebox{-.5\height}{\includegraphics[scale=0.9]{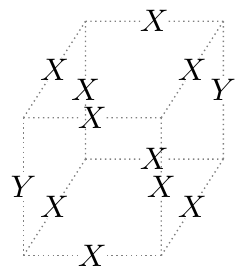}}
\\
\hline
\raisebox{-.5\height}{\includegraphics[scale=0.9]{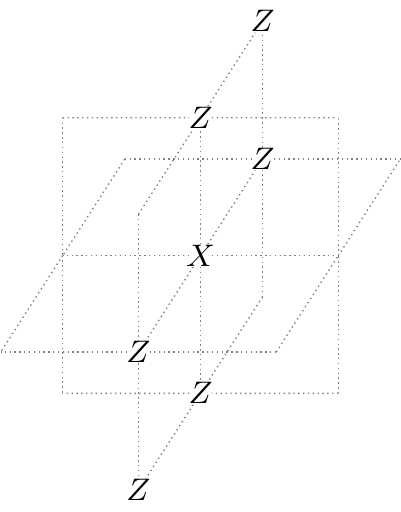}}
&
\raisebox{-.5\height}{\includegraphics[scale=0.9]{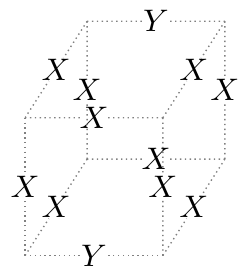}}
\\
\hline
\raisebox{-.5\height}{\includegraphics[scale=0.9]{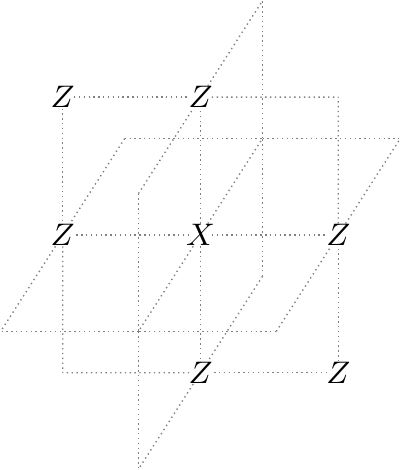}}
&
\raisebox{-.5\height}{\includegraphics[scale=0.9]{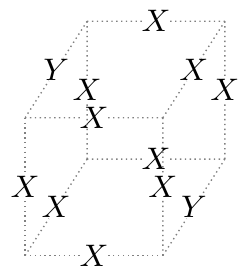}}
\\
 \hline
 \end{tabular}
\label{tab:Xcube}
\end{table}

\subsubsection{Without time reversal}
Without time-reversal, we find $\mathbb Z_2^4$ phases. The first generator is a weak phase that decorates a charge sitting at every vertex. It is dual to the X-cube model with a fracton sitting in every unit cell.

The next three generators correspond to higher dimensional decorations. Namely, they correspond to stacks of 2D cluster states in the $(100)$, $(010)$, and $(001)$ planes, as illustrated in Table \ref{tab:Xcube}. Since this is a stack of 2D SSPT phases that we found in Section \ref{2Dsquare}, the three models are weak SSPT's. That is, when translation is broken, we are able to trivialize the SSPT by annihilating the stack of 2D cluster states in pairs. As an example, consider the gate
\begin{align}
\raisebox{-.5\height}{\includegraphics[]{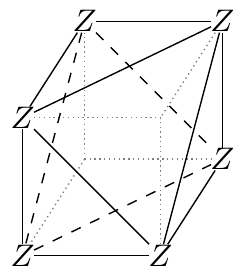}},
\end{align}
 where each the thick and dashed lines represents $CZ$ gates between two sites. This gate commutes with all the planar symmetries, and the product of such a gate in the $xy$ plane creates 2D cluster states on the top and bottom layers up to a sign of the stabilizer. By breaking translational symmetry in the $z$ direction, applying the above gate in every other $xy$ plane trivializes the second SSPT of Table \ref{tab:Xcube}. Similarly, a product of this gate in the $xz$ and $yz$ planes trivializes the remaining two generators.

Our crystalline SSPT solution naturally generalizes the decoration construction \cite{HuangSongHuangHermele2017,ThorngrenElse2018,ElseThorngren2019} of crystalline SPT's with global symmetry, where SPT phases of a lower dimension can be decorated to give non-trivial crystalline phases. To probe such SSPT's, a dislocation parallel to the decoration plane introduced will have a gapless dislocation edge, with degeneracy that scales exponentially with its length.

Alternatively, let us now study the non-trivial properties of the SSPT from its dual, which is a translation-enriched fracton phase. It is known that in the gauging duality, a lower dimensional SPT is dual to a transparent gapped domain wall that permutes topological excitations that pass through \cite{Yoshida2015,Yoshida2016,Yoshida2017,PotterMorimoto2017,PoFidkowskiVishwanathPotter2017,KubicaYoshida2018,FidkowskiPoPotterVishwanath2019}. In the construction of crystalline topological phases, similarly, the unit cell can be decorated with lower dimensional invertible defects\cite{ElseThorngren2019,SongFangQi2018,Songetal2018,ShiozakiXiongGomi2018}, so that topological excitations get permuted when they move from one unit cell to the next. Let us demonstrate this property for the dual fracton model.

Consider the X-cube model, whose cube term is now 
\begin{equation}
\raisebox{-.5\height}{\includegraphics[]{Xcube3.pdf}},
\end{equation}
 which is dual to the SSPT generator in the second row in Table \ref{tab:Xcube}. Four fractons can still be excited with a $Z$ operator. However, when exciting the lineon excitation with $X$ in the vertical direction, it also creates two cube excitations, which can then move along with the lineon as two fracton dipoles. Thus, we see that as the lineon passes through the transparent domain wall, it gets attached with two fracton dipoles. This feature is robust even after breaking translation to an odd multiple.

\begin{figure}[h!]
    \centering
\raisebox{-.5\height}{\includegraphics[]{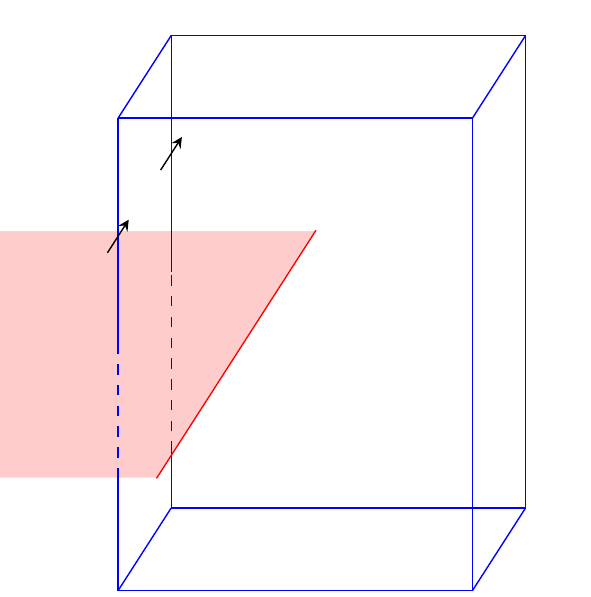}}
    \caption{The translation-enriched X-cube model, which is dual to stacks of 2D cluster states in the horizontal plane, can be probed via a dislocation defect. A lineon cage (blue) going around a dislocation line (red) as shown creates two fracton-dipoles (arrows) parallel to the dislocation line.}
    \label{fig:disclination}
\end{figure}

To probe the translation enrichment of this model, we introduce a dislocation into the modified X-cube, where the dislocation edge is along say the $x$ direction as shown in Figure \ref{fig:disclination}. Consider a pair of lineons mobile in the $yz$ plane. Moving the lineon pair around the dislocation edge, each lineon will pass the domain wall once and each get attached with two fractons. The four fractons effectively form two fracton dipoles that also can move in the $yz$ plane. To conclude, the dislocation in the crystalline enriched model allows us to pull out pair of fracton dipoles parallel to the dislocation edge.
\begin{table*}[t!]
\caption{SSPT's protected by planar symmetry and time-reversal on a cubic lattice. The first three generators are stacks of 2D SSPT phases and therefore weak, while the last two are strong. Their duals are X-cube-like models where lineons transform non-trivially under time-reversal.}
\begin{tabular}{|c|c|}
\hline
SSPT($\tilde X$) & Fracton($M$) \\
 \hline
 \raisebox{-.5\height}{\includegraphics[scale=0.89]{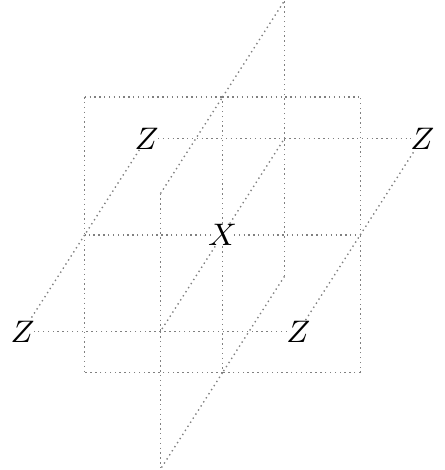}}
&
 \raisebox{-.5\height}{\includegraphics[scale=0.89]{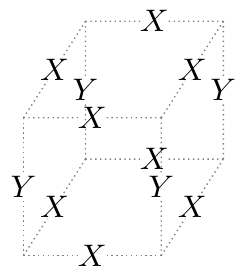}}
\\
\hline
 \raisebox{-.5\height}{\includegraphics[scale=0.89]{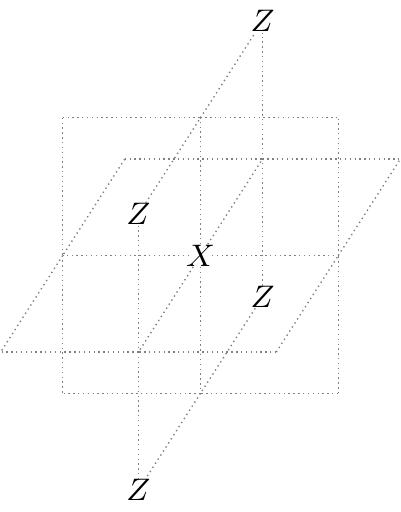}}
&
 \raisebox{-.5\height}{\includegraphics[scale=0.89]{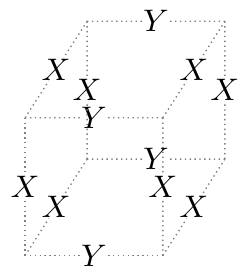}}
\\
\hline
 \raisebox{-.5\height}{\includegraphics[scale=0.89]{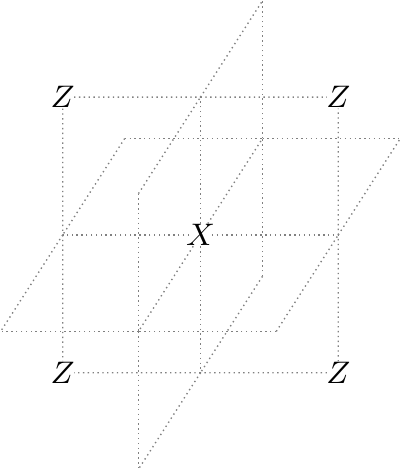}}
&
 \raisebox{-.5\height}{\includegraphics[scale=0.89]{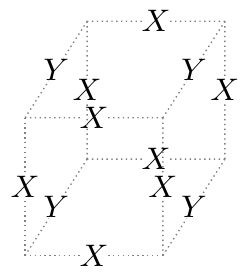}}
\\
\hline
\raisebox{-.5\height}{\includegraphics[scale=0.89]{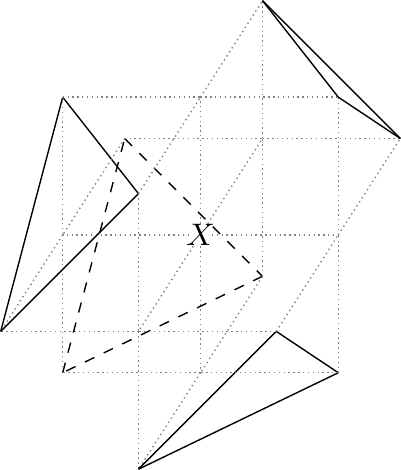}}
$\times$
\raisebox{-.5\height}{\includegraphics[scale=0.89]{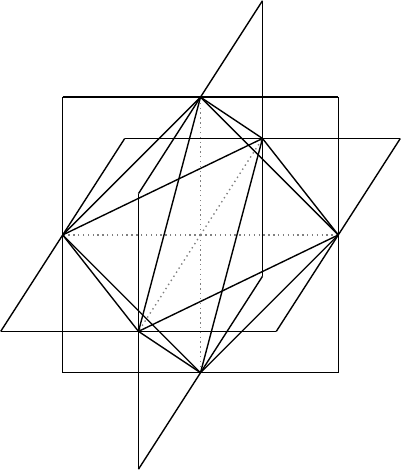}}
&
\raisebox{-.5\height}{\includegraphics[scale=0.89]{Xcube1.pdf}} $\times$
\raisebox{-.5\height}{\includegraphics[scale=0.89]{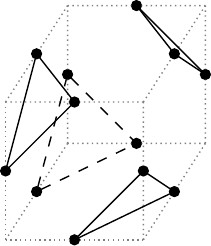}}
\\
\hline
\raisebox{-.5\height}{\includegraphics[scale=0.89]{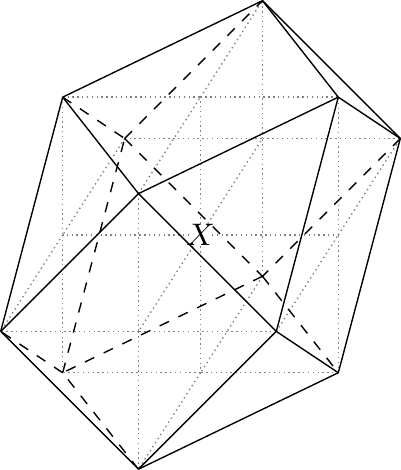}}
&
\raisebox{-.5\height}{\includegraphics[scale=0.89]{Xcube1.pdf}}
$\times$
\raisebox{-.5\height}{\includegraphics[scale=0.89]{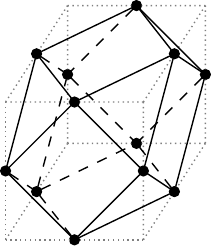}}
\\
\hline
 \end{tabular}
\label{tab:Xcubetimereversal}
\end{table*}

\subsubsection{With time reversal}
With time-reversal, the gauged side is now an X-cube model enriched by translation and time-reversal. Because gauging planar symmetries gives a gauge theory in 3D, the hopping matrix $M$ only needs to commute up to gauge constraint terms. Furthermore, time reversal only needs to acts as complex conjugation up to gauge constraints.

We find an addition of five SSPT generators with time reversal, as shown in Table \ref{tab:Xcubetimereversal}. The first three generators are the time-reversal SSPT's in 2D discussed previously in Section \ref{2Dsquare}, again decorated in the $xy$, $yz$, and $xz$ planes, respectively. To see that they are also weak crystalline phases, the following gate
\begin{equation}
\raisebox{-.5\height}{\includegraphics[]{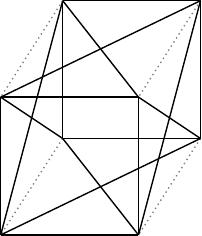}}
\end{equation}
commutes with all planar symmetries and time-reversal. By breaking translation in say the $z$ direction, applying this gate to every other $xy$ layer disentangles the stack of 2D time-reversal SSPT in pairs.

The last two generators are rather intriguing. They are in fact hypergraph states on a regular lattice, which we will from now naturally call \textit{hypercluster} states, and the Hamiltonian is no longer a stabilizer code. The unitaries that entangles the ground-state of each SSPT from the product state can be expressed as
\begin{widetext}
\begin{align}
    \label{equ:XcubeT1unitary}
    U_1 =\prod_i& \exp   \frac{\pi i}{8} \left[
\raisebox{-.5\height}{\includegraphics[]{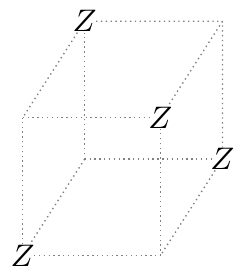}}
    -2\raisebox{-.5\height}{\includegraphics[]{Z1.pdf}}  -2\raisebox{-.5\height}{\includegraphics[]{Z3.pdf}}
    -2 \raisebox{-.5\height}{\includegraphics[]{Z5.pdf}} \right] \\
=\prod_i&  CCZ_{x,y,z}CCZ_{ x,\bar y,\bar z}CCZ_{\bar x,\bar y,z}CCZ_{\bar x, y,\bar z} CCZ_{1,x,y}CCZ_{1,x,\bar y}CCZ_{1,\bar x,y}CCZ_{1,\bar x,\bar y} \nonumber\\
& \times CCZ_{1,y,z}CCZ_{1,y,\bar z}CCZ_{1,\bar y,z}CCZ_{1,\bar y,\bar z}CCZ_{1,z,x}CCZ_{1,z,\bar x}CCZ_{1,\bar z,x}CCZ_{1,\bar z,\bar x}\\
U_2 = \prod_i& \exp  \frac{\pi i}{8} \left[\raisebox{-.5\height}{\includegraphics[]{Z7.pdf}} - 
\raisebox{-.5\height}{\includegraphics[]{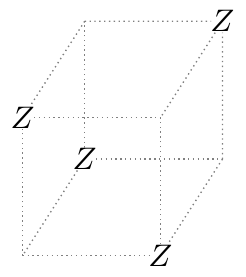}}
 \right] \\
 = \prod_i& CCZ_{x,y,z}CCZ_{ x,y,\bar z}CCZ_{x,\bar y,z}CCZ_{x,\bar y,\bar z}CCZ_{\bar x,y,z}CCZ_{\bar x,y,\bar z}CCZ_{\bar x,\bar y,z}CCZ_{\bar x,\bar y,\bar z}
 \label{equ:XcubeT2unitary}
\end{align}
\end{widetext}
Here, $CCZ_{x,y,z}$ is a shorthand for $CCZ_{\vec r_i+\hat x,\vec r_i+\hat y,\vec r_i+\hat z}$ and $\bar x$ means $-\hat x$ etc. The two representations of the unitaries is to emphasize that the SSPT can be trivialized by breaking either time-reversal or the planar symmetries. The result of conjugating $X$ with the unitaries is given in the last two rows of Table \ref{tab:Xcubetimereversal}.

To see the non-trivialness of these SSPT's, we look at its property from the fracton side. To perform the duality on the unitary, we use the exponentiated form and choose the following mapping of operators
\begin{align}
\raisebox{-.5\height}{\includegraphics[]{Z7.pdf}}
& \longleftrightarrow
\raisebox{-.5\height}{\includegraphics[]{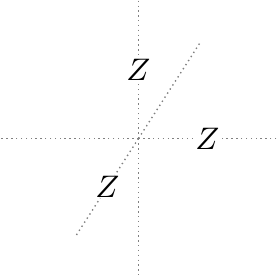}}
,
\end{align}

\begin{align}
\raisebox{-.5\height}{\includegraphics[]{Z9.pdf}}
& \longleftrightarrow
\raisebox{-.5\height}{\includegraphics[]{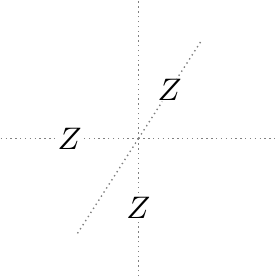}}
. 
\end{align}
One can show that conjugating the cube term in the X-Cube model with the dual unitaries $ U_{1,\text{gauged}}$ and $ U_{2,\text{gauged}}$ is the same as the cube terms shown in Table \ref{tab:Xcubetimereversal} up to vertex terms. Furthermore, the lineons, which were originally excited by a string of $X$ operators, are now excited by a string of $U_\text{gauged} X \tilde U_\text{gauged}^\dagger$. Explicitly, the lineon operators in the $y$ direction  for the two time-reversal models are
\begin{widetext}
\begin{align}
  U_{1,\text{gauged}} \left (
  \raisebox{-.5\height}{\includegraphics[]{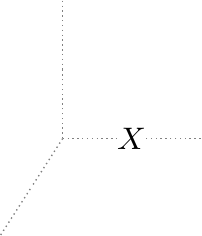}}
\right )
    U_{1,\text{gauged}}^\dagger &=
  \raisebox{-.5\height}{\includegraphics[]{Xlineon1.pdf}}
      \times
      \raisebox{-.5\height}{\includegraphics[]{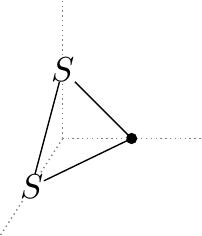}}
\label{equ:T-lineon1}
\\
     U_{2,\text{gauged}} \left (
       \raisebox{-.5\height}{\includegraphics[]{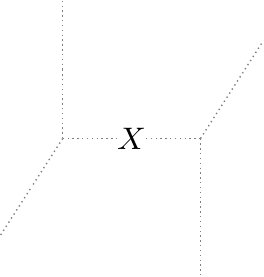}} 
\right)
     U_{2,\text{gauged}}^\dagger &= \raisebox{-.5\height}{\includegraphics[]{Xlineon2.pdf}}  \times 
     \raisebox{-.5\height}{\includegraphics[]{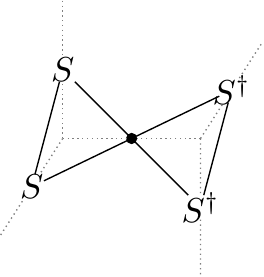}}
\end{align}
\end{widetext}

The lineon operators in the other directions can be obtained by the appropriate rotations. One can check that other than anticommuting with the vertex terms at the end points, they commute with the cube terms in the middle of the strings, and are annihilated by projectors to the gauge-invariant subspace at the end of the strings. 

We now show that the models we have written down are protected by time reversal. In particular, we demonstrate that the excited states of both models form Kramers pairs. For the first model, denote $U_{1,\text{gauged}}XU_{1,\text{gauged}}^\dagger \equiv X_1$. Consider a pair of lineon excitations obtained by acting an open string of $\prod X_1$ on the ground-state $\ket{0}$.  We would like to show that  $\bra{0} \prod X_1 \mathcal{T} \prod X_1\ket{0}=0$, meaning that $\prod X_1 \ket{0}$ is orthogonal to its time-reversal partner. Using Eq. \eqref{equ:T-lineon1}, we find that
\begin{equation}
\prod X_1 \mathcal{T} \prod X_1 = \raisebox{-.5\height}{\includegraphics[]{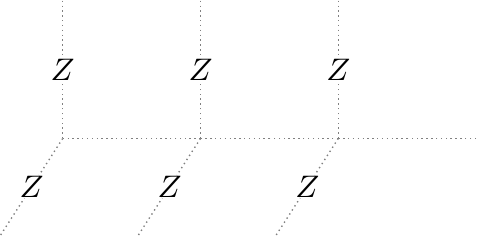}}
\cdots
\end{equation}
Since $\prod X_1 \mathcal{T} \prod X_1\ket{0}$ is a state with two fractons excited at each end point of the string, it must be orthogonal to $\ket{0}$. Thus, $\bra{0} \prod X_1 \mathcal{T} \prod X_1\ket{0}=0$ as desired.

We would like to emphasize that the action of $\mathcal{T}$ on $\prod X_1\ket{0}$ does \textit{not} attach a pair of fractons to the end of the lineon string. The states $\prod X_1\ket{0}$ and $\mathcal{T}\prod X_1\ket{0}$ are degenerate. This is seen from the fact that the cube operators are always annihilated at the end points of the lineon string, so fracton excitations cannot be created by acting with time-reversal.

The second generator can be analyzed in an identical manner. Defining, $U_{2,\text{gauged}}XU_{2,\text{gauged}}^\dagger \equiv X_2$, we see that
\begin{equation}
\prod X_2 \mathcal{T} \prod X_2 = \raisebox{-.5\height}{\includegraphics[]{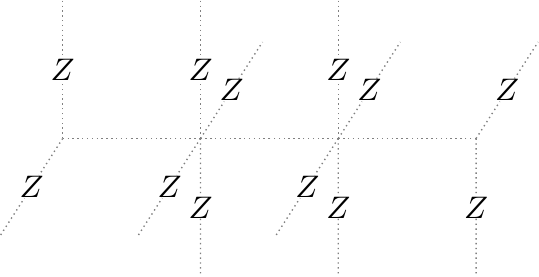}}
\end{equation}
The state $\prod X_1 \mathcal{T} \prod X_1\ket{0}$ has four excited fractons at each end point, and is therefore orthogonal to $\ket{0}$.

\begin{table*}[!t]
\caption{The generators of SSPT's protected by planar symmetries in the FCC lattice. The first three generators are without time reversal, and the remaining two is with time-reversal.}
\begin{tabular}{|c|c|c|c|c|}
\hline
 SSPT ($\tilde X$) & (hyper)cluster State & Comment & Fracton ($M$) &Comment\\
 \hline
$-X$
& 
N/A& Charge decorated (weak)
& $-$ \raisebox{-.5\height}{\includegraphics[]{3Dline2.pdf}} &
\\
\hline
\raisebox{-.5\height}{\includegraphics[]{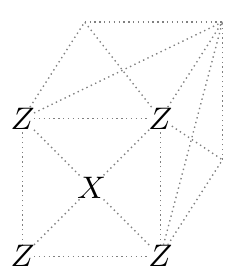}} 
& 
\raisebox{-.5\height}{\includegraphics[]{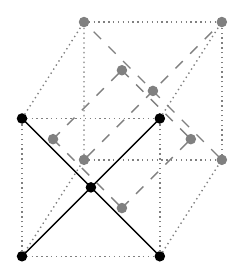}} 
& 2D SSPT stack in (100)
&
\raisebox{-.5\height}{\includegraphics[]{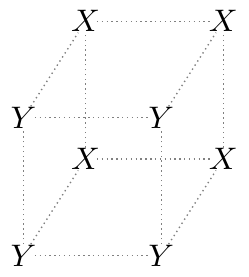}} 
& 
\\
\hline 
\raisebox{-.5\height}{\includegraphics[]{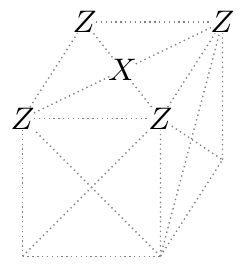}} 
& 
\raisebox{-.5\height}{\includegraphics[]{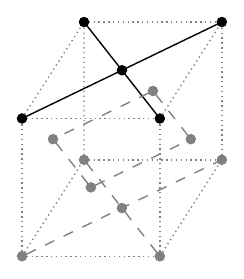}} 
& 2D SSPT stack in (010) &
\raisebox{-.5\height}{\includegraphics[]{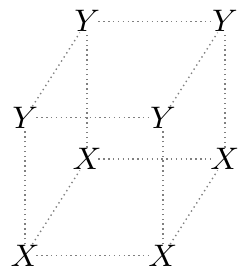}} 
&
\\
\hline 
\raisebox{-.5\height}{\includegraphics[]{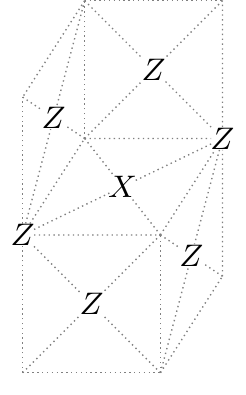}} 
& 
\raisebox{-.5\height}{\includegraphics[]{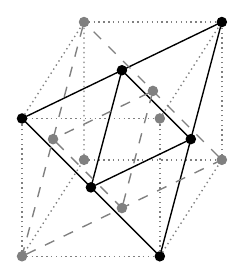}} 
& 2D SSPT stack in (111)&
\raisebox{-.5\height}{\includegraphics[]{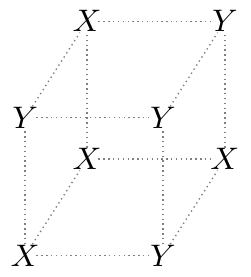}} 
&
\\
\hhline{=====}
\raisebox{-.5\height}{\includegraphics[]{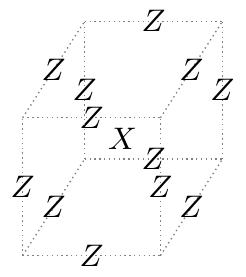}} 
&
\raisebox{-.5\height}{\includegraphics[]{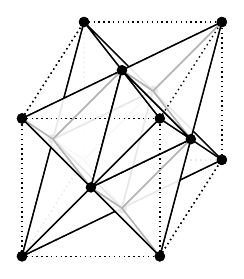}} 
& Cluster state on FCC&
\raisebox{-.5\height}{\includegraphics[]{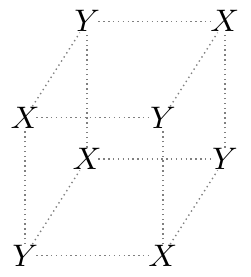}} 
&
\\
\hline
\raisebox{-.5\height}{\includegraphics[]{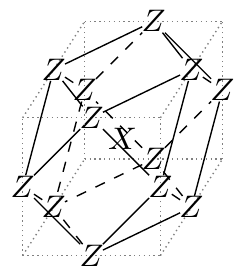}} 
&
\raisebox{-.5\height}{\includegraphics[]{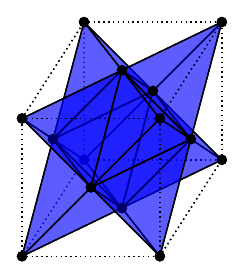}} 
& Hypercluster state on FCC&
\raisebox{-.5\height}{\includegraphics[]{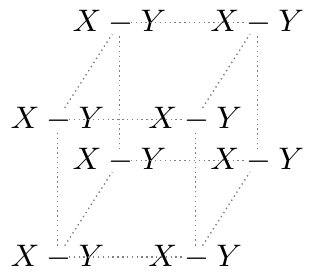}} 
&
Fracton permutation by $\mathcal T$
\\
\hline
\end{tabular}
\label{tab:checkerboard}
\end{table*}

\subsection{3D FCC lattice with planar symmetry}\label{checkerboard}
We consider the dual of the checkerboard model \cite{VijayHaahFu2016}, which is an FCC lattice with planar symmetries spanning the three faces of the cubes. The three translation vectors are given by translating a vertex to its nearest face centers.

The SSPT's are given in  Table \ref{tab:checkerboard}. Without time reversal, we find four generators. The first is a weak phase, dual to the checkerboard model where an ``$X$''-type fracton sits at each cube center of the checkerboard.  The next two generators are stacks of 2D cluster states on a square lattice sitting on planes that are parallel to the sides of the cube. We remark that the cluster state in the third direction can be realized as a combination of these two generators under a symmetric unitary transformation. In other words, the cluster state in all three planes can be simultaneously (but not independently) ``nucleated" from the vacuum symmetrically. In contrast, the stack of 2D cluster states on a cubic lattice were all distinct. This demonstrates the dependence of the classification on the underlying geometry as per usual crystalline phases.

The final generator is a stack of 2D cluster states on the triangular lattice, which lives on the $(111)$ planes. The intersection of the three planar symmetries with this plane forms the three line symmetries that protect the 2D cluster state on that plane.

Ignoring the first generator, the remaining three generator contains eight SSPT phases and has a nice interpretation in terms of stacked 2D cluster states. Apart from the trivial product state, there are three planes ($(100)$, $(010)$, and $(001)$) to put 2D cluster states on the square lattice. Furthermore, there are four planes ($(111)$, $(\bar111)$, $(1\bar11)$, and $(11\bar1)$) to put the cluster state on the triangular lattice. %We suspect that putting 2D cluster state on any other choice of planes will break at least one of the translation symmetries. 

To trivialize the above weak phases, it is easiest to see this by trivializing the dual model. Indeed, properly applying a certain product of $S$ gates that breaks translation can change the $Y$ operators back to $X$ operators so that we recover the original checkerboard model.

With time-reversal, we find two more generators. The first is a cluster state where the edges are connecting all the vertices to its nearest face centers.

The second generator is a hypercluster state, created by acting with $CCZ$ on all the triangles that are boundaries of tetrahedra. In terms of symmetry defects, we notice that the number of defects in any configuration are always multiples of four, so each defect can be attached with a phase factor $e^{\frac{\pi i}{4}}$ while keeping the overall wavefunction real.

The dual of the second generator is the following modified Checkerboard model. 
\begin{widetext}
\begin{equation}
H'_\text{gauged} = -\frac{1}{32}\sum_i \left [
\raisebox{-.5\height}{\includegraphics[]{Checkerboard15.pdf}} 
+
\raisebox{-.5\height}{\includegraphics[]{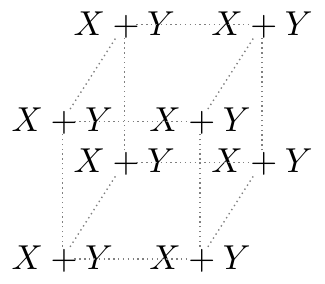}} 
\right].
\label{equ:Tenrichedcheckerboard}
\end{equation}
\end{widetext}
The operator $(X\pm Y)/\sqrt{2}$ acting on a vertex excites four fractons for each type of cube term. We see that time-reversal permutes these two types of fracton excitations. Indeed, if time-reversal is broken, then it is equivalent to the usual checkerboard model via a transversal $T = \text{diag}(1,\exp \frac{\pi i}{4})$ gate.

It might be at first confusing why a time-reversal model of order two can be disentangled with an operator of order eight. Consider squaring the disentangler, so that it is now a transversal $S$ gate. Then, conjugating the cube of $X$'s transforms it into a cube of $Y$'s. However, this is just the product of a cube of $X$'s and a cube of $Z$'s. Thus, in the gauge-invariant subspace, the action of the transversal $T$ gate actually has only order 2.

In the SSPT side, the transversal $T$ gate ungauges to the following finite depth unitary
\begin{align}
    U = \prod_i& \exp \left[\frac{\pi i}{8} (Z_1Z_xZ_yZ_z + Z_1Z_{\bar x}Z_{\bar y} Z_{\bar z})\right]\\
    = \prod_i& CCZ_{1,x,y}CCZ_{1,y,z}CCZ_{1,x,z}CCZ_{x,y,z}\nonumber \\
   & \times CCZ_{1,\bar x,\bar y}CCZ_{1,\bar y,\bar z}CCZ_{1,\bar x,\bar z}CCZ_{\bar x,\bar y,\bar z}\nonumber \\
    &\times  CZ_{1,x}CZ_{1,y}CZ_{1,z}CZ_{x,y}CZ_{y,z}CZ_{x,z}\nonumber\\
    &\times CZ_{1,\bar x}CZ_{1,\bar y}CZ_{1,\bar z}CZ_{\bar x,\bar y}CZ_{\bar y,\bar z}CZ_{\bar x,\bar z}
\end{align}
\begin{table*}[!t]
\caption{SSPT's on a cubic lattice with four planar symmetries. All solutions found are weak. The dual of the trivial SSPT corresponds to the Double of the Chamon model.}
\begin{tabular}{|c|c|c|}
\hline
SSPT($\tilde X$) & Comment & Fracton($M$) \\
 \hline
  $-X$ & Charge decorated&
  $-$\raisebox{-.5\height}{\includegraphics[scale=0.9]{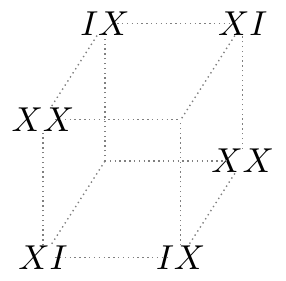}}\\
\hline
\raisebox{-.5\height}{\includegraphics[scale=0.9]{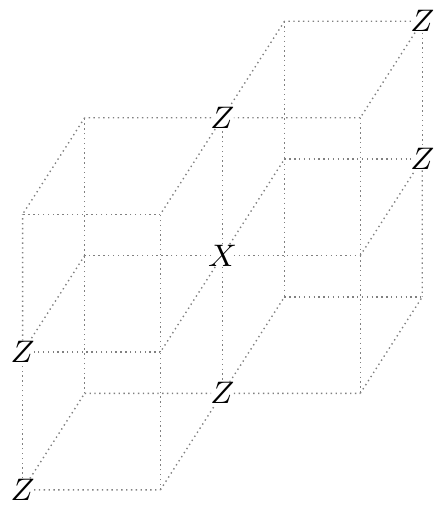}}
& 2D SSPT stacks in (110) planes &
\raisebox{-.5\height}{\includegraphics[scale=0.9]{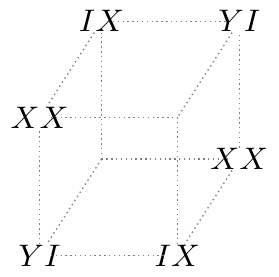}}
\\
\hline
\raisebox{-.5\height}{\includegraphics[scale=0.9]{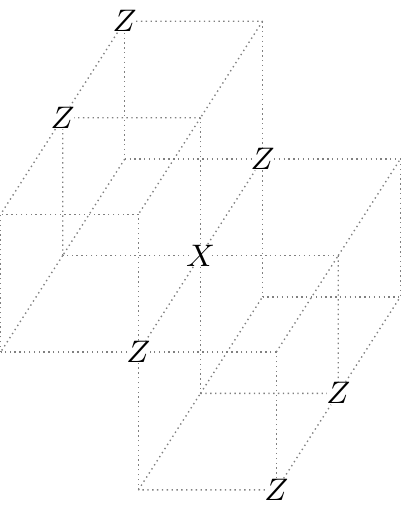}}
& 2D SSPT stacks in (011) planes
&\raisebox{-.5\height}{\includegraphics[scale=0.9]{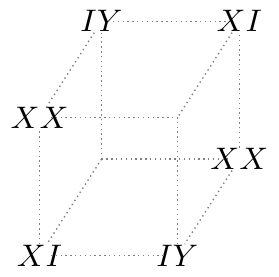}}
\\
\hhline{===}
\raisebox{-.5\height}{\includegraphics[scale=0.9]{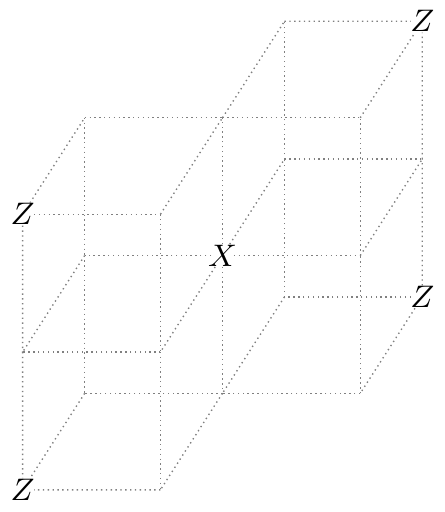}}
& 2D $\mathcal T$ SSPT stacks in (110) planes&
\raisebox{-.5\height}{\includegraphics[scale=0.9]{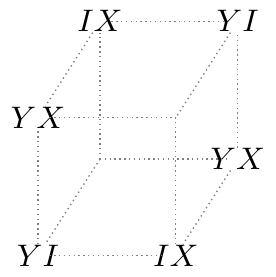}}
 \\
\hline
\raisebox{-.5\height}{\includegraphics[scale=0.9]{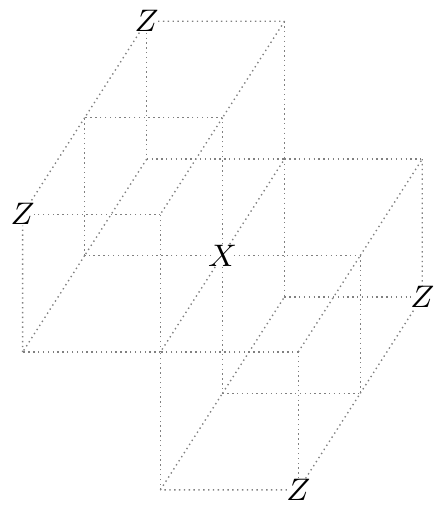}}
& 2D $\mathcal T$ SSPT stacks in (011) planes&
\raisebox{-.5\height}{\includegraphics[scale=0.9]{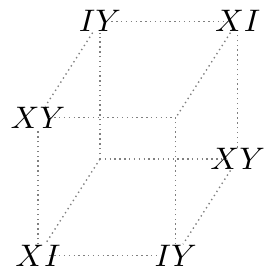}}
\\
 \hline
 \end{tabular}
 \label{tab:chamondouble}
\end{table*}
where $x,y,z$ denote the primitive vectors of the FCC lattice which points from a corner to the nearest face centers. This show that the state can be trivialized by breaking either time-reversal symmetry or planar symmetries. The product of four $Z$'s above are just the two Ising terms of the tetrahedral Ising model. By squaring this unitary, one can show that it is equal to the identity up to boundary terms. Thus, on a closed manifold, this unitary trivializes the SPT by breaking time-reversal.

 We remark that a cousin of this model was constructed in Ref. \onlinecite{YouLitinskivonOppen2018}, where reflection swaps $X$ and $Z$-type fractons in the checkerboard model. This provides some evidence that despite lacking Lorentz symmetry, the ``Crystalline correspondence" \cite{ThorngrenElse2018} could still be valid in fracton phases as well.

\subsection{Chamon Double}\label{Chamon}
We study the cubic lattice with planar symmetries in the $(100)$, $(010)$, $(001)$, and  $(111)$ planes. The Ising terms are given by
\begin{equation}
\raisebox{-.5\height}{\includegraphics[]{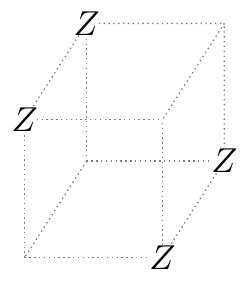}}
,
\raisebox{-.5\height}{\includegraphics[]{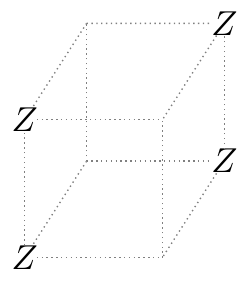}}
\end{equation}
which commutes with the planar symmetries. After gauging, the stabilizers are given by
\begin{equation}
\raisebox{-.5\height}{\includegraphics[]{Chamon3.pdf}}
,
\raisebox{-.5\height}{\includegraphics[]{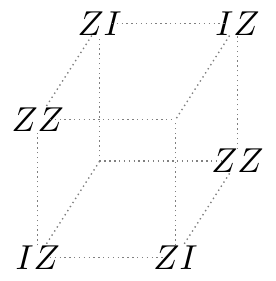}}
\end{equation}
They correspond to the double of Chamon model\cite{Chamon2005,BravyiLeemhuisTerhal2011}, which can be written on the cubic lattice as
\begin{align}
\raisebox{-.5\height}{\includegraphics[]{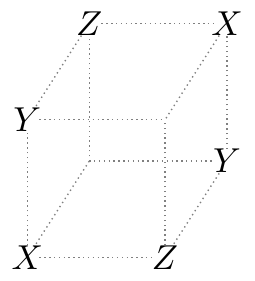}}.
\end{align}

The SSPT's are shown in Table \ref{tab:chamondouble}. We find 3 generators without time-reversal, two of which are stacks of the 2D SSPT state in the $(110)$ and $(011)$ planes.  With time-reversal, we find two more solutions, which are the time-reversal 2D SSPT living in the same planes.

\subsection{Haah's Code}\label{Haah}
We consider the dual of Haah's code\cite{Haah2011,VijayHaahFu2016}, which is an Ising model with fractal symmetries. The Ising terms are given by

\begin{align}
\raisebox{-.5\height}{\includegraphics[]{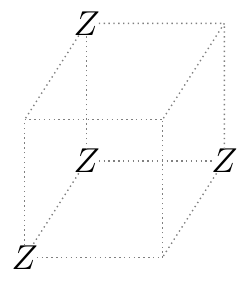}}
&,
\raisebox{-.5\height}{\includegraphics[scale=0.9]{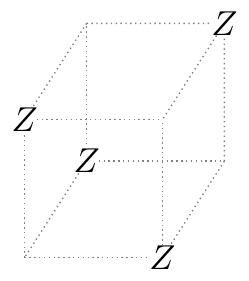}}
\end{align}

The solutions are listed in Table \ref{tab:Haah}. Without time reversal, we find only the weak solution which decorates a charge in every unit cell. With time reversal, we find two solutions. In the dual model, they can be both be trivialized by acting with a transversal $S$ gate in the first and second species, respectively when time-reversal is broken.

We remark that the second generator is also an SSPT protected by planar symmetries. It is a combination of the generators $1,2$ and $3$ in table \ref{tab:Xcubetimereversal}. An example of a fixed point wavefunction which is protected by both fractal and non-fractal symmetries has been discussed for 2D SSPT's in Ref. \onlinecite{Stephenetal2019}.

\begin{table}[!t]
\caption{Fractal SSPT's in 3D. The trivial SSPT is dual to Haah's code.}
\begin{tabular}{|c|c|}
\hline
SSPT($\tilde X$) & Fracton($M$) \\
 \hline
 $-X$& $-$
 \raisebox{-.5\height}{\includegraphics[scale=0.9]{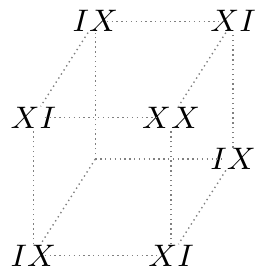}}
\\
  \hhline{==}
  \raisebox{-.5\height}{\includegraphics[scale=0.9]{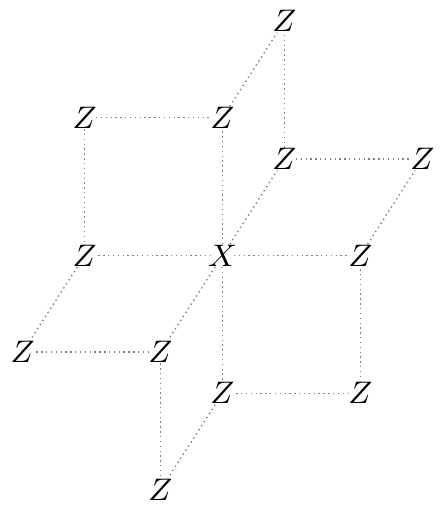}}
&
\raisebox{-.5\height}{\includegraphics[scale=0.9]{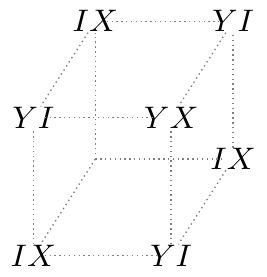}}
\\
\hline
\raisebox{-.5\height}{\includegraphics[scale=0.9]{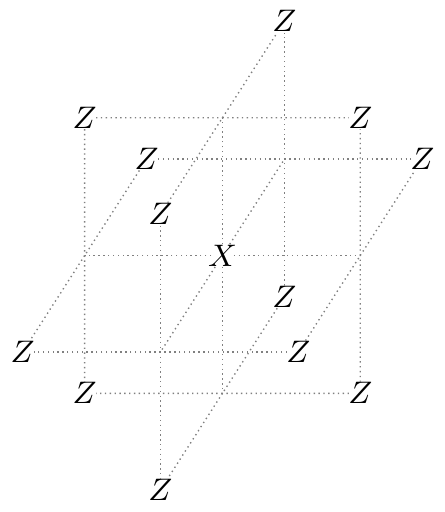}}
&
\raisebox{-.5\height}{\includegraphics[scale=0.9]{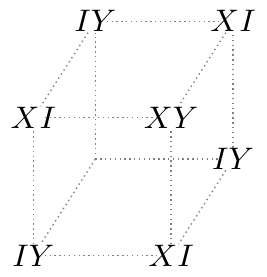}} 
\\
\hline
 \end{tabular}
 \label{tab:Haah}
\end{table}

\section{Discussion}
Using the defect homology equations, we have performed an extensive search for self-consistent translation-invariant ground-state wavefunctions that preserve subsystem symmetries. The equations can be interpreted as consistency conditions of hopping symmetry defects

Solving these equations, we obtain solutions to well-known SSPT phases, as well as new ones protected by translation and/or time-reversal symmetry. The dual of these phases give rise to new symmetry-enriched fracton phases in 3+1D.

Looking forward, we hope to use more general space group symmetries to simplify the defect homology equations, which may allow us to obtain new solutions with longer ranges of interactions that might realize new phases without the need of additional translation or time-reversal symmetry. It would also be interesting to consider non-Euclidean geometries such as hyperbolic spaces, where the translation group is non-Abelian and see whether such geometry affects the types of models that arise. One can also consider topological orders with a mix of both mobile and immobile excitations.

Perhaps most importantly, it would be crucial to find the mathematical structure that captures the consistency conditions which resemble the $F$ and $R$ symbols in 2+1D topological orders. A way to study this would be via the fusion and braiding of such excitations. Developing such a framework might allow for new models which cannot be realized as a commuting projector models, such as chiral theories.

\begin{acknowledgments}
We thank Trithep Devakul, Dominic Else, Kyle Kawagoe, Ryan Thorngren, and Ashvin Vishwanath for helpful discussions. We also acknowledge stimulating discussions with the participants of the Simons Collaboration on Ultra Quantum Matter Workshop at Harvard University, which was supported by a grant from the Simons Foundation (651440). NT acknowledges the support of the Natural Sciences and Engineering Research Council of Canada (NSERC). SV is supported by the Harvard Society of Fellows.

\end{acknowledgments}

\bibliography{references}

%apsrev4-2.bst 2019-01-14 (MD) hand-edited version of apsrev4-1.bst
%Control: key (0)
%Control: author (8) initials jnrlst
%Control: editor formatted (1) identically to author
%Control: production of article title (0) allowed
%Control: page (0) single
%Control: year (1) truncated
%Control: production of eprint (0) enabled
\begin{thebibliography}{59}%
\makeatletter
\providecommand \@ifxundefined [1]{%
 \@ifx{#1\undefined}
}%
\providecommand \@ifnum [1]{%
 \ifnum #1\expandafter \@firstoftwo
 \else \expandafter \@secondoftwo
 \fi
}%
\providecommand \@ifx [1]{%
 \ifx #1\expandafter \@firstoftwo
 \else \expandafter \@secondoftwo
 \fi
}%
\providecommand \natexlab [1]{#1}%
\providecommand \enquote  [1]{``#1''}%
\providecommand \bibnamefont  [1]{#1}%
\providecommand \bibfnamefont [1]{#1}%
\providecommand \citenamefont [1]{#1}%
\providecommand \href@noop [0]{\@secondoftwo}%
\providecommand \href [0]{\begingroup \@sanitize@url \@href}%
\providecommand \@href[1]{\@@startlink{#1}\@@href}%
\providecommand \@@href[1]{\endgroup#1\@@endlink}%
\providecommand \@sanitize@url [0]{\catcode `\\12\catcode `\$12\catcode
  `\&12\catcode `\#12\catcode `\^12\catcode `\_12\catcode `\%12\relax}%
\providecommand \@@startlink[1]{}%
\providecommand \@@endlink[0]{}%
\providecommand \url  [0]{\begingroup\@sanitize@url \@url }%
\providecommand \@url [1]{\endgroup\@href {#1}{\urlprefix }}%
\providecommand \urlprefix  [0]{URL }%
\providecommand \Eprint [0]{\href }%
\providecommand \doibase [0]{https://doi.org/}%
\providecommand \selectlanguage [0]{\@gobble}%
\providecommand \bibinfo  [0]{\@secondoftwo}%
\providecommand \bibfield  [0]{\@secondoftwo}%
\providecommand \translation [1]{[#1]}%
\providecommand \BibitemOpen [0]{}%
\providecommand \bibitemStop [0]{}%
\providecommand \bibitemNoStop [0]{.\EOS\space}%
\providecommand \EOS [0]{\spacefactor3000\relax}%
\providecommand \BibitemShut  [1]{\csname bibitem#1\endcsname}%
\let\auto@bib@innerbib\@empty
%</preamble>
\bibitem [{\citenamefont {Levin}\ and\ \citenamefont
  {Wen}(2005)}]{LevinWen2005}%
  \BibitemOpen
  \bibfield  {author} {\bibinfo {author} {\bibfnamefont {M.~A.}\ \bibnamefont
  {Levin}}\ and\ \bibinfo {author} {\bibfnamefont {X.-G.}\ \bibnamefont
  {Wen}},\ }\bibfield  {title} {\bibinfo {title} {String-net condensation: A
  physical mechanism for topological phases},\ }\href
  {https://doi.org/10.1103/PhysRevB.71.045110} {\bibfield  {journal} {\bibinfo
  {journal} {Phys. Rev. B}\ }\textbf {\bibinfo {volume} {71}},\ \bibinfo
  {pages} {045110} (\bibinfo {year} {2005})}\BibitemShut {NoStop}%
\bibitem [{\citenamefont {Haah}(2011)}]{Haah2011}%
  \BibitemOpen
  \bibfield  {author} {\bibinfo {author} {\bibfnamefont {J.}~\bibnamefont
  {Haah}},\ }\bibfield  {title} {\bibinfo {title} {Local stabilizer codes in
  three dimensions without string logical operators},\ }\href
  {https://doi.org/10.1103/PhysRevA.83.042330} {\bibfield  {journal} {\bibinfo
  {journal} {Phys. Rev. A}\ }\textbf {\bibinfo {volume} {83}},\ \bibinfo
  {pages} {042330} (\bibinfo {year} {2011})}\BibitemShut {NoStop}%
\bibitem [{\citenamefont {Vijay}\ \emph {et~al.}(2016)\citenamefont {Vijay},
  \citenamefont {Haah},\ and\ \citenamefont {Fu}}]{VijayHaahFu2016}%
  \BibitemOpen
  \bibfield  {author} {\bibinfo {author} {\bibfnamefont {S.}~\bibnamefont
  {Vijay}}, \bibinfo {author} {\bibfnamefont {J.}~\bibnamefont {Haah}},\ and\
  \bibinfo {author} {\bibfnamefont {L.}~\bibnamefont {Fu}},\ }\bibfield
  {title} {\bibinfo {title} {Fracton topological order, generalized lattice
  gauge theory, and duality},\ }\href
  {https://doi.org/10.1103/PhysRevB.94.235157} {\bibfield  {journal} {\bibinfo
  {journal} {Phys. Rev. B}\ }\textbf {\bibinfo {volume} {94}},\ \bibinfo
  {pages} {235157} (\bibinfo {year} {2016})}\BibitemShut {NoStop}%
\bibitem [{\citenamefont {Bravyi}\ and\ \citenamefont
  {Haah}(2013)}]{BravyiHaah2013}%
  \BibitemOpen
  \bibfield  {author} {\bibinfo {author} {\bibfnamefont {S.}~\bibnamefont
  {Bravyi}}\ and\ \bibinfo {author} {\bibfnamefont {J.}~\bibnamefont {Haah}},\
  }\bibfield  {title} {\bibinfo {title} {Quantum self-correction in the 3d
  cubic code model},\ }\href {https://doi.org/10.1103/PhysRevLett.111.200501}
  {\bibfield  {journal} {\bibinfo  {journal} {Phys. Rev. Lett.}\ }\textbf
  {\bibinfo {volume} {111}},\ \bibinfo {pages} {200501} (\bibinfo {year}
  {2013})}\BibitemShut {NoStop}%
\bibitem [{\citenamefont {Haah}(2013)}]{Haah2013}%
  \BibitemOpen
  \bibfield  {author} {\bibinfo {author} {\bibfnamefont {J.}~\bibnamefont
  {Haah}},\ }\bibfield  {title} {\bibinfo {title} {Commuting pauli hamiltonians
  as maps between free modules},\ }\href@noop {} {\bibfield  {journal}
  {\bibinfo  {journal} {Communications in Mathematical Physics}\ }\textbf
  {\bibinfo {volume} {324}},\ \bibinfo {pages} {351} (\bibinfo {year}
  {2013})}\BibitemShut {NoStop}%
\bibitem [{\citenamefont {Vijay}\ \emph {et~al.}(2015)\citenamefont {Vijay},
  \citenamefont {Haah},\ and\ \citenamefont {Fu}}]{VijayHaahFu2015}%
  \BibitemOpen
  \bibfield  {author} {\bibinfo {author} {\bibfnamefont {S.}~\bibnamefont
  {Vijay}}, \bibinfo {author} {\bibfnamefont {J.}~\bibnamefont {Haah}},\ and\
  \bibinfo {author} {\bibfnamefont {L.}~\bibnamefont {Fu}},\ }\bibfield
  {title} {\bibinfo {title} {A new kind of topological quantum order: A
  dimensional hierarchy of quasiparticles built from stationary excitations},\
  }\href {https://doi.org/10.1103/PhysRevB.92.235136} {\bibfield  {journal}
  {\bibinfo  {journal} {Phys. Rev. B}\ }\textbf {\bibinfo {volume} {92}},\
  \bibinfo {pages} {235136} (\bibinfo {year} {2015})}\BibitemShut {NoStop}%
\bibitem [{\citenamefont {Pretko}(2017)}]{Pretko2017}%
  \BibitemOpen
  \bibfield  {author} {\bibinfo {author} {\bibfnamefont {M.}~\bibnamefont
  {Pretko}},\ }\bibfield  {title} {\bibinfo {title} {Subdimensional particle
  structure of higher rank $u(1)$ spin liquids},\ }\href
  {https://doi.org/10.1103/PhysRevB.95.115139} {\bibfield  {journal} {\bibinfo
  {journal} {Phys. Rev. B}\ }\textbf {\bibinfo {volume} {95}},\ \bibinfo
  {pages} {115139} (\bibinfo {year} {2017})}\BibitemShut {NoStop}%
\bibitem [{\citenamefont {Ma}\ \emph {et~al.}(2017)\citenamefont {Ma},
  \citenamefont {Lake}, \citenamefont {Chen},\ and\ \citenamefont
  {Hermele}}]{MaLakeChenHermele2017}%
  \BibitemOpen
  \bibfield  {author} {\bibinfo {author} {\bibfnamefont {H.}~\bibnamefont
  {Ma}}, \bibinfo {author} {\bibfnamefont {E.}~\bibnamefont {Lake}}, \bibinfo
  {author} {\bibfnamefont {X.}~\bibnamefont {Chen}},\ and\ \bibinfo {author}
  {\bibfnamefont {M.}~\bibnamefont {Hermele}},\ }\bibfield  {title} {\bibinfo
  {title} {Fracton topological order via coupled layers},\ }\href
  {https://doi.org/10.1103/PhysRevB.95.245126} {\bibfield  {journal} {\bibinfo
  {journal} {Phys. Rev. B}\ }\textbf {\bibinfo {volume} {95}},\ \bibinfo
  {pages} {245126} (\bibinfo {year} {2017})}\BibitemShut {NoStop}%
\bibitem [{\citenamefont {Vijay}(2017)}]{Vijay2017}%
  \BibitemOpen
  \bibfield  {author} {\bibinfo {author} {\bibfnamefont {S.}~\bibnamefont
  {Vijay}},\ }\bibfield  {title} {\bibinfo {title} {Isotropic layer
  construction and phase diagram for fracton topological phases},\ }\href@noop
  {} {\bibfield  {journal} {\bibinfo  {journal} {arXiv preprint
  arXiv:1701.00762}\ } (\bibinfo {year} {2017})}\BibitemShut {NoStop}%
\bibitem [{\citenamefont {Shirley}\ \emph {et~al.}(2018)\citenamefont
  {Shirley}, \citenamefont {Slagle}, \citenamefont {Wang},\ and\ \citenamefont
  {Chen}}]{ShirleySlagleWangChen2018}%
  \BibitemOpen
  \bibfield  {author} {\bibinfo {author} {\bibfnamefont {W.}~\bibnamefont
  {Shirley}}, \bibinfo {author} {\bibfnamefont {K.}~\bibnamefont {Slagle}},
  \bibinfo {author} {\bibfnamefont {Z.}~\bibnamefont {Wang}},\ and\ \bibinfo
  {author} {\bibfnamefont {X.}~\bibnamefont {Chen}},\ }\bibfield  {title}
  {\bibinfo {title} {Fracton models on general three-dimensional manifolds},\
  }\href {https://doi.org/10.1103/PhysRevX.8.031051} {\bibfield  {journal}
  {\bibinfo  {journal} {Phys. Rev. X}\ }\textbf {\bibinfo {volume} {8}},\
  \bibinfo {pages} {031051} (\bibinfo {year} {2018})}\BibitemShut {NoStop}%
\bibitem [{\citenamefont {Raussendorf}\ \emph {et~al.}(2019)\citenamefont
  {Raussendorf}, \citenamefont {Okay}, \citenamefont {Wang}, \citenamefont
  {Stephen},\ and\ \citenamefont {Nautrup}}]{Raussendorfetal2019}%
  \BibitemOpen
  \bibfield  {author} {\bibinfo {author} {\bibfnamefont {R.}~\bibnamefont
  {Raussendorf}}, \bibinfo {author} {\bibfnamefont {C.}~\bibnamefont {Okay}},
  \bibinfo {author} {\bibfnamefont {D.-S.}\ \bibnamefont {Wang}}, \bibinfo
  {author} {\bibfnamefont {D.~T.}\ \bibnamefont {Stephen}},\ and\ \bibinfo
  {author} {\bibfnamefont {H.~P.}\ \bibnamefont {Nautrup}},\ }\bibfield
  {title} {\bibinfo {title} {Computationally universal phase of quantum
  matter},\ }\href {https://doi.org/10.1103/PhysRevLett.122.090501} {\bibfield
  {journal} {\bibinfo  {journal} {Phys. Rev. Lett.}\ }\textbf {\bibinfo
  {volume} {122}},\ \bibinfo {pages} {090501} (\bibinfo {year}
  {2019})}\BibitemShut {NoStop}%
\bibitem [{\citenamefont {Devakul}\ and\ \citenamefont
  {Williamson}(2018)}]{DevakulWilliamson2018}%
  \BibitemOpen
  \bibfield  {author} {\bibinfo {author} {\bibfnamefont {T.}~\bibnamefont
  {Devakul}}\ and\ \bibinfo {author} {\bibfnamefont {D.~J.}\ \bibnamefont
  {Williamson}},\ }\bibfield  {title} {\bibinfo {title} {Universal quantum
  computation using fractal symmetry-protected cluster phases},\ }\href
  {https://doi.org/10.1103/PhysRevA.98.022332} {\bibfield  {journal} {\bibinfo
  {journal} {Phys. Rev. A}\ }\textbf {\bibinfo {volume} {98}},\ \bibinfo
  {pages} {022332} (\bibinfo {year} {2018})}\BibitemShut {NoStop}%
\bibitem [{\citenamefont {Stephen}\ \emph {et~al.}(2019)\citenamefont
  {Stephen}, \citenamefont {Nautrup}, \citenamefont {Bermejo-Vega},
  \citenamefont {Eisert},\ and\ \citenamefont {Raussendorf}}]{Stephenetal2019}%
  \BibitemOpen
  \bibfield  {author} {\bibinfo {author} {\bibfnamefont {D.~T.}\ \bibnamefont
  {Stephen}}, \bibinfo {author} {\bibfnamefont {H.~P.}\ \bibnamefont
  {Nautrup}}, \bibinfo {author} {\bibfnamefont {J.}~\bibnamefont
  {Bermejo-Vega}}, \bibinfo {author} {\bibfnamefont {J.}~\bibnamefont
  {Eisert}},\ and\ \bibinfo {author} {\bibfnamefont {R.}~\bibnamefont
  {Raussendorf}},\ }\bibfield  {title} {\bibinfo {title} {Subsystem symmetries,
  quantum cellular automata, and computational phases of quantum matter},\
  }\href {https://doi.org/10.22331/q-2019-05-20-142} {\bibfield  {journal}
  {\bibinfo  {journal} {Quantum}\ }\textbf {\bibinfo {volume} {3}},\ \bibinfo
  {pages} {142} (\bibinfo {year} {2019})}\BibitemShut {NoStop}%
\bibitem [{\citenamefont {Daniel}\ \emph {et~al.}(2020)\citenamefont {Daniel},
  \citenamefont {Alexander},\ and\ \citenamefont
  {Miyake}}]{DanielAlexanderMiyake2019}%
  \BibitemOpen
  \bibfield  {author} {\bibinfo {author} {\bibfnamefont {A.~K.}\ \bibnamefont
  {Daniel}}, \bibinfo {author} {\bibfnamefont {R.~N.}\ \bibnamefont
  {Alexander}},\ and\ \bibinfo {author} {\bibfnamefont {A.}~\bibnamefont
  {Miyake}},\ }\bibfield  {title} {\bibinfo {title} {Computational universality
  of symmetry-protected topologically ordered cluster phases on 2d archimedean
  lattices},\ }\href {https://doi.org/10.22331/q-2020-02-10-228} {\bibfield
  {journal} {\bibinfo  {journal} {Quantum}\ }\textbf {\bibinfo {volume} {4}},\
  \bibinfo {pages} {228} (\bibinfo {year} {2020})}\BibitemShut {NoStop}%
\bibitem [{\citenamefont {Kugel}\ and\ \citenamefont
  {Khomski{\u\i}}(1982)}]{KugelKhomski1982}%
  \BibitemOpen
  \bibfield  {author} {\bibinfo {author} {\bibfnamefont {K.~I.}\ \bibnamefont
  {Kugel}}\ and\ \bibinfo {author} {\bibfnamefont {D.}~\bibnamefont
  {Khomski{\u\i}}},\ }\bibfield  {title} {\bibinfo {title} {The jahn-teller
  effect and magnetism: transition metal compounds},\ }\href@noop {} {\bibfield
   {journal} {\bibinfo  {journal} {Soviet Physics Uspekhi}\ }\textbf {\bibinfo
  {volume} {25}},\ \bibinfo {pages} {231} (\bibinfo {year} {1982})}\BibitemShut
  {NoStop}%
\bibitem [{\citenamefont {Levin}\ and\ \citenamefont {Gu}(2012)}]{LevinGu2012}%
  \BibitemOpen
  \bibfield  {author} {\bibinfo {author} {\bibfnamefont {M.}~\bibnamefont
  {Levin}}\ and\ \bibinfo {author} {\bibfnamefont {Z.-C.}\ \bibnamefont {Gu}},\
  }\bibfield  {title} {\bibinfo {title} {Braiding statistics approach to
  symmetry-protected topological phases},\ }\href
  {https://doi.org/10.1103/PhysRevB.86.115109} {\bibfield  {journal} {\bibinfo
  {journal} {Phys. Rev. B}\ }\textbf {\bibinfo {volume} {86}},\ \bibinfo
  {pages} {115109} (\bibinfo {year} {2012})}\BibitemShut {NoStop}%
\bibitem [{\citenamefont {Else}\ and\ \citenamefont
  {Thorngren}(2019)}]{ElseThorngren2019}%
  \BibitemOpen
  \bibfield  {author} {\bibinfo {author} {\bibfnamefont {D.~V.}\ \bibnamefont
  {Else}}\ and\ \bibinfo {author} {\bibfnamefont {R.}~\bibnamefont
  {Thorngren}},\ }\bibfield  {title} {\bibinfo {title} {Crystalline topological
  phases as defect networks},\ }\href
  {https://doi.org/10.1103/PhysRevB.99.115116} {\bibfield  {journal} {\bibinfo
  {journal} {Phys. Rev. B}\ }\textbf {\bibinfo {volume} {99}},\ \bibinfo
  {pages} {115116} (\bibinfo {year} {2019})}\BibitemShut {NoStop}%
\bibitem [{\citenamefont {Song}\ \emph {et~al.}(2018)\citenamefont {Song},
  \citenamefont {Fang},\ and\ \citenamefont {Qi}}]{SongFangQi2018}%
  \BibitemOpen
  \bibfield  {author} {\bibinfo {author} {\bibfnamefont {Z.}~\bibnamefont
  {Song}}, \bibinfo {author} {\bibfnamefont {C.}~\bibnamefont {Fang}},\ and\
  \bibinfo {author} {\bibfnamefont {Y.}~\bibnamefont {Qi}},\ }\bibfield
  {title} {\bibinfo {title} {Real-space recipes for general topological
  crystalline states},\ }\href@noop {} {\bibfield  {journal} {\bibinfo
  {journal} {arXiv preprint arXiv:1810.11013}\ } (\bibinfo {year}
  {2018})}\BibitemShut {NoStop}%
\bibitem [{\citenamefont {Song}\ \emph {et~al.}(2019)\citenamefont {Song},
  \citenamefont {Huang}, \citenamefont {Qi}, \citenamefont {Fang},\ and\
  \citenamefont {Hermele}}]{Songetal2018}%
  \BibitemOpen
  \bibfield  {author} {\bibinfo {author} {\bibfnamefont {Z.}~\bibnamefont
  {Song}}, \bibinfo {author} {\bibfnamefont {S.-J.}\ \bibnamefont {Huang}},
  \bibinfo {author} {\bibfnamefont {Y.}~\bibnamefont {Qi}}, \bibinfo {author}
  {\bibfnamefont {C.}~\bibnamefont {Fang}},\ and\ \bibinfo {author}
  {\bibfnamefont {M.}~\bibnamefont {Hermele}},\ }\bibfield  {title} {\bibinfo
  {title} {Topological states from topological crystals},\ }\href
  {https://doi.org/10.1126/sciadv.aax2007} {\bibfield  {journal} {\bibinfo
  {journal} {Science Advances}\ }\textbf {\bibinfo {volume} {5}},\ \bibinfo
  {pages} {eaax2007} (\bibinfo {year} {2019})}\BibitemShut {NoStop}%
\bibitem [{\citenamefont {Shiozaki}\ \emph {et~al.}(2018)\citenamefont
  {Shiozaki}, \citenamefont {Xiong},\ and\ \citenamefont
  {Gomi}}]{ShiozakiXiongGomi2018}%
  \BibitemOpen
  \bibfield  {author} {\bibinfo {author} {\bibfnamefont {K.}~\bibnamefont
  {Shiozaki}}, \bibinfo {author} {\bibfnamefont {C.~Z.}\ \bibnamefont
  {Xiong}},\ and\ \bibinfo {author} {\bibfnamefont {K.}~\bibnamefont {Gomi}},\
  }\bibfield  {title} {\bibinfo {title} {Generalized homology and
  atiyah-hirzebruch spectral sequence in crystalline symmetry protected
  topological phenomena},\ }\href@noop {} {\bibfield  {journal} {\bibinfo
  {journal} {arXiv preprint arXiv:1810.00801}\ } (\bibinfo {year}
  {2018})}\BibitemShut {NoStop}%
\bibitem [{\citenamefont {Hu}\ \emph {et~al.}(2013)\citenamefont {Hu},
  \citenamefont {Wan},\ and\ \citenamefont {Wu}}]{HuWanWu2013}%
  \BibitemOpen
  \bibfield  {author} {\bibinfo {author} {\bibfnamefont {Y.}~\bibnamefont
  {Hu}}, \bibinfo {author} {\bibfnamefont {Y.}~\bibnamefont {Wan}},\ and\
  \bibinfo {author} {\bibfnamefont {Y.-S.}\ \bibnamefont {Wu}},\ }\bibfield
  {title} {\bibinfo {title} {Twisted quantum double model of topological phases
  in two dimensions},\ }\href {https://doi.org/10.1103/PhysRevB.87.125114}
  {\bibfield  {journal} {\bibinfo  {journal} {Phys. Rev. B}\ }\textbf {\bibinfo
  {volume} {87}},\ \bibinfo {pages} {125114} (\bibinfo {year}
  {2013})}\BibitemShut {NoStop}%
\bibitem [{\citenamefont {Dijkgraaf}\ and\ \citenamefont
  {Witten}(1990)}]{DijkgraafWitten}%
  \BibitemOpen
  \bibfield  {author} {\bibinfo {author} {\bibfnamefont {R.}~\bibnamefont
  {Dijkgraaf}}\ and\ \bibinfo {author} {\bibfnamefont {E.}~\bibnamefont
  {Witten}},\ }\bibfield  {title} {\bibinfo {title} {{Topological Gauge
  Theories and Group Cohomology}},\ }\href {https://doi.org/10.1007/BF02096988}
  {\bibfield  {journal} {\bibinfo  {journal} {Commun. Math. Phys.}\ }\textbf
  {\bibinfo {volume} {129}},\ \bibinfo {pages} {393} (\bibinfo {year}
  {1990})}\BibitemShut {NoStop}%
%%CITATION = CMPHA,129,393;%%
\bibitem [{\citenamefont {Chen}\ \emph {et~al.}(2013)\citenamefont {Chen},
  \citenamefont {Gu}, \citenamefont {Liu},\ and\ \citenamefont
  {Wen}}]{Chenetal2013}%
  \BibitemOpen
  \bibfield  {author} {\bibinfo {author} {\bibfnamefont {X.}~\bibnamefont
  {Chen}}, \bibinfo {author} {\bibfnamefont {Z.-C.}\ \bibnamefont {Gu}},
  \bibinfo {author} {\bibfnamefont {Z.-X.}\ \bibnamefont {Liu}},\ and\ \bibinfo
  {author} {\bibfnamefont {X.-G.}\ \bibnamefont {Wen}},\ }\bibfield  {title}
  {\bibinfo {title} {Symmetry protected topological orders and the group
  cohomology of their symmetry group},\ }\href
  {https://doi.org/10.1103/PhysRevB.87.155114} {\bibfield  {journal} {\bibinfo
  {journal} {Phys. Rev. B}\ }\textbf {\bibinfo {volume} {87}},\ \bibinfo
  {pages} {155114} (\bibinfo {year} {2013})}\BibitemShut {NoStop}%
\bibitem [{\citenamefont {Chamon}(2005)}]{Chamon2005}%
  \BibitemOpen
  \bibfield  {author} {\bibinfo {author} {\bibfnamefont {C.}~\bibnamefont
  {Chamon}},\ }\bibfield  {title} {\bibinfo {title} {Quantum glassiness in
  strongly correlated clean systems: An example of topological
  overprotection},\ }\href {https://doi.org/10.1103/PhysRevLett.94.040402}
  {\bibfield  {journal} {\bibinfo  {journal} {Phys. Rev. Lett.}\ }\textbf
  {\bibinfo {volume} {94}},\ \bibinfo {pages} {040402} (\bibinfo {year}
  {2005})}\BibitemShut {NoStop}%
\bibitem [{\citenamefont {Bravyi}\ \emph {et~al.}(2011)\citenamefont {Bravyi},
  \citenamefont {Leemhuis},\ and\ \citenamefont
  {Terhal}}]{BravyiLeemhuisTerhal2011}%
  \BibitemOpen
  \bibfield  {author} {\bibinfo {author} {\bibfnamefont {S.}~\bibnamefont
  {Bravyi}}, \bibinfo {author} {\bibfnamefont {B.}~\bibnamefont {Leemhuis}},\
  and\ \bibinfo {author} {\bibfnamefont {B.~M.}\ \bibnamefont {Terhal}},\
  }\bibfield  {title} {\bibinfo {title} {Topological order in an exactly
  solvable 3d spin model},\ }\href {https://doi.org/10.1016/j.aop.2010.11.002}
  {\bibfield  {journal} {\bibinfo  {journal} {Annals of Physics}\ }\textbf
  {\bibinfo {volume} {326}},\ \bibinfo {pages} {839} (\bibinfo {year}
  {2011})}\BibitemShut {NoStop}%
\bibitem [{\citenamefont {You}\ \emph {et~al.}(2018{\natexlab{a}})\citenamefont
  {You}, \citenamefont {Devakul}, \citenamefont {Burnell},\ and\ \citenamefont
  {Sondhi}}]{Youetal2018}%
  \BibitemOpen
  \bibfield  {author} {\bibinfo {author} {\bibfnamefont {Y.}~\bibnamefont
  {You}}, \bibinfo {author} {\bibfnamefont {T.}~\bibnamefont {Devakul}},
  \bibinfo {author} {\bibfnamefont {F.~J.}\ \bibnamefont {Burnell}},\ and\
  \bibinfo {author} {\bibfnamefont {S.~L.}\ \bibnamefont {Sondhi}},\ }\bibfield
   {title} {\bibinfo {title} {Subsystem symmetry protected topological order},\
  }\href {https://doi.org/10.1103/PhysRevB.98.035112} {\bibfield  {journal}
  {\bibinfo  {journal} {Phys. Rev. B}\ }\textbf {\bibinfo {volume} {98}},\
  \bibinfo {pages} {035112} (\bibinfo {year} {2018}{\natexlab{a}})}\BibitemShut
  {NoStop}%
\bibitem [{\citenamefont {Devakul}\ \emph {et~al.}(2018)\citenamefont
  {Devakul}, \citenamefont {Williamson},\ and\ \citenamefont
  {You}}]{DevakulWilliamsonYou2018}%
  \BibitemOpen
  \bibfield  {author} {\bibinfo {author} {\bibfnamefont {T.}~\bibnamefont
  {Devakul}}, \bibinfo {author} {\bibfnamefont {D.~J.}\ \bibnamefont
  {Williamson}},\ and\ \bibinfo {author} {\bibfnamefont {Y.}~\bibnamefont
  {You}},\ }\bibfield  {title} {\bibinfo {title} {Classification of subsystem
  symmetry-protected topological phases},\ }\href
  {https://doi.org/10.1103/PhysRevB.98.235121} {\bibfield  {journal} {\bibinfo
  {journal} {Phys. Rev. B}\ }\textbf {\bibinfo {volume} {98}},\ \bibinfo
  {pages} {235121} (\bibinfo {year} {2018})}\BibitemShut {NoStop}%
\bibitem [{\citenamefont {Devakul}\ \emph {et~al.}(2020)\citenamefont
  {Devakul}, \citenamefont {Shirley},\ and\ \citenamefont
  {Wang}}]{DevakulShirleyWang2019}%
  \BibitemOpen
  \bibfield  {author} {\bibinfo {author} {\bibfnamefont {T.}~\bibnamefont
  {Devakul}}, \bibinfo {author} {\bibfnamefont {W.}~\bibnamefont {Shirley}},\
  and\ \bibinfo {author} {\bibfnamefont {J.}~\bibnamefont {Wang}},\ }\bibfield
  {title} {\bibinfo {title} {Strong planar subsystem symmetry-protected
  topological phases and their dual fracton orders},\ }\href
  {https://doi.org/10.1103/PhysRevResearch.2.012059} {\bibfield  {journal}
  {\bibinfo  {journal} {Phys. Rev. Research}\ }\textbf {\bibinfo {volume}
  {2}},\ \bibinfo {pages} {012059} (\bibinfo {year} {2020})}\BibitemShut
  {NoStop}%
\bibitem [{\citenamefont {Pai}\ and\ \citenamefont
  {Hermele}(2019)}]{PaiHermele2019}%
  \BibitemOpen
  \bibfield  {author} {\bibinfo {author} {\bibfnamefont {S.}~\bibnamefont
  {Pai}}\ and\ \bibinfo {author} {\bibfnamefont {M.}~\bibnamefont {Hermele}},\
  }\bibfield  {title} {\bibinfo {title} {Fracton fusion and statistics},\
  }\href {https://doi.org/10.1103/PhysRevB.100.195136} {\bibfield  {journal}
  {\bibinfo  {journal} {Phys. Rev. B}\ }\textbf {\bibinfo {volume} {100}},\
  \bibinfo {pages} {195136} (\bibinfo {year} {2019})}\BibitemShut {NoStop}%
\bibitem [{\citenamefont {Wegner}(1971)}]{Wegner1971}%
  \BibitemOpen
  \bibfield  {author} {\bibinfo {author} {\bibfnamefont {F.~J.}\ \bibnamefont
  {Wegner}},\ }\bibfield  {title} {\bibinfo {title} {Duality in generalized
  ising models and phase transitions without local order parameters},\ }\href
  {https://doi.org/10.1063/1.1665530} {\bibfield  {journal} {\bibinfo
  {journal} {J. Math. Phys.}\ }\textbf {\bibinfo {volume} {12}},\ \bibinfo
  {pages} {2259} (\bibinfo {year} {1971})}\BibitemShut {NoStop}%
\bibitem [{\citenamefont {Kogut}(1979)}]{Kogut1979}%
  \BibitemOpen
  \bibfield  {author} {\bibinfo {author} {\bibfnamefont {J.~B.}\ \bibnamefont
  {Kogut}},\ }\bibfield  {title} {\bibinfo {title} {An introduction to lattice
  gauge theory and spin systems},\ }\href
  {https://doi.org/10.1103/RevModPhys.51.659} {\bibfield  {journal} {\bibinfo
  {journal} {Rev. Mod. Phys.}\ }\textbf {\bibinfo {volume} {51}},\ \bibinfo
  {pages} {659} (\bibinfo {year} {1979})}\BibitemShut {NoStop}%
\bibitem [{\citenamefont {Kubica}\ and\ \citenamefont
  {Yoshida}(2018)}]{KubicaYoshida2018}%
  \BibitemOpen
  \bibfield  {author} {\bibinfo {author} {\bibfnamefont {A.}~\bibnamefont
  {Kubica}}\ and\ \bibinfo {author} {\bibfnamefont {B.}~\bibnamefont
  {Yoshida}},\ }\bibfield  {title} {\bibinfo {title} {Ungauging quantum
  error-correcting codes},\ }\href@noop {} {\bibfield  {journal} {\bibinfo
  {journal} {arXiv preprint arXiv:1805.01836}\ } (\bibinfo {year}
  {2018})}\BibitemShut {NoStop}%
\bibitem [{\citenamefont {Williamson}(2016)}]{Williamson2016}%
  \BibitemOpen
  \bibfield  {author} {\bibinfo {author} {\bibfnamefont {D.~J.}\ \bibnamefont
  {Williamson}},\ }\bibfield  {title} {\bibinfo {title} {Fractal symmetries:
  Ungauging the cubic code},\ }\href
  {https://doi.org/10.1103/PhysRevB.94.155128} {\bibfield  {journal} {\bibinfo
  {journal} {Phys. Rev. B}\ }\textbf {\bibinfo {volume} {94}},\ \bibinfo
  {pages} {155128} (\bibinfo {year} {2016})}\BibitemShut {NoStop}%
\bibitem [{Note1()}]{Note1}%
  \BibitemOpen
  \bibinfo {note} {In general, $M_{i,i+1}^2=e^{i\theta }$, but we can redefine
  $M_{i,i+1} \rightarrow M_{i,i+1}e^{i\theta /2}$}\BibitemShut {NoStop}%
\bibitem [{\citenamefont {Huang}\ \emph {et~al.}(2017)\citenamefont {Huang},
  \citenamefont {Song}, \citenamefont {Huang},\ and\ \citenamefont
  {Hermele}}]{HuangSongHuangHermele2017}%
  \BibitemOpen
  \bibfield  {author} {\bibinfo {author} {\bibfnamefont {S.-J.}\ \bibnamefont
  {Huang}}, \bibinfo {author} {\bibfnamefont {H.}~\bibnamefont {Song}},
  \bibinfo {author} {\bibfnamefont {Y.-P.}\ \bibnamefont {Huang}},\ and\
  \bibinfo {author} {\bibfnamefont {M.}~\bibnamefont {Hermele}},\ }\bibfield
  {title} {\bibinfo {title} {Building crystalline topological phases from
  lower-dimensional states},\ }\href
  {https://doi.org/10.1103/PhysRevB.96.205106} {\bibfield  {journal} {\bibinfo
  {journal} {Phys. Rev. B}\ }\textbf {\bibinfo {volume} {96}},\ \bibinfo
  {pages} {205106} (\bibinfo {year} {2017})}\BibitemShut {NoStop}%
\bibitem [{\citenamefont {Chen}\ \emph {et~al.}(2014)\citenamefont {Chen},
  \citenamefont {Lu},\ and\ \citenamefont {Vishwanath}}]{ChenLuVishwanath2014}%
  \BibitemOpen
  \bibfield  {author} {\bibinfo {author} {\bibfnamefont {X.}~\bibnamefont
  {Chen}}, \bibinfo {author} {\bibfnamefont {Y.-M.}\ \bibnamefont {Lu}},\ and\
  \bibinfo {author} {\bibfnamefont {A.}~\bibnamefont {Vishwanath}},\ }\bibfield
   {title} {\bibinfo {title} {Symmetry-protected topological phases from
  decorated domain walls},\ }\href {https://www.nature.com/articles/ncomms4507}
  {\bibfield  {journal} {\bibinfo  {journal} {Nature communications}\ }\textbf
  {\bibinfo {volume} {5}},\ \bibinfo {pages} {3507} (\bibinfo {year}
  {2014})}\BibitemShut {NoStop}%
\bibitem [{\citenamefont {Chen}\ \emph
  {et~al.}(2011{\natexlab{a}})\citenamefont {Chen}, \citenamefont {Gu},\ and\
  \citenamefont {Wen}}]{ChenGuWen2011}%
  \BibitemOpen
  \bibfield  {author} {\bibinfo {author} {\bibfnamefont {X.}~\bibnamefont
  {Chen}}, \bibinfo {author} {\bibfnamefont {Z.-C.}\ \bibnamefont {Gu}},\ and\
  \bibinfo {author} {\bibfnamefont {X.-G.}\ \bibnamefont {Wen}},\ }\bibfield
  {title} {\bibinfo {title} {Classification of gapped symmetric phases in
  one-dimensional spin systems},\ }\href
  {https://doi.org/10.1103/PhysRevB.83.035107} {\bibfield  {journal} {\bibinfo
  {journal} {Phys. Rev. B}\ }\textbf {\bibinfo {volume} {83}},\ \bibinfo
  {pages} {035107} (\bibinfo {year} {2011}{\natexlab{a}})}\BibitemShut
  {NoStop}%
\bibitem [{\citenamefont {Chen}\ \emph
  {et~al.}(2011{\natexlab{b}})\citenamefont {Chen}, \citenamefont {Gu},\ and\
  \citenamefont {Wen}}]{ChenGuWen2011-2}%
  \BibitemOpen
  \bibfield  {author} {\bibinfo {author} {\bibfnamefont {X.}~\bibnamefont
  {Chen}}, \bibinfo {author} {\bibfnamefont {Z.-C.}\ \bibnamefont {Gu}},\ and\
  \bibinfo {author} {\bibfnamefont {X.-G.}\ \bibnamefont {Wen}},\ }\bibfield
  {title} {\bibinfo {title} {Complete classification of one-dimensional gapped
  quantum phases in interacting spin systems},\ }\href
  {https://doi.org/10.1103/PhysRevB.84.235128} {\bibfield  {journal} {\bibinfo
  {journal} {Phys. Rev. B}\ }\textbf {\bibinfo {volume} {84}},\ \bibinfo
  {pages} {235128} (\bibinfo {year} {2011}{\natexlab{b}})}\BibitemShut
  {NoStop}%
\bibitem [{\citenamefont {Schuch}\ \emph {et~al.}(2011)\citenamefont {Schuch},
  \citenamefont {P\'erez-Garc\'{\i}a},\ and\ \citenamefont
  {Cirac}}]{Schuchelat2011}%
  \BibitemOpen
  \bibfield  {author} {\bibinfo {author} {\bibfnamefont {N.}~\bibnamefont
  {Schuch}}, \bibinfo {author} {\bibfnamefont {D.}~\bibnamefont
  {P\'erez-Garc\'{\i}a}},\ and\ \bibinfo {author} {\bibfnamefont
  {I.}~\bibnamefont {Cirac}},\ }\bibfield  {title} {\bibinfo {title}
  {Classifying quantum phases using matrix product states and projected
  entangled pair states},\ }\href {https://doi.org/10.1103/PhysRevB.84.165139}
  {\bibfield  {journal} {\bibinfo  {journal} {Phys. Rev. B}\ }\textbf {\bibinfo
  {volume} {84}},\ \bibinfo {pages} {165139} (\bibinfo {year}
  {2011})}\BibitemShut {NoStop}%
\bibitem [{\citenamefont {Else}\ and\ \citenamefont
  {Nayak}(2014)}]{ElseNayak2014}%
  \BibitemOpen
  \bibfield  {author} {\bibinfo {author} {\bibfnamefont {D.~V.}\ \bibnamefont
  {Else}}\ and\ \bibinfo {author} {\bibfnamefont {C.}~\bibnamefont {Nayak}},\
  }\bibfield  {title} {\bibinfo {title} {Classifying symmetry-protected
  topological phases through the anomalous action of the symmetry on the
  edge},\ }\href {https://doi.org/10.1103/PhysRevB.90.235137} {\bibfield
  {journal} {\bibinfo  {journal} {Phys. Rev. B}\ }\textbf {\bibinfo {volume}
  {90}},\ \bibinfo {pages} {235137} (\bibinfo {year} {2014})}\BibitemShut
  {NoStop}%
\bibitem [{\citenamefont {Xu}\ and\ \citenamefont {Moore}(2004)}]{XuMoore2004}%
  \BibitemOpen
  \bibfield  {author} {\bibinfo {author} {\bibfnamefont {C.}~\bibnamefont
  {Xu}}\ and\ \bibinfo {author} {\bibfnamefont {J.~E.}\ \bibnamefont {Moore}},\
  }\bibfield  {title} {\bibinfo {title} {Strong-weak coupling self-duality in
  the two-dimensional quantum phase transition of $p+ip$ superconducting
  arrays},\ }\href {https://doi.org/10.1103/PhysRevLett.93.047003} {\bibfield
  {journal} {\bibinfo  {journal} {Phys. Rev. Lett.}\ }\textbf {\bibinfo
  {volume} {93}},\ \bibinfo {pages} {047003} (\bibinfo {year}
  {2004})}\BibitemShut {NoStop}%
\bibitem [{\citenamefont {Wen}(2003)}]{Wen2003}%
  \BibitemOpen
  \bibfield  {author} {\bibinfo {author} {\bibfnamefont {X.-G.}\ \bibnamefont
  {Wen}},\ }\bibfield  {title} {\bibinfo {title} {Quantum orders in an exact
  soluble model},\ }\href {https://doi.org/10.1103/PhysRevLett.90.016803}
  {\bibfield  {journal} {\bibinfo  {journal} {Phys. Rev. Lett.}\ }\textbf
  {\bibinfo {volume} {90}},\ \bibinfo {pages} {016803} (\bibinfo {year}
  {2003})}\BibitemShut {NoStop}%
\bibitem [{\citenamefont {Gaiotto}\ \emph {et~al.}(2015)\citenamefont
  {Gaiotto}, \citenamefont {Kapustin}, \citenamefont {Seiberg},\ and\
  \citenamefont {Willett}}]{Gaiottoetal2015}%
  \BibitemOpen
  \bibfield  {author} {\bibinfo {author} {\bibfnamefont {D.}~\bibnamefont
  {Gaiotto}}, \bibinfo {author} {\bibfnamefont {A.}~\bibnamefont {Kapustin}},
  \bibinfo {author} {\bibfnamefont {N.}~\bibnamefont {Seiberg}},\ and\ \bibinfo
  {author} {\bibfnamefont {B.}~\bibnamefont {Willett}},\ }\bibfield  {title}
  {\bibinfo {title} {{Generalized Global Symmetries}},\ }\href
  {https://doi.org/10.1007/JHEP02(2015)172} {\bibfield  {journal} {\bibinfo
  {journal} {JHEP}\ }\textbf {\bibinfo {volume} {02}},\ \bibinfo {pages}
  {172}}\BibitemShut {NoStop}%
%%CITATION = ARXIV:1412.5148;%%
\bibitem [{\citenamefont {Devakul}\ \emph {et~al.}(2019)\citenamefont
  {Devakul}, \citenamefont {You}, \citenamefont {Burnell},\ and\ \citenamefont
  {Sondhi}}]{Devakuletal2018}%
  \BibitemOpen
  \bibfield  {author} {\bibinfo {author} {\bibfnamefont {T.}~\bibnamefont
  {Devakul}}, \bibinfo {author} {\bibfnamefont {Y.}~\bibnamefont {You}},
  \bibinfo {author} {\bibfnamefont {F.~J.}\ \bibnamefont {Burnell}},\ and\
  \bibinfo {author} {\bibfnamefont {S.~L.}\ \bibnamefont {Sondhi}},\ }\bibfield
   {title} {\bibinfo {title} {{Fractal Symmetric Phases of Matter}},\ }\href
  {https://doi.org/10.21468/SciPostPhys.6.1.007} {\bibfield  {journal}
  {\bibinfo  {journal} {SciPost Phys.}\ }\textbf {\bibinfo {volume} {6}},\
  \bibinfo {pages} {7} (\bibinfo {year} {2019})}\BibitemShut {NoStop}%
\bibitem [{\citenamefont {Newman}\ and\ \citenamefont
  {Moore}(1999)}]{NewmanMoore1999}%
  \BibitemOpen
  \bibfield  {author} {\bibinfo {author} {\bibfnamefont {M.~E.~J.}\
  \bibnamefont {Newman}}\ and\ \bibinfo {author} {\bibfnamefont
  {C.}~\bibnamefont {Moore}},\ }\bibfield  {title} {\bibinfo {title} {Glassy
  dynamics and aging in an exactly solvable spin model},\ }\href
  {https://doi.org/10.1103/PhysRevE.60.5068} {\bibfield  {journal} {\bibinfo
  {journal} {Phys. Rev. E}\ }\textbf {\bibinfo {volume} {60}},\ \bibinfo
  {pages} {5068} (\bibinfo {year} {1999})}\BibitemShut {NoStop}%
\bibitem [{\citenamefont {Devakul}(2019)}]{Devakul2019}%
  \BibitemOpen
  \bibfield  {author} {\bibinfo {author} {\bibfnamefont {T.}~\bibnamefont
  {Devakul}},\ }\bibfield  {title} {\bibinfo {title} {Classifying local fractal
  subsystem symmetry-protected topological phases},\ }\href
  {https://doi.org/10.1103/PhysRevB.99.235131} {\bibfield  {journal} {\bibinfo
  {journal} {Phys. Rev. B}\ }\textbf {\bibinfo {volume} {99}},\ \bibinfo
  {pages} {235131} (\bibinfo {year} {2019})}\BibitemShut {NoStop}%
\bibitem [{\citenamefont {Wang}\ \emph {et~al.}(2019)\citenamefont {Wang},
  \citenamefont {Shirley},\ and\ \citenamefont {Chen}}]{WangShirleyChen2019}%
  \BibitemOpen
  \bibfield  {author} {\bibinfo {author} {\bibfnamefont {T.}~\bibnamefont
  {Wang}}, \bibinfo {author} {\bibfnamefont {W.}~\bibnamefont {Shirley}},\ and\
  \bibinfo {author} {\bibfnamefont {X.}~\bibnamefont {Chen}},\ }\bibfield
  {title} {\bibinfo {title} {Foliated fracton order in the majorana
  checkerboard model},\ }\href {https://doi.org/10.1103/PhysRevB.100.085127}
  {\bibfield  {journal} {\bibinfo  {journal} {Phys. Rev. B}\ }\textbf {\bibinfo
  {volume} {100}},\ \bibinfo {pages} {085127} (\bibinfo {year}
  {2019})}\BibitemShut {NoStop}%
\bibitem [{\citenamefont {Prem}\ \emph {et~al.}(2019)\citenamefont {Prem},
  \citenamefont {Huang}, \citenamefont {Song},\ and\ \citenamefont
  {Hermele}}]{PremHuangSongHermele2019}%
  \BibitemOpen
  \bibfield  {author} {\bibinfo {author} {\bibfnamefont {A.}~\bibnamefont
  {Prem}}, \bibinfo {author} {\bibfnamefont {S.-J.}\ \bibnamefont {Huang}},
  \bibinfo {author} {\bibfnamefont {H.}~\bibnamefont {Song}},\ and\ \bibinfo
  {author} {\bibfnamefont {M.}~\bibnamefont {Hermele}},\ }\bibfield  {title}
  {\bibinfo {title} {Cage-net fracton models},\ }\href
  {https://doi.org/10.1103/PhysRevX.9.021010} {\bibfield  {journal} {\bibinfo
  {journal} {Phys. Rev. X}\ }\textbf {\bibinfo {volume} {9}},\ \bibinfo {pages}
  {021010} (\bibinfo {year} {2019})}\BibitemShut {NoStop}%
\bibitem [{\citenamefont {Thorngren}\ and\ \citenamefont
  {Else}(2018)}]{ThorngrenElse2018}%
  \BibitemOpen
  \bibfield  {author} {\bibinfo {author} {\bibfnamefont {R.}~\bibnamefont
  {Thorngren}}\ and\ \bibinfo {author} {\bibfnamefont {D.~V.}\ \bibnamefont
  {Else}},\ }\bibfield  {title} {\bibinfo {title} {Gauging spatial symmetries
  and the classification of topological crystalline phases},\ }\href
  {https://doi.org/10.1103/PhysRevX.8.011040} {\bibfield  {journal} {\bibinfo
  {journal} {Phys. Rev. X}\ }\textbf {\bibinfo {volume} {8}},\ \bibinfo {pages}
  {011040} (\bibinfo {year} {2018})}\BibitemShut {NoStop}%
\bibitem [{\citenamefont {Yoshida}(2015)}]{Yoshida2015}%
  \BibitemOpen
  \bibfield  {author} {\bibinfo {author} {\bibfnamefont {B.}~\bibnamefont
  {Yoshida}},\ }\bibfield  {title} {\bibinfo {title} {Topological color code
  and symmetry-protected topological phases},\ }\href
  {https://doi.org/10.1103/PhysRevB.91.245131} {\bibfield  {journal} {\bibinfo
  {journal} {Phys. Rev. B}\ }\textbf {\bibinfo {volume} {91}},\ \bibinfo
  {pages} {245131} (\bibinfo {year} {2015})}\BibitemShut {NoStop}%
\bibitem [{\citenamefont {Yoshida}(2016)}]{Yoshida2016}%
  \BibitemOpen
  \bibfield  {author} {\bibinfo {author} {\bibfnamefont {B.}~\bibnamefont
  {Yoshida}},\ }\bibfield  {title} {\bibinfo {title} {Topological phases with
  generalized global symmetries},\ }\href
  {https://doi.org/10.1103/PhysRevB.93.155131} {\bibfield  {journal} {\bibinfo
  {journal} {Phys. Rev. B}\ }\textbf {\bibinfo {volume} {93}},\ \bibinfo
  {pages} {155131} (\bibinfo {year} {2016})}\BibitemShut {NoStop}%
\bibitem [{\citenamefont {Yoshida}(2017)}]{Yoshida2017}%
  \BibitemOpen
  \bibfield  {author} {\bibinfo {author} {\bibfnamefont {B.}~\bibnamefont
  {Yoshida}},\ }\bibfield  {title} {\bibinfo {title} {Gapped boundaries, group
  cohomology and fault-tolerant logical gates},\ }\href@noop {} {\bibfield
  {journal} {\bibinfo  {journal} {Annals of Physics}\ }\textbf {\bibinfo
  {volume} {377}},\ \bibinfo {pages} {387} (\bibinfo {year}
  {2017})}\BibitemShut {NoStop}%
\bibitem [{\citenamefont {Potter}\ and\ \citenamefont
  {Morimoto}(2017)}]{PotterMorimoto2017}%
  \BibitemOpen
  \bibfield  {author} {\bibinfo {author} {\bibfnamefont {A.~C.}\ \bibnamefont
  {Potter}}\ and\ \bibinfo {author} {\bibfnamefont {T.}~\bibnamefont
  {Morimoto}},\ }\bibfield  {title} {\bibinfo {title} {Dynamically enriched
  topological orders in driven two-dimensional systems},\ }\href
  {https://doi.org/10.1103/PhysRevB.95.155126} {\bibfield  {journal} {\bibinfo
  {journal} {Phys. Rev. B}\ }\textbf {\bibinfo {volume} {95}},\ \bibinfo
  {pages} {155126} (\bibinfo {year} {2017})}\BibitemShut {NoStop}%
\bibitem [{\citenamefont {Po}\ \emph {et~al.}(2017)\citenamefont {Po},
  \citenamefont {Fidkowski}, \citenamefont {Vishwanath},\ and\ \citenamefont
  {Potter}}]{PoFidkowskiVishwanathPotter2017}%
  \BibitemOpen
  \bibfield  {author} {\bibinfo {author} {\bibfnamefont {H.~C.}\ \bibnamefont
  {Po}}, \bibinfo {author} {\bibfnamefont {L.}~\bibnamefont {Fidkowski}},
  \bibinfo {author} {\bibfnamefont {A.}~\bibnamefont {Vishwanath}},\ and\
  \bibinfo {author} {\bibfnamefont {A.~C.}\ \bibnamefont {Potter}},\ }\bibfield
   {title} {\bibinfo {title} {Radical chiral floquet phases in a periodically
  driven kitaev model and beyond},\ }\href
  {https://doi.org/10.1103/PhysRevB.96.245116} {\bibfield  {journal} {\bibinfo
  {journal} {Phys. Rev. B}\ }\textbf {\bibinfo {volume} {96}},\ \bibinfo
  {pages} {245116} (\bibinfo {year} {2017})}\BibitemShut {NoStop}%
\bibitem [{\citenamefont {Fidkowski}\ \emph
  {et~al.}(2019{\natexlab{a}})\citenamefont {Fidkowski}, \citenamefont {Po},
  \citenamefont {Potter},\ and\ \citenamefont
  {Vishwanath}}]{FidkowskiPoPotterVishwanath2019}%
  \BibitemOpen
  \bibfield  {author} {\bibinfo {author} {\bibfnamefont {L.}~\bibnamefont
  {Fidkowski}}, \bibinfo {author} {\bibfnamefont {H.~C.}\ \bibnamefont {Po}},
  \bibinfo {author} {\bibfnamefont {A.~C.}\ \bibnamefont {Potter}},\ and\
  \bibinfo {author} {\bibfnamefont {A.}~\bibnamefont {Vishwanath}},\ }\bibfield
   {title} {\bibinfo {title} {Interacting invariants for floquet phases of
  fermions in two dimensions},\ }\href
  {https://doi.org/10.1103/PhysRevB.99.085115} {\bibfield  {journal} {\bibinfo
  {journal} {Phys. Rev. B}\ }\textbf {\bibinfo {volume} {99}},\ \bibinfo
  {pages} {085115} (\bibinfo {year} {2019}{\natexlab{a}})}\BibitemShut
  {NoStop}%
\bibitem [{\citenamefont {You}\ \emph {et~al.}(2018{\natexlab{b}})\citenamefont
  {You}, \citenamefont {Litinski},\ and\ \citenamefont {von
  Oppen}}]{YouLitinskivonOppen2018}%
  \BibitemOpen
  \bibfield  {author} {\bibinfo {author} {\bibfnamefont {Y.}~\bibnamefont
  {You}}, \bibinfo {author} {\bibfnamefont {D.}~\bibnamefont {Litinski}},\ and\
  \bibinfo {author} {\bibfnamefont {F.}~\bibnamefont {von Oppen}},\ }\bibfield
  {title} {\bibinfo {title} {Higher order topological superconductors as
  generators of quantum codes},\ }\href@noop {} {\bibfield  {journal} {\bibinfo
   {journal} {arXiv preprint arXiv:1810.10556}\ } (\bibinfo {year}
  {2018}{\natexlab{b}})}\BibitemShut {NoStop}%
\bibitem [{\citenamefont {Freedman}\ and\ \citenamefont
  {Hastings}(2016)}]{FreedmanHastings2015}%
  \BibitemOpen
  \bibfield  {author} {\bibinfo {author} {\bibfnamefont {M.~H.}\ \bibnamefont
  {Freedman}}\ and\ \bibinfo {author} {\bibfnamefont {M.~B.}\ \bibnamefont
  {Hastings}},\ }\bibfield  {title} {\bibinfo {title} {Double semions in
  arbitrary dimension},\ }\href {https://doi.org/10.1007/s00220-016-2604-0}
  {\bibfield  {journal} {\bibinfo  {journal} {Commun. Math. Phys.}\ }\textbf
  {\bibinfo {volume} {347}},\ \bibinfo {pages} {389} (\bibinfo {year}
  {2016})},\ \Eprint {https://arxiv.org/abs/1507.05676} {arXiv:1507.05676}
  \BibitemShut {NoStop}%
\bibitem [{\citenamefont {Fidkowski}\ \emph
  {et~al.}(2019{\natexlab{b}})\citenamefont {Fidkowski}, \citenamefont {Haah},
  \citenamefont {Hastings},\ and\ \citenamefont
  {Tantivasadakarn}}]{Fidkowskietal2019}%
  \BibitemOpen
  \bibfield  {author} {\bibinfo {author} {\bibfnamefont {L.}~\bibnamefont
  {Fidkowski}}, \bibinfo {author} {\bibfnamefont {J.}~\bibnamefont {Haah}},
  \bibinfo {author} {\bibfnamefont {M.~B.}\ \bibnamefont {Hastings}},\ and\
  \bibinfo {author} {\bibfnamefont {N.}~\bibnamefont {Tantivasadakarn}},\
  }\bibfield  {title} {\bibinfo {title} {Disentangling the generalized double
  semion model},\ }\href@noop {} {\bibfield  {journal} {\bibinfo  {journal}
  {arXiv preprint arXiv:1906.04188}\ } (\bibinfo {year}
  {2019}{\natexlab{b}})}\BibitemShut {NoStop}%
\bibitem [{\citenamefont {Kitaev}(2003)}]{Kitaev2003}%
  \BibitemOpen
  \bibfield  {author} {\bibinfo {author} {\bibfnamefont {A.~Y.}\ \bibnamefont
  {Kitaev}},\ }\bibfield  {title} {\bibinfo {title} {Fault-tolerant quantum
  computation by anyons},\ }\href
  {https://doi.org/10.1016/S0003-4916(02)00018-0} {\bibfield  {journal}
  {\bibinfo  {journal} {Annals of Physics}\ }\textbf {\bibinfo {volume}
  {303}},\ \bibinfo {pages} {2} (\bibinfo {year} {2003})}\BibitemShut {NoStop}%
\end{thebibliography}%

\appendix

\section{SPT Calculation using Symmetry Defect Homology}\label{globalresults}

In this appendix, we calculate the symmetry defect homology for additional SPT phases with global symmetry. We confirm that the results without time-reversal agree with the classification using group cohomology. A general result in 1D is discussed in Appendix \ref{app:1DgeneralG}. The results are summarized in Table \ref{tab:global}.
\begin{table*}
\caption{Summary of SPT phases with global symmetry calculated using symmetry defect homology.}
\begin{tabular}{|c|c|c|c|c|c|c|c|}
\hline
\multirow{ 2}{*}{$d$} & \multirow{ 2}{*}{Lattice} & \multirow{ 2}{*}{Symmetry}  & \multicolumn{2}{c|}{Without $\mathcal T$} & \multicolumn{2}{c|}{With $\mathcal T$}   &Dual Topological Order\\
\cline{4-7}
&&& Weak &Strong & Weak & Strong & of trivial SPT\\
\hline
1 & &$\mathbb Z_2$  & $\mathbb Z_2$ & 0 &  $\mathbb Z_2$ &$\mathbb Z_2$ & --  \\
1 & &$\mathbb Z_2 \times \mathbb Z_2$   & $\mathbb Z_2^2 $&$\mathbb Z_2$ & $\mathbb Z_2^2 $ & $\mathbb Z_2^3$ & -- \\
2 & Triangular& $\mathbb Z_2$  & $\mathbb Z_2 $&$ \mathbb Z_2$ & $\mathbb Z_2^4$ &$\mathbb Z_2 $ & Toric Code\\
\hline
\end{tabular}
\label{tab:global}
\end{table*}

\subsection{1D chain with global $\mathbb Z_2 \times \mathbb Z_2$ symmetry}
Consider a 1D chain where the local Hilbert space consists of two qubits, each transforming under a $\mathbb Z_2$ symmetry, which we label as $(a)$ and $(b)$, respectively.

Without time-reversal, we find  $\mathbb Z_2^3$ phases, which is consistent with the cohomology group $H^2(\mathbb Z_2 \times \mathbb Z_2,U(1)) \times H^1(\mathbb Z_2 \times \mathbb Z_2,U(1))$. The first two generators correspond to weak phases where each unit cell is charged by one of the $\mathbb Z_2$ symmetries. The third generator can be represented by $M^{(a)}=XZ;XI$ and $M^{(b)}=IX;ZX$. This choice gives rise to hopping operators where a minus sign is gained whenever an $(a)$ domain wall passes through a $(b)$ domain wall, or a pair of domain wall gets created/annihilated with the other species in between. Its dual is given by $\tilde X^{(a)}=IZ;XZ$ and $\tilde X^{(b)}=ZX;ZI$. This is exactly the stabilizer for the 1D cluster state.

With an additional time-reversal, we find $\mathbb Z_2^2$ more phases. This corresponds to the ``Haldane chain" for each diagonal of the $\mathbb Z_2 \times \mathbb Z_2^T$ phase. Indeed, if time reversal is broken, then the hopping operators for the two generators can be trivialized with transversal $SI$ and $IS$ gates respectively.  For completion, the generators are summarized in terms of stabilizers in Table \ref{tab:1Dcluster}.

\begin{table*}[t]
\caption{Generators of translational invariant $\mathbb Z_2 \times \mathbb Z_2 \times \mathbb Z_2^T$ SPT phases in 1+1D from the consistency conditions. This does not include the intrinsically $\mathbb Z_2^T$ phases.}
\begin{tabular}{|c|c|c|c|c|c|}
\hline
\multicolumn{2}{|c|}{SPT} &\multirow{2}{*}{Comment} & \multicolumn{2}{c|}{SPT Dual} & \multirow{2}{*}{Comment}\\
\cline{1-2} \cline{4-5}
$\tilde X^{(a)}$ & $\tilde X^{(b)}$ & &  $M^{(a)}$ & $M^{(b)}$ & \\
\hline
$-XI$ & $IX$  & $\mathbb Z_2^{(a)}$ Crystalline &$-XI;XI$ & $IX;IX$& $\mathbb Z_2^{(a)}$ Crystalline SSB\\
$XI$ & $-IX$  & $\mathbb Z_2^{(b)}$ Crystalline &$XI;XI$ & -$IX;IX$& $\mathbb Z_2^{(b)}$ Crystalline SSB\\
$IZ;XZ$ & $ZX;ZI$ & $\mathbb Z_2^{(a)} \times \mathbb Z_2^{(b)}$ cluster state &$XZ;XI$ & $ZX; IX$& self-dual\\
\hhline{======}
 $-ZI;XI;ZI$ &$IX$ & $\mathbb Z_2^{(a)} \times \mathbb Z_2^T$ cluster state & $YI;YI$ & $IX;IX$ & $\mathbb Z_2^{(a)} \times \mathbb Z_2^T$ SSB to diagonal\\
 $XI$ &$-IZ;IX;IZ$ & $\mathbb Z_2^{(b)} \times \mathbb Z_2^T$ cluster state & $XI;XI$ & $IY; IY$ & $\mathbb Z_2^{(b)} \times \mathbb Z_2^T$ SSB to diagonal \\
\hline
\end{tabular}
\label{tab:1Dcluster}
\end{table*}

\subsection{2D global $\mathbb Z_2$}

We consider a product state on a triangular lattice with global $\mathbb Z_2$ symmetry. Under duality, this is mapped to a honeycomb lattice with sites on edges. The symmetry defects are domain walls, which are dual to strings, created by
\begin{equation}
M=
\raisebox{-.5\height}{\includegraphics[]{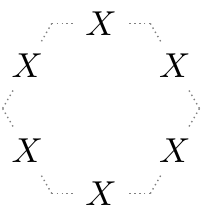}},
\end{equation}
 which excites six sites around each plaquette. We now search for all hopping terms of this nature, where plaquette terms must commute, but only in the constraint Hilbert space where a product of three $Z$ operators around a vertex is one. The results are shown in Table \ref{tab:2Dglobal}. Without time reversal, we find two generators. The first generator is a crystalline phase with $\mathbb Z_2$ charge in every unit cell. It is dual to a toric code where every unit cell contains a $\mathbb Z_2$ gauge charge $e$. The second generator is the Levin-Gu SPT. It is dual to the Double Semion model, although we remark that the form found in this search has no external ``legs", and is the form discussed in Refs. \cite{FreedmanHastings2015,Fidkowskietal2019}.
 
 With time-reversal, we find three more generators. They correspond to stacks of vertical 1D Haldane chains under the $\mathbb Z_2 \times \mathbb Z_2^T$ symmetries, and their $60^\circ$ and $120^\circ$ rotations.
 
 \begin{table*}
\caption{Generators of translation invariant $\mathbb Z_2\times \mathbb Z_2^T$ SPT phases in 2+1D from the consistency conditions.}
\begin{tabular}{|c|c|c|c|}
\hline
SPT ($\tilde X$) &Comment & SPT Dual ($M$) & Comment\\
\hline
$-X$ & charge-decorated & $-$ \raisebox{-.5\height}{\includegraphics[]{TC1.pdf}}
& $e$-decorated crystalline Toric Code\\
\hline
 $-$
 \raisebox{-.5\height}{\includegraphics[]{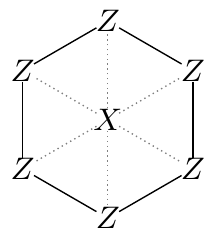}}
& Levin-Gu\cite{LevinGu2012} &
\raisebox{-.5\height}{\includegraphics[]{TC1.pdf}}
$\times$
\raisebox{-.5\height}{\includegraphics[]{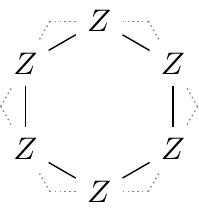}}
& Double Semion model\cite{LevinWen2005,FreedmanHastings2015}\\
\hhline{====}
$-$
\raisebox{-.5\height}{\includegraphics[]{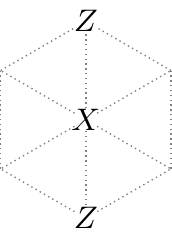}}
& Haldane chain decoration &
\raisebox{-.5\height}{\includegraphics[]{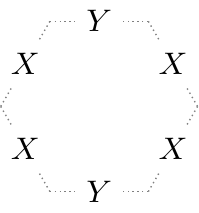}}
& \\
\hline
$-$
\raisebox{-.5\height}{\includegraphics[]{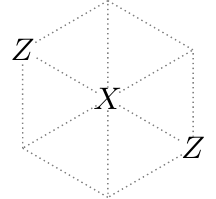}}
&  Haldane chain decoration &
\raisebox{-.5\height}{\includegraphics[]{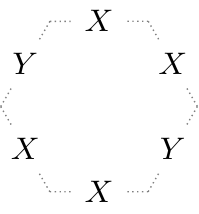}}
& \\
\hline
$-$
\raisebox{-.5\height}{\includegraphics[]{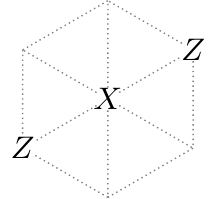}}
& Haldane chain decoration &
\raisebox{-.5\height}{\includegraphics[]{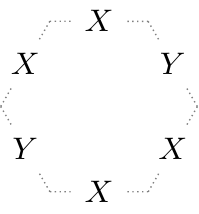}}
& \\
\hline
\end{tabular}
\label{tab:2Dglobal}
\end{table*}

\section{Consistency with Group Cohomology in 1D}\label{app:1DgeneralG}
In this Appendix, we extend the formalism to a general finite group $G$ in 1D. We show that the group cohomology classification for translation invariant SPT's given by $H^2(G,U(1)) \times H^1(G,U(1))$ is contained as a solution of the symmetry defect homology \eqref{equ:symdefecthomology} by explicitly constructing a hopping matrix $M$ which satisfies the consitency conditions. The duality presented here is identical to that used to obtain the quantum double model in 2D\cite{Kitaev2003}. See Ref. \onlinecite{Yoshida2017} for a review.

First, the Hilbert space is labeled by group elements $g\in G$, with symmetry $L^g_+$ acting via the left regular representation 
\begin{equation}
   L^g_+ = \sum_h \ket{gh}\bra{h}
\end{equation}
The product state is an equal superposition of all states
\begin{equation}
   \ket{\psi_0} = \frac{1}{\sqrt{|G|}} \sum_g \ket{g}
\end{equation}
and so the product state Hamiltonian is given by
\begin{equation}
   H= -\sum_i \ket{\psi_0}_i \bra{\psi_0} =-\sum_i \left(\frac{1}{|G|}\sum_g L^g_+\right)_i.
\end{equation}
The duality at the level of wavefunctions is given by the gauging map
\begin{equation}
\ket{g}_i \otimes \ket{h}_{i+1} \longleftrightarrow \ket{gh^{-1}}_{i+1},
\end{equation}
which induces the following map of operators
\begin{equation}
\left(L^g_+ \right)_i \longleftrightarrow \left(L^g_- \right)_i \otimes \left(L^g_+ \right)_{i+1},
\end{equation}
where $ L^g_- = \sum_h \ket{hg^{-1}}\bra{h}$.

Hence, acting with a symmetry $g$ on a site creates and fuses a $g^{-1}$ domain wall to the left site, and a $g$ domain wall to the right site in the dual picture. The dual Hamiltonian is
\begin{align}
H_\text{gauged} &= -\sum_i \left[ \frac{1}{|G|} \sum_g \left(L^g_- \right)_i \otimes \left(L^g_+ \right)_{i+1} \right]\\
&=-\sum_i \left[ \frac{1}{|G|} \sum_g \left (\sum_{h,k} \ket{hg^{-1}}_i\bra{h} \otimes \ket{gk}_{i+1}\bra{k} \right) \right] \nonumber
\end{align}

To construct other symmetric states, we define the more general hopping matrix of domain walls
\begin{align}
M^g_{i,i+1} &=  \sum_{h,k} e^{i\phi_{h,k \rightarrow hg^{-1},gk}}  \ket{hg^{-1}}_i\bra{h} \otimes \ket{gk}_{i+1}\bra{k}.
\end{align}

The corresponding gauged Hamiltonian is
\begin{align}
H'_\text{gauged} &= -\sum_i P_i,
\end{align}
where
\begin{align}
  P_i =\frac{1}{|G|} \sum_g M^g_{i,i+1}.
\end{align}
We further constrain the hopping matrix to make $H'_\text{gauged}$ a commuting projector, and also to make the problem tractable. We consider the following conditions
\begin{align}
\label{equ:constrain1}
M^g_{i,i+1}M^h_{i,i+1}&=M^{gh}_{i,i+1},\\
[M^g_{i-1,i},M^h_{i,i+1}]&=0.
\label{equ:constrain2}
\end{align}
Constraint \eqref{equ:constrain1} implies that  $P^2_i=P_i$, while constraint \eqref{equ:constrain2} implies $[P_{i-1},P_i]=0$. Thus, $H'_\text{gauged}$ is a commuting projector Hamiltonian.

Now, given a class $[\omega] \in H^2(G,U(1))$ and $[\alpha] \in H^1(G,U(1))$, we can always pick representative ``canonical" cocycles such that
\begin{align}
 \omega(1,1)&=\omega(1,g)=\omega(g,1)=1,\\
 \alpha(1)&=1.
\end{align}
A proof of the above statement can be found in Appendix J of Ref. \onlinecite{Chenetal2013}.

We will now show that for each $M^g$, the phase factors defined as
\begin{align}
e^{i\phi_{h,k \rightarrow hg^{-1},gk}} =\frac{ \omega(h,g^{-1})\omega(g,k)\alpha(g)}{\omega(g^{-1},g)}
\label{equ:phitococycles}
\end{align}
satisfies all the constraints given above. Physically, this sign is obtained from creating domain walls $g^{-1}$ and $g$ from the vacuum, fusing $g^{-1}$ with $h$ from the right, and hopping $g$ to the right to fuse with $k$ from the left.

To show constraint \eqref{equ:constrain1}, it suffices to show that on a general state $\ket{a}_i \otimes \ket{b}_{i+1}$,
\begin{equation}
M^g_{i,i+1}M^h_{i,i+1} \ket{a}_i \otimes \ket{b}_{i+1}=M^{gh}_{i,i+1} \ket{a}_i \otimes \ket{b}_{i+1}.
\end{equation}
This requires
\begin{equation}
e^{i\phi_{a,b \rightarrow a h^{-1},h b}} e^{i\phi_{ a h^{-1},h b \rightarrow a h^{-1}g^{-1},g h b}} = e^{i\phi_{a,b \rightarrow  a h^{-1}g^{-1},g h b}}
\end{equation}
Substituting Eq. \eqref{equ:phitococycles}, we find
\begin{align}
&\frac{ \omega(a,h^{-1})\omega(h,b)\alpha(h)}{\omega(h^{-1},h)} \frac{ \omega(ah^{-1},g^{-1})\omega(g,hb)\alpha(g)}{\omega(g^{-1},g)} \nonumber \\
&= \frac{ \omega(a,h^{-1}g^{-1})\omega(gh,b)\alpha(gh)}{\omega(h^{-1}g^{-1},gh)}
\end{align}
Using the cocycle conditions of $\omega$ and $\alpha$ in Eqs. \eqref{equ:cocyclecondition} and \eqref{equ:cocyclealpha}, one finds that the above equation is indeed satisfied.

To show constraint \eqref{equ:constrain2}, consider a general state $\ket{a}_{i-1} \otimes \ket{b}_{i} \otimes \ket{c}_{i+1}$. It suffices to show that
\begin{align}
&e^{i\phi_{a,b \rightarrow a g^{-1},g b}} e^{i\phi_{ gb ,c \rightarrow gbh^{-1},hc}} \nonumber\\
&= e^{i\phi_{b,c \rightarrow  b h^{-1},h c}} e^{i\phi_{a,bh^{-1} \rightarrow  ag^{-1}, gbh^{-1}}}
\end{align}

Substituting Eq. \eqref{equ:phitococycles}, we find that it needs to satisfy
\begin{equation}
\omega(g,b)\omega(gb,h^{-1})=\omega(b,h^{-1})\omega(g,bh^{-1}),
\end{equation}
which is just an instance of the cocycle condition \eqref{equ:cocyclecondition}.

\section{Ungauging Phase Gates}
A general multiply controlled-phase gate over $N$ qubits has the form
\begin{equation}
\exp \left[ 2\pi i \phi \prod_{i=1}^N \frac{1-Z_i}{2}  \right]
\end{equation}
Since it is written in terms of Pauli $Z$ operators, we can apply the ungauging map to map $Z$ into a product of $Z$'s. We can then expand $Z_j = 1-2g_j$ where $g_j=0,1$ to rewrite it as a product of phase gates.

When the controlled phase gate is in the $N^\text{th}$ level of the Clifford hierarchy $\mathcal C_N$, ungauging it will only give a product of $\mathcal C_N$ phase gates.

As an example, let us ungauge the $S$ gate in the Xu-Moore duality Eq. \eqref{equ:XuMooreDuality}. First, we write the $S$ gate in terms of $Z$ as
\begin{equation}
    S= e^{\frac{\pi i}{2}(1-Z)}
\end{equation}
So ungauging gives
\begin{equation}
\exp \frac{\pi i}{2} \left [1-\begin{array}{cc}
Z_1&Z_2\\
Z_3&Z_4
\end{array} \right ]
\end{equation}
To properly expand them as controlled phase gates, we use $Z_j = 1-2g_j$ and expand. This gives
\begin{equation}
\exp \frac{\pi i}{2} \left[ \sum_{i=1}^4 g_i + 2 \sum_{i<j} g_ig_j \right] = \prod_{i=1}^4 S_i \prod_{i<j} CZ_{ij}
\end{equation}
Thus, pictorially, the gauging map in the Xu-Moore duality is given by
\begin{equation}
\raisebox{-.5\height}{\begin{tikzpicture}
\node[label=center:$S$] (1) at (0,0) {};
\node[label=center:$S$] (9) at (1,1) {};
\node[label=center:$S$] (12) at (1,0) {};
\node[label=center:$S$] (14) at (0,1) {};
\draw[-] (1) -- (9) node[midway, below] {};
\draw[-] (1) -- (12) node[midway, below] {};
\draw[-] (1) -- (14) node[midway, below] {};
\draw[-] (12) -- (9) node[midway, below] {};
\draw[-] (14) -- (9) node[midway, below] {};
\draw[-] (12) -- (14) node[midway, below] {};
\end{tikzpicture}}\longleftrightarrow S.
\end{equation}

By nature of the duality, one can verify that this ungauged $S$ gate commutes with the horizontal and vertical line symmetries.

As an application, consider the generating SSPT phase in the last row of Table \ref{tab:2D}, which is protected by line symmetries in and a global time-reversal in 2D. The hopping matrix $M$ can be disentangled to the trivial one using a transversal $S$ gate in the absence of time-reversal. By ungauging this, we obtain a symmetric local unitary, which disentangles the corresponding cluster state.

\section{Degeneracy of a 3D SSPT with Line Symmetries}\label{app:3Dlinecalc}
In this Appendix, we use algebraic techniques, developed by Ref. \onlinecite{Haah2013} to study the degeneracy of a 3D SSPT introduced in the main text.  The Hamiltonian that describes this SSPT is a stabilizer code, which can be described by the generating map $\sigma \in \mathbb{F}_{2}[x,y,z]^{2}$ given by
\begin{align}
\sigma = \begin{pmatrix}f(x,y,z)\\g(x,y,z)\end{pmatrix},
\end{align}
where
\begin{align}
f(x,y,z) &\equiv 1 + x + y +z+ xy + yz + xz + xyz,\\
g(x,y,z) &\equiv 1 + xyz.
\end{align}
Since this model has a single interaction term per lattice spin, by Corollary 4.5 of Ref. \onlinecite{Haah2013}, the ground-state degeneracy on the three-torus with length $L$ is $D = 2^{k}$, where
\begin{align}
k = \mathrm{dim}_{\mathbb{F}_{2}}\left[ \frac{\mathbb{F}_{2}[x,y,z]}{\langle f(x,y,z), g(x,y,z), x^{L}-1, y^{L} - 1, z^{L}-1\rangle}\right]\nonumber
\end{align}
We calculate this quantity by computing the Gr\"{o}bner basis for the ideal $\langle f(x,y,z), g(x,y,z), x^{L}-1, y^{L} - 1, z^{L}-1\rangle$.  In the lexicographic order, where $x>y>z$, this basis is
\begin{align}
\mathrm{GB} = &\langle 1+z^{L}, 1+y^{L}, 1 + yz + y^{2}(1+z) + z^{L-1}(1+y), \nonumber\\
&x + y + z + yz + y^{L-1} + z^{L-1}\rangle
\end{align}
The leading terms in each generator are then given by $z^{L}$, $y^{L}$, $y^{2}z$, and $x$, respectively.  Each monomial that lies outside of the ideal generated by the leading terms contributes a factor of two to the ground-state degeneracy.  As a result, we conclude that
\begin{align}
k = \log_{2}D =  3L-2.
\end{align}

\end{document}